\documentclass[aps,pre,reprint,twocolumn,superscriptaddress,showpacs]{revtex4-2}
\usepackage{amssymb,amsmath}
\usepackage[table]{xcolor}
\usepackage{graphicx}
\usepackage{mathtools}
\usepackage[export]{adjustbox}
\usepackage{overpic}
\usepackage[colorlinks=true,
linkcolor=black,
urlcolor=blue,
citecolor=blue]{hyperref}
\usepackage{nicefrac}
\usepackage{multirow}
\usepackage{lipsum}
\usepackage[normalem]{ulem}
\usepackage{pbox}
\newcolumntype{C}[1]{>{\centering\let\newline\\\arraybackslash\hspace{0pt}}m{#1}}
\usepackage{verbatim}
\usepackage{enumitem}

\usepackage{framed,color}
\definecolor{shadecolor}{rgb}{0.85,0.80,0.80}

\definecolor{myorange}{RGB}{253, 184, 99}
\definecolor{mypurple}{RGB}{178, 171, 210}

\newcommand{\comments}[1]{}
\usepackage{multirow}

\usepackage{float}

\newcommand{\avg}[1]{\left\langle{#1}\right\rangle}

\newcommand{\beq}{\begin{equation}}
	\newcommand{\eeq}{\end{equation}}
\newcommand{\bal}{\begin{aligned}}
	\newcommand{\eal}{\end{aligned}}

\newcommand{\be}{\begin{equation}}
	\newcommand{\ee}{\end{equation}}
\newcommand{\bd}{\begin{displaymath}}
	\newcommand{\ed}{\end{displaymath}}
\newcommand{\BE}{\begin{eqnarray}}
	\newcommand{\EE}{\end{eqnarray}}

\allowdisplaybreaks

\begin{document}
	\title{The disordered logistic map}
	\author{Joseph W. Baron}
	\email{jwb96@bath.ac.uk}
	\affiliation{Department of Mathematical Sciences, University of Bath, Bath, BA2 7AY, UK}
	
	\author{Tobias Galla}
	\email{tobias.galla@ifisc.uib-csic.es}
	\affiliation{Instituto de F{\' i}sica Interdisciplinar y Sistemas Complejos IFISC (CSIC-UIB), 07122 Palma de Mallorca, Spain}

	\begin{abstract}
		 The logistic map is a quintessential model in the study of low-dimensional chaos. High-dimensional chaos, on the other hand, presents itself in disordered systems with many interacting and heterogeneously coupled components. Here, we formulate a system of many logistic maps, interacting through disordered couplings. Using a combination of dynamic mean-field theory, random matrix theory and numerical simulations, we show that even the smallest amount of disorder can remove the period-doubling cascade in the conventional logistic map. Instead we find a transition to high-dimensional chaos, marked by an oscillatory instability not previously reported for disordered systems.  We also show that with sufficiently strong homogeneous coupling between the maps one recovers elements of the conventional period-doubling cascade. Our findings indicate that well-known phenomena in dynamical systems can be fragile in the face of disorder. At the same time, new phenomena emerge that are neither found in simple low-dimensional dynamics nor in high-dimensional disordered systems.
	\end{abstract}

	\maketitle

	The realisation that simple mathematical models with only one or a few degrees of freedom can display complicated behaviour and sensitivity to initial conditions came as surprise in the 1960s  \cite{lorenz1963,may1976simple}. The implications for the prediction of natural processes are profound, and the theory of dynamical systems and chaos now pervades a vast number of fields including physics, the atmospheric sciences, engineering, chemistry, population dynamics, economics and the social sciences \cite{kiel1997chaos, gleick2008chaos,strogatz2024nonlinear, mandelbrot2013fractals}. Perhaps the simplest model showing chaos is the paradigmatic logistic map $x(t+1)=rx(t)[1-x(t)]$. As the bifurcation parameter $0<r<4$ is increased, the map undergoes a period-doubling cascade, and ultimately chaos sets in at $r=r_c \approx 3.56995$ \cite{GrossmannThomae+1977+1353+1363}. The period-doubling sequence is quantitatively `universal' across unimodal maps, and is characterised by Feigenbaum's constant \cite{feigenbaum1978, strogatz2024nonlinear}.  

	Instability and chaos can also occur in high-dimensional disordered dynamical systems. These are systems in which interaction coefficients are drawn at random in the beginning and then remain fixed throughout the dynamics. Models of this type are used to describe spin glasses, neural networks, and financial markets among others \cite{mezard1987, charbonneau2023spin}. 
    
    One famous and influential example for a disordered system is that given by Robert May in ecology. In his seminal paper titled ``Will a large complex system be stable?''~\cite{may}, May was able to determine the stability of a model ecosystem with fixed random interactions about its fixed-point equilibrium.  Using results from the then-nascent field of random matrix theory (RMT) \cite{mehta2004random, tao2012topics, taovu2010, taovukrishnapur2010}, he made the salient observation that a sharp dynamic transition was approached as the variance of the interaction coefficients was increased, whereupon the fixed point of the system became unstable.
    
    The random matrix theory ideas exploited by May go back to Wigner in nuclear physics \cite{wigner1958distribution, wigner1967random}. RMT characterises the eigenvalue spectra of large random matrices in terms of the statistics of the matrix entries, and thus permits the stability analysis of systems like May's. The abrupt dynamical transition at a critical variance of interaction coefficients is now known as the May--Wigner transition~\cite{hastings1982may}. This type of transition has been found in a number of nonlinear disordered systems across disciplines, including neural networks \cite{sompolinsky1988chaos, molgedey1992suppressing}, ecology and game dynamics \cite{rieger1989solvable, opper1992phase, coolen2005mathematical, galla2013complex}, and machine learning \cite{couillet2022random, pennington2017nonlinear}. 
    
    Importantly, the dynamics of these disordered systems beyond the instability is often chaotic with attractors of extensive dimension \cite{Engelken2023, martorell2025ergodicity}, and is characterised by the dominance of slow modes \cite{opper1992phase}. This is in stark contrast to the type of chaos exhibited by the logistic map, which is necessarily low-dimensional, and has a broad spectrum of modes, including rapid oscillations \cite{awrejcewicz2018quantifying}. The onset of chaos in disordered dynamical systems can be identified using techniques from the theory of disordered systems, in particular dynamic mean field theory \cite{dedominicis1978dynamics, kirkpatrick1987, sompolinsky1982relaxational} and the replica method \cite{edwardsjones}. 
    
	In this letter, we introduce the disordered logistic map. By linking a large number of logistic map processes through a disordered coupling term, we combine the characteristics of a low-dimensional chaotic map with the possibility of disorder-induced chaos in high dimensions. While the possibility of a May--Wigner transition to high-dimensional chaos persists, we also observe a number of interesting effects that are possessed by neither the traditional logistic map nor previously studied disordered systems. We find that even the smallest amount of disordered coupling drastically changes the stability behaviour of the logistic map, and that it can remove its characteristic period-doubling cascade. Instead of the period doubling, we find a direct transition to a new high-dimensional high-frequency chaotic state, which is in a sense a combination of the high-dimensionality characteristic of disordered systems and the rapid oscillations of the logistic map. Elements of the period-doubling cascade are only restored when a sufficient degree of homogeneous coupling is introduced. 
    
    The disordered logistic map therefore constitutes a simple yet phenomenologically rich model, exhibiting three distinct chaotic behaviours:  period doubling, May-Wigner slow-mode chaos, and a third phase that incorporates aspects of the other two. This suggests that there could yet be a plethora of chaotic behaviours in models with disorder that have yet to be characterised, but now they can, owing to recent advances in disordered-systems theory and RMT.

	\medskip
	
	\noindent{\em The disordered logistic map.}
	We study the following system of coupled discrete-time maps, 
	\begin{align}
	x_i(t+1) = H \left[ r x_i(t) \bigg\{1-x_i(t)+ \sum_{j} a_{ij} x_j(t) \bigg\}\right], \label{eq:dis_log_map}
	\end{align}
 where $i=1,\dots, N$. The $a_{ij}$ are the interaction coefficients. We set $H(u)=u$ for $0\leq u \leq 1$ so that the dynamics for each of the $x_i$ reduces to the conventional logistic map if all $a_{ij}$ are equal to zero. To constrain the $x_i$ to the unit interval, we also set $H(u)=0$ for $u\leq 0$, and $H(u)=1$ for $u\geq 1$. We call the variables $x_i$ the components of the coupled-map system.  The initial conditions for the map are always chosen such that $0<x_i(t=0)<1$ for all $i$. 
	
	The $a_{ij}$ are random interaction coefficients that remain fixed throughout the dynamics, and are thus the quenched disorder in the problem. As is common for disordered systems, it suffices only to know the first and second moments of the $a_{ij}$. Writing $\avg{\cdots}$ for the average over the disorder, we first focus on the situation in which $\avg{a_{ij}}=0$ , and we introduce
	\be\label{eq:variance}
	 \langle a_{ij}^2 \rangle =\frac{\sigma^2}{N}.
	\ee
 The parameter $\sigma>0$ sets the magnitude of the disorder. For $\sigma=0$, we have $a_{ij}=0$ for all $i,j$, and Eq.~(\ref{eq:dis_log_map}) reduces to $N$ uncoupled copies of the conventional logistic map. The scaling with system size in Eq.~(\ref{eq:variance}) ensures a sensible limit $N \to \infty$ \cite{mezard1987}. For simplicity, we assume that there are no correlations among the $a_{ij}$. 

\medskip

\begin{figure}[t]
		\centering
		\includegraphics[width=1\linewidth]{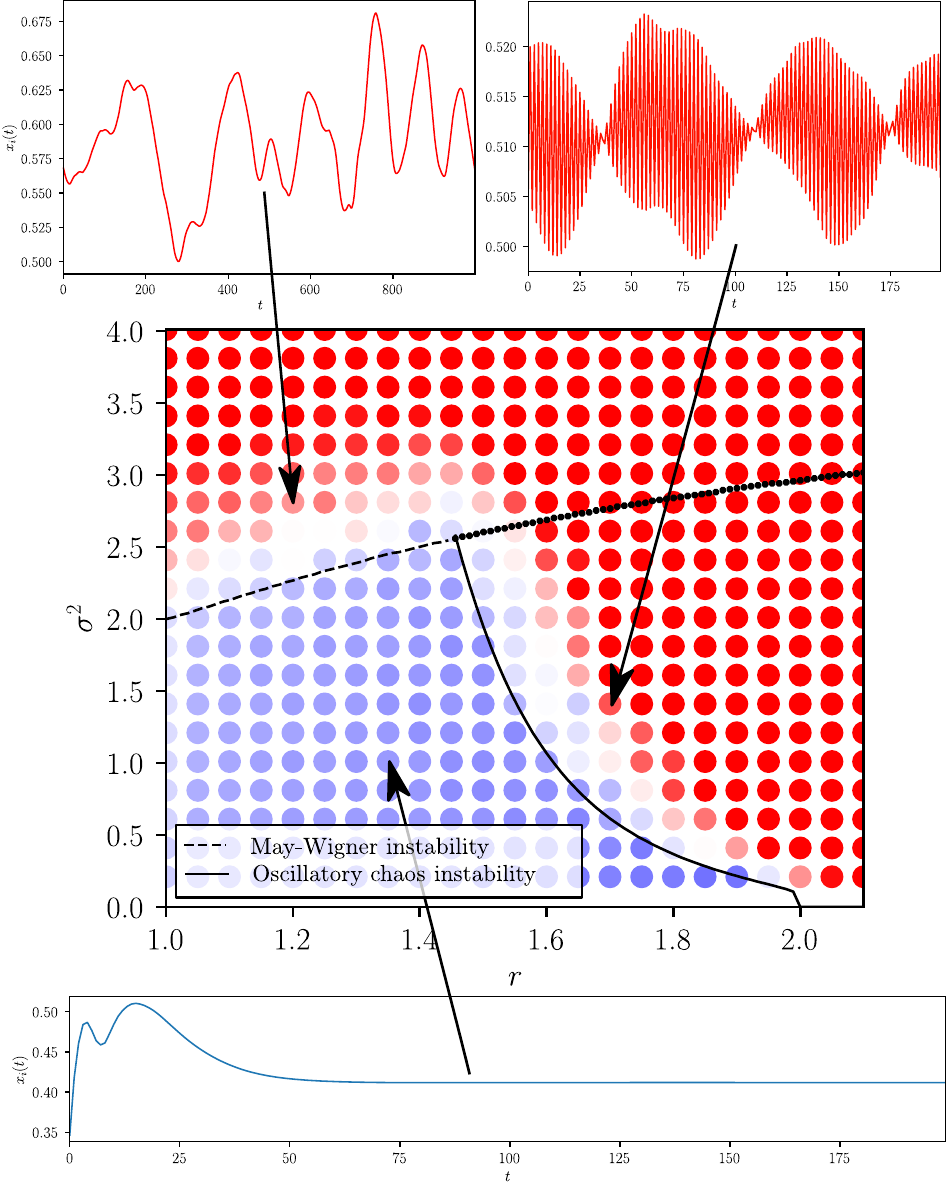}
		\caption{Stability diagram for the disordered logistic map in Eq.~(\ref{eq:dis_log_map}). The dashed line is the onset of the May--Wigner instability, where the eigenvalue spectrum of the reduced Jacobian spreads beyond $\lambda=1$. We continue this line as a series of dots away from the fixed-point phase, where it is no longer strictly valid, but is used to distinguish regions of the diagram with different behaviours. The solid line indicates the onset of the oscillatory chaotic (OC) instability (left-most bulk eigenvalue crossing $-1$). We find a unique stable fixed point for fixed realisations of the disorder in the stable phase to the lower left (blue region). Typical trajectories beyond the two instabilities are illustrated in the insets. The sign of the leading Lyapunov exponent $\Lambda_1$ (obtained from simulations with $N=4000$) is indicated by colour (red corresponding to $\Lambda_1>0$, blue to $\Lambda_1<0$ and white to $\Lambda_1=0$). We atttribute the differences between theory and simulation just above the predicted instability lines to finite-size effects.} 
		\label{fig:phase_diagram}
	\end{figure}

\noindent {\em Simulations and phase diagram.} We have simulated the disordered logistic map, varying the bifurcation parameter $r$ and the strength of the disorder $\sigma$. To identify chaotic behaviour we measured the largest Lyapunov exponent (LLE). Details of the numerical methods can be found in the End Matter. Results are summarised in Fig.~\ref{fig:phase_diagram}, in which we find three different types of dynamics. 

In a clearly defined region in parameter space, the map converges to stable fixed points. This is indicated by the blue colour in Fig.~\ref{fig:phase_diagram}, and occurs at low values of the bifurcation parameter and the disorder strength. As described further below, we are able to characterise this region analytically, and to determine when stability breaks down. 

We also observe two types of chaotic behaviour. The first is found when we increase the bifurcation parameter $r$ sufficiently (i.e., moving horizontally in Fig.~\ref{fig:phase_diagram}). The system transitions to a state with chaotic dynamics (positive LLE, red colour in the figure), characterised by high-frequency oscillations. We refer to this as the oscillatory chaotic instability from here on. A different type of chaotic behaviour is found if we start at sufficiently low $r$ in the stable phase and then increase the disorder strength $\sigma$ to move vertically in the stability diagram. In the upper-left part of the figure the LLE is again positive, but typical trajectories are more smooth, and we see no signs of rapid oscillations.

	\medskip

    \begin{figure}[t]
		\centering
		\includegraphics[width=0.75\linewidth]{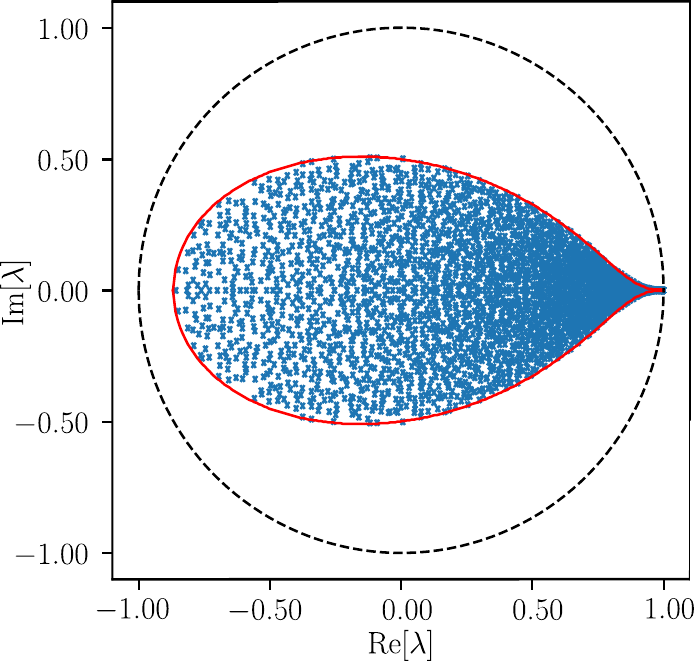}
		\caption{Eigenvalue spectrum of the reduced Jacobian matrix in the fixed-point regime. Blue crosses are the result of a single numerical experiment ($N=4000$), the solid red line is from Eq.~(\ref{eq:main_boundary}). In the stable phase the spectrum of the reduced Jacobian comes arbitrarily close to the point $+1$ in the complex plane (right-hand side tip of the spectrum). The fixed-point solution becomes unstable when eigenvalues stray outside the unit circle (dashed black line). System parameters: $r = 1.6$, $\sigma = 0.8$.  }
		\label{fig:eigenvaluespectrum}
	\end{figure}
	
	\noindent {\em Analytical characterisation.}
	We now summarise how the fixed points and the breakdown of their stability can be studied analytically. Full details of those calculations can be found in the Supplemental Material (SM) \cite{sm}.
    
    Fixed points of the disordered logistic map can be shown to fulfill $x_i^\star=H\left(1-\frac{1}{r}+\sum_j a_{ij}x_j^\star\right)$. Thus, at a fixed point, only those components for which the argument inside $H$ is between zero and one take values such that $0<x_i^\star<1$. Otherwise, the components are confined to the boundaries of this range ($x_i^\star=0$ or $x_i^\star=1$). We will refer to these latter components as `frozen', as these degrees of freedom do not contribute to any instabilities.  A similar `freezing' of degrees of freedom occurs through species extinction in disordered Lotka--Volterra systems \cite{bunin2017,Galla_2018}, or through the saturation of interactions with functional response \cite{sidhomgalla}. We write $\phi$ for the fraction of components that are not frozen, and $P(x^\star)$ for the distribution of the non-frozen components ($0<x^\star<1$), such that $\int_0^1 dx^\star\,P(x^\star)=\phi$. Via the central limit theorem, the distribution $P(x^\star)$ of non-frozen components can be seen to be Gaussian, truncated at $x^\star=0$ and $x^\star=1$ (see End Matter).
	
	As was observed in Refs. \cite{stone, baron2023breakdown}, stability of the fixed-point solution is governed by the eigenvalues of the \textit{reduced} Jacobian matrix (the Jacobian restricted to the $\phi N$ non-frozen components). The fixed point is stable if all eigenvalues of the reduced Jacobian are within the unit disk in the complex plane, and unstable otherwise. The spectrum forms a drop-shaped continuous region in the complex plane. An example  is shown in Fig.~\ref{fig:eigenvaluespectrum}.

    Using established methods from RMT  \cite{baron2020dispersal, baron2022eigenvalue, sommers, haake} we can compute the boundary of the eigenvalue spectrum in the limit $N\to\infty$. Writing the eigenvalues as $\lambda=\lambda_x+i\lambda_y$, this boundary is given by
	\begin{align}
		r^2 \sigma^2\int_0^1 dx^\star \,P(x^\star) \frac{ (x^\star)^2}{(\lambda_x + r x^\star -1)^2 + \lambda_y^2} = 1. \label{eq:main_boundary}
	\end{align}

    From this condition, we are able to find the sets of parameters $r$ and $\sigma^2$ for which the spectrum leaves the unit disk either at $\lambda = 1$ or at $\lambda=-1$, and thus a breakdown of stability occurs. We find no evidence of an eigenvalue leaving at any other point on the unit circle. We conclude that there are two different types of instability, in-line with the two types of chaotic behaviour in Fig.~\ref{fig:phase_diagram}. We now briefly discuss the distinct natures of these two types of instability.
    
    A detailed analysis (reported in the SM) shows that the spectrum of the reduced Jacobian comes arbitrarily close (from the left) to the point $\lambda=1$, for all values of $r$ and $\sigma^2$ in the stable phase, as seen in Fig.~\ref{fig:eigenvaluespectrum}. The condition for the onset of linear instability at $\lambda=1$ is found to be $\phi\sigma^2=1$. Beyond this the spectrum strays past the point $\lambda=1$. This is the discrete-time analogue of the May--Wigner instability.
	
	The condition for the left-most eigenvalue to exit the unit circle can be obtained by setting $\lambda=-1$ in Eq.~(\ref{eq:main_boundary}). Unlike the right-hand side of the eigenvalue spectrum, the left edge does not touch unit circle for parameters in the stable phase. When the instability at $-1$ occurs, deviations from the fixed point change sign on every time step, in contrast to the smooth deviations of the May--Wigner instability. That is, the system has oscillations with the highest possible angular frequency ($\omega=\pi$) for a discrete-time map. This is the \textit{oscillatory} chaotic (OC) instability.
    
    The instability lines obtained by setting $\lambda = \pm 1$ in Eq.~(\ref{eq:main_boundary}) are plotted in Fig.~\ref{fig:phase_diagram}, and match well with the numerical data for the LLE. The two transitions can also be found from dynamic mean-field theory (DMFT) \cite{dedominicis1978dynamics}. Following the lines of \cite{opper1992phase}, if one adds noise to the dynamics, the transitions can be detected as divergences of the power spectrum of fluctuations (see End Matter). We find that the fluctuation spectrum diverges at frequency $\omega=0$ near the May--Wigner transition, and at $\omega=\pi$ near the OC transition. Notably, the power spectrum near the May--Wigner instability can diverge with different exponents $\omega^{-\alpha}$ depending on the input noise. At the OC instability, the divergence is instead always of the type $(\omega-\pi)^{-2}$.

\medskip

    {\em Minimal disorder disrupts the period-doubling cascade of the logistic map.} Simulations and analytical theory both show that the OC phase line approaches the point $r=2$ as  $\sigma \to 0^+$ (see Fig.~\ref{fig:phase_diagram}). Thus, the disordered map at any $\sigma>0$ is chaotic as soon as $r>2$. This is markedly different from the behaviour of the the conventional logistic map which converges to a stable fixed point up to $r=3$, and then undergoes its well-known period-doubling cascade. The disordered logistic map system instead transitions directly from fixed points to chaos, as is confirmed by the LLE measurements in Fig. \ref{fig:phase_diagram}. More detailed simulations reported in the SM provide further evidence for the absence of cycles for large $N$. We conclude that even minimal disorder removes the period-doubling cascade in the thermodynamic limit. 
    
    To further characterise the transition to chaos, we have measured the Kaplan--Yorke dimension of the chaotic attractor. We find that the attractor dimension is extensive ($D_{KY}\propto N$) when either the May--Wigner or the OC instabilities are crossed, and thus the observed chaos is high-dimensional. We attribute the absence of the period-doubling cascade, and the high dimensionality of the attractor to the fact that it is a bulk eigenvalue spectrum that exits the unit disk, as opposed to an isolated eigenvalue. We now discuss the difference in behaviour when it is a single eigenvalue that controls the instability.

    \medskip
	
	\noindent {\em Homogeneous coupling restores period doubling.} So far, due to the introduction of the disordered couplings between the maps, we have seen no hint of the usual period-doubling behaviour that is expected of the conventional logistic map. By including an homogeneous coupling, which one might expect to promote collective behaviour and ameliorate the effects of disorder, we see the re-emergence of some elements of period doubling. Specifically, we now consider
	\begin{align}
		x_i(t+1)& = H \bigg[ r x_i(t) \bigg\{1-(1-\mu) x_i(t)\nonumber \\
		&- \frac{\mu}{N}\sum_{j} x_j + \sum_{j} a_{ij} x_j(t) \bigg\}\bigg].\label{eq:dis_log_map_homogeneous}
	\end{align}
The term proportional to $N^{-1}\sum_j x_j$ is the homogeneous coupling, and $0\leq \mu<1$  characterises its strength. The above form has been chosen such that, for any $\mu$, the dynamics reduces to the standard logistic map when $\sigma=0$ and when all components are started from the same initial value.

	\begin{figure}[t!]
		\centering  
		\includegraphics[width=1.0\linewidth]{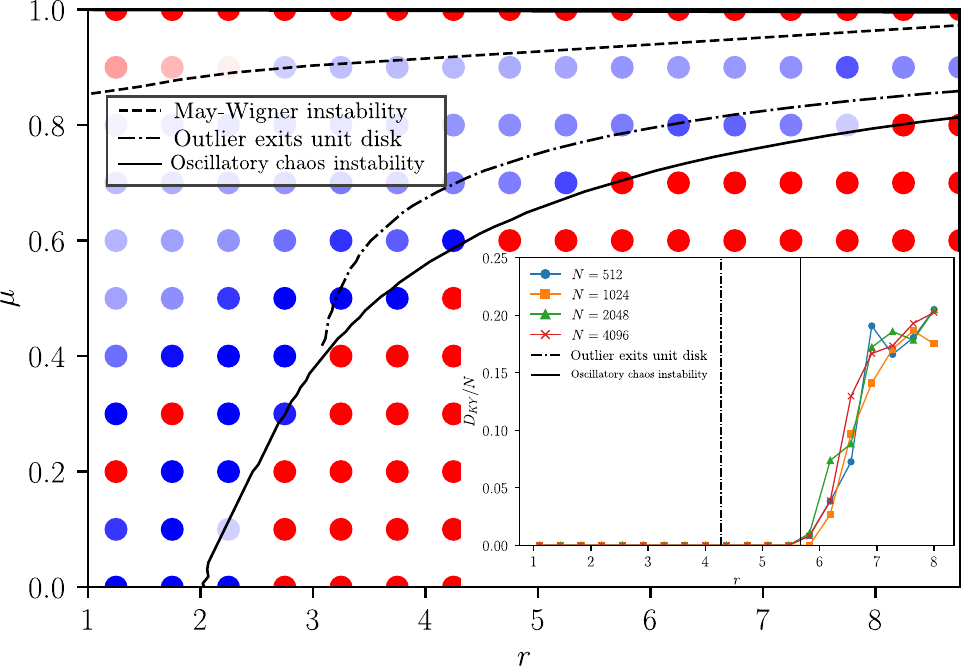}
		\caption{Stability diagram for the model with additional homogeneous coupling [Eq.~(\ref{eq:dis_log_map_homogeneous})], at fixed $\sigma = 0.2$. The dashed line is the May--Wigner instability as before, and the solid line is the OC instability (bulk eigenvalue leaving the unit disk at $\lambda=-1$). The dot-dashed line indicates where in parameter space the outlier eigenvalue exits the unit disk (at $\lambda=-1$). The LLE $\Lambda_1$ is shown in relief (blue $\Lambda_1<0$, red $\Lambda_1>0$). We find no chaos as a result of the outlier eigenvalue leaving the unit disk ($\Lambda_1$ is found to be negative in the region between the dot-dashed and solid lines). Inset: Kaplan-Yorke dimension vs $r$ at fixed $\mu = 0.7$, further confirming that chaos is only seen when the bulk spectrum leaves the unit disk but not when only the outlier has exited. The chaos beyond the OC instability is high-dimensional with extensive attractor dimension $D_{KY}\propto N$.}
		\label{fig:modified_phase_diagram}
	\end{figure}

      An analysis of the eigenvalues of the reduced Jacobian now reveals the possibility of an isolated outlier eigenvalue. If such an outlier exists it is located to the left of the bulk spectrum for $\mu>0$. Therefore, there are now three types of possible instability. As before, the bulk spectrum can exit the unit disk at $\lambda=1$ (May--Wigner instability) or at $\lambda=-1$ (OC instability). Additionally, the outlier can leave the disk at $-1$. We can determine the conditions for this to happen analytically, see End Matter. The outlier eigenvalue is isolated from the bulk, and we therefore expect, and indeed find, coherent oscillatory behaviour when the outlier passes $-1$. This is in contrast to the excitation of many dynamical modes that occurs when the bulk continuum of eigenvalues exits the unit circle at the OC instability.
      
    We plot a stability diagram for this model in the $(r,\mu)$ plane at fixed $\sigma=0.2$ in Fig.~\ref{fig:modified_phase_diagram}. The lines are from the theory and show the onset of the three types of instability. The coloured markers in the background indicate the LLE as before. The dynamics is chaotic in the red region, and stable in the blue region.

    For sufficiently small values of $\mu$ ($\mu\lesssim 0.4$ in the example), the behaviour is similar to what we described before. At low values of the bifurcation parameter $r$ we find stable fixed points. Increasing $r$ the system crosses the OC instability, i.e. the continuous bulk spectrum strays past $-1$. When this happens the dynamics transitions directly to high-dimensional chaos, and there are no signs of period doubling. 

    For larger values of $\mu$ ($0.4\lesssim \mu\lesssim 0.85$ in Fig.~\ref{fig:modified_phase_diagram}) we again find the system to be in the stable phase when $r$ is sufficiently small. As $r$ is increased (horizontal cut in the diagram) instability sets in, but now marked by the outlier eigenvalue leaving the unit disk (dot-dashed line). Beyond this point the fixed-point theory is no longer valid, but we can still use it to approximate the point at which the bulk eigenvalue spectrum also exits the unit disk. This gives us the part of the solid line in Fig.~\ref{fig:modified_phase_diagram} located to the right of the dot-dashed line.

  \begin{figure}[t]
		\centering  
		\includegraphics[width=1\linewidth]{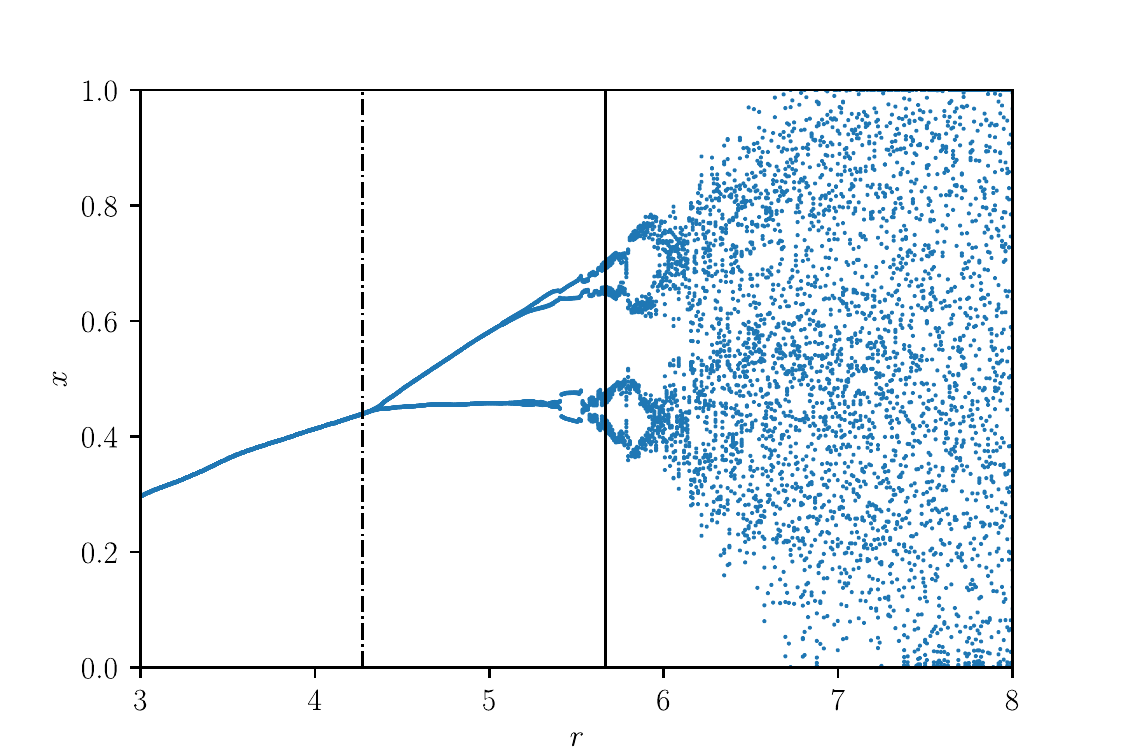}
		\caption{Bifurcation diagram, $\sigma = 0.2$, $\mu = 0.7$, $N=1000$. For various $r$, we plot the values that a single component takes over the course of its trajectory as a set of blue dots. The vertical dot-dashed line on the left indicates the period-doubling instability as predicted from theory, whereas the solid vertical line indicates the onset of oscillatory chaos. }
		\label{fig:bifurcation_diagram_mu0point7}
	\end{figure}
    
    Simulations show that the LLE remains negative for parameter combinations between the dot-dashed and the solid line in Fig.~\ref{fig:modified_phase_diagram}. In this region the outlier eigenvalue has left the unit disk, but the bulk spectrum has not. The inset of the figure further shows that the attractor dimension also remains zero in this situation, and becomes extensive only when the bulk transition has been crossed. This means that the system does not transition to a chaotic state when the outlier eigenvalue leaves the unit disk. A more detailed inspection of simulations (see End Matter) shows that we recover a partial period-doubling cascade.  The bifurcation diagram in Fig.~\ref{fig:bifurcation_diagram_mu0point7} provides further evidence of this.
    
  We highlight that the point at which the outlier leaves the unit disk (dot-dashed line in Fig.~\ref{fig:modified_phase_diagram}) is restricted to values above $r=3$ for any $\sigma>0$. Period-doubling onset in the disordered map thus occurs for larger $r$ than in the conventional logistic map (or not at all if $\mu$ is not sufficiently high). Therefore, we can say that, depending on the strength of any homogeneous coupling, the disorder either removes the period-doubling cascade, or postpones it.

	\medskip
    
	\noindent {\em Discussion.} To summarise, adding disorder to the logistic map significantly alters its dynamics, and produces three distinct transitions to chaos, two of which are not present in the usual logistic map. One is similar to the well-known May--Wigner instability of continuous-time disordered systems \cite{wigner1958distribution, may, allesina2015stability}. In this transition high-dimensional chaos ensues, dominated by low-frequency modes. A second transition in the disordered logistic map is instead marked by the dominance of high-frequency oscillatory modes, but again leading to chaos with a high-dimensional attractor. As far as we are aware, this transition has not been reported in existing disordered-systems literature. Overall, we conclude that any arbitrarily small amount of disorder removes the period-doubling cascade of the logistic map. Only with the introduction of an additional sufficiently strong homogeneous coupling across components do we recover elements of this cascade.

    	The period-doubling cascade of the conventional logistic map has been observed in multiple experimental systems \cite{strogatz2024nonlinear}. Our results suggest that these systems must be described either by genuinely low-dimensional equations, by dynamics without disorder, or there must be a significant homogeneous coupling. Our findings also indicate that chaos may be more likely than regular motion in high-dimensional systems with heterogeneity, such as for example those used to describe financial markets \cite{bouchaud2023application} or ecological communities \cite{bunin2017,may1974biological, hassell1976patterns, turchin1992complex}. This could help to understand observations that chaos is not rare in natural ecosystems \cite{rogers2022chaos}.

    To test if our findings apply more generally, we have also performed an analysis of a disordered H\'enon map. This is reported in the SM, and reveals behaviour that is very similar to that of the disordered logistic map. We therefore hypothesise that other maps with period doubling will display similar phenomena if disorder is added. 
    
    Future work could focus on the effects of disorder on the continuous-time R\"ossler or Lorenz systems. It would be interesting to understand how dynamical systems exhibiting different routes to chaos are affected by disordered couplings (e.g. Ruelle--Takens--Newhouse route to chaos via quasiperiodicity, or the Pomeau--Manneville scenario via intermittency). Similarly it would be worthwhile investigating the effects of disorder on excitable systems, or dynamical systems with oscillatory behaviour for example in cyclic games \cite{mobilia2010oscillatory,giral2025stabilization,giral2025interplay} 
	 
The natural world is complex, and stylised low-dimensional models often neglect heterogeneity across components. A pertinent question to ask is whether the behaviour observed in these simple models survives the introduction of heterogeneity captured by disordered couplings.

Our findings show that even the most iconic phenomena exhibited by simple models (such as the period-doubling cascade of the logistic map) can be fragile in the face of disorder. At the same time, interesting new behaviours can emerge that are found neither in simple low-dimensional models nor in previously studied high-dimensional disordered systems. Our work thus highlights the need for caution when supposing homogeneous couplings in stylised models of complex systems, but it also hints at a wealth of new phenomena that can be unveiled through modern techniques for handling disordered interactions.

	\acknowledgements
	We thank Jean-Philippe Bouchaud for discussions, and for pertinent suggestions which helped to shape this work. JWB thanks the Leverhulme Trust for support through the Leverhulme Early Career Fellowship scheme. TG acknowledges partial financial support from the Agencia Estatal de Investigación and Fondo Europeo de Desarrollo Regional (FEDER, UE) under project COSASTI (PID2024-157493NB-C22) funded by MICIU/AEI/10.13039/501100011033 and ERDF/EU, and the María de Maeztu Program for units of Excellence in R\&D, Grant No. CEX2021-001164-M. 

\section*{Data and code availability}
Data and/or codes for the figures in the main paper and the SM can be found at \cite{repository}.

	\clearpage
	\section*{End matter}
   We give additional information about the analytical calculations and numerical methods, describing the overall strategy, some of the main steps and technical results. A full account can be found in the SM \cite{sm}. The End Matter also contains additional numerical results.
\subsection*{Fixed-point solution}
Fixed points of Eq.~(\ref{eq:dis_log_map_homogeneous}) can be seen to fulfill $
    x_i^\star=H\left[\frac{1}{1-\mu}\big(1-\frac{1}{r}-\mu M^\star+\sum_j a_{ij}x_j^*\big)\right]$, where $M^*=N^{-1}\sum_j x_j^*$. Via the central limit theorem, the term $\sum_{j} a_{ij}x_j^*$ is a Gaussian random variable for large $N$, with variance $\sigma^2 q$, where $q=N^{-1}\sum_j (x_j^*)^2$. This can formally be derived using the standard steps of dynamic mean-field theory (DMFT) analysis (see e.g. \cite{galla2024generating, froy}). Therefore, the distribution of the fixed-point values of $x_i$ that are strictly between zero and one  is a truncated Gaussian with density
    \begin{align}
    P(x^\star) = \sqrt{\frac{(1-\mu)^2}{2\pi\sigma^2 q}} e^{-\frac{\left[x^\star(1-\mu)-\left(1-\frac{1}{r}-\mu M^*\right)\right]^2}{2\sigma^2 q}}. \label{abundancedistribution_mu}
\end{align}
The total probability weight of these components is $\phi=\int_0^1 dx^\star P(x^\star)$. There are also `frozen' components taking values $x^\star=0$ or $x^\star=1$. We write the probabilities for this as $\psi_0$ and $\psi_1$ respectively, with $\psi_1 = \int_1^\infty dx^\star \,P(x^\star)$, and $\phi+\psi_0+\psi_1=1$. We have the self-consistency relations $
    M^*=\int_0^1 dx^\star\, P(x^\star) x^\star + \psi_1$, $q= \int_0^1 dx^\star\, P(x^\star) (x^\star)^2+\psi_1$. These equations can be solved numerically to find $M^\star$, $q$, $\psi_1$ and $\phi$ as a function of the model parameters $r$, $\sigma$ and $\mu$. 
\subsection*{Eigenvalue spectrum of the reduced Jacobian matrix} 
\subsubsection*{Bulk eigenvalue spectrum}
The elements of the reduced Jacobian are given by $J_{ij}= 
        \delta_{ij} + r x_i^\star \left[a_{ij} - \frac{\mu}{N} -\delta_{ij}\left(1-\mu \right)\right]$,
where $i$ and $j$ are restricted to  non-frozen components.
To find the region of the complex plane containing the spectrum we use the replica method for non-hermitian matrices \cite{sommers, haake, baron2026lecture}. 

Two important observations simplify this computation. Firstly, the $\{x_i^\star\}$ and the interaction matrix elements $a_{ij}$ can be treated as independent random variables for $N \to \infty$ \cite{baron2023breakdown}. Secondly, while there will be non-trivial correlations among the $a_{ij}$ (conditioned on $i$ and $j$ being non-frozen components) similar to \cite{bunin2017}, these correlations do not affect the bulk spectrum \cite{baron2022eigenvalue, baron2023breakdown, birolibunin}. We may therefore take $a_{ij}$ to be centered Gaussian random variables with variance $\sigma^2/N$ and the $x_i^\star$ to be drawn as {\em iid} random variables from the distribution $P(x^*)$. With these simplifications the calculation follows standard steps described for example in Refs. \cite{baron2026lecture, baron2020dispersal, sommers, haake}. Ultimately, we find that the boundary of the bulk eigenvalue spectrum fulfills
\begin{align}
    \frac{1}{\sigma^2} = \int_0^1 dx^* \,P(x^*) \frac{r^2 x^2}{\lambda_y^2 + [1 - r (1-\mu) x^* - \lambda_x]^2} .
\end{align}
For $\mu=0$ this reduces to Eq.~(\ref{eq:main_boundary}).

\subsubsection*{Outlier eigenvalue for $\mu\neq 0$}
For $\mu\neq 0$ the reduced Jacobian can have an isolated outlier eigenvalue. We use existing generating-functional methods \cite{baron2023breakdown} to determine its location. To do this, correlations between the elements $a_{ij}$ in the interaction matrix between non-frozen components must be taken into account. These correlations are not present in the original $N\times N$ interaction matrix, but appear as a consequence of the dynamics \cite{bunin2016interaction, bunin2017, baron2023breakdown}.

\begin{figure}[t] 
\includegraphics[width=0.45\textwidth]{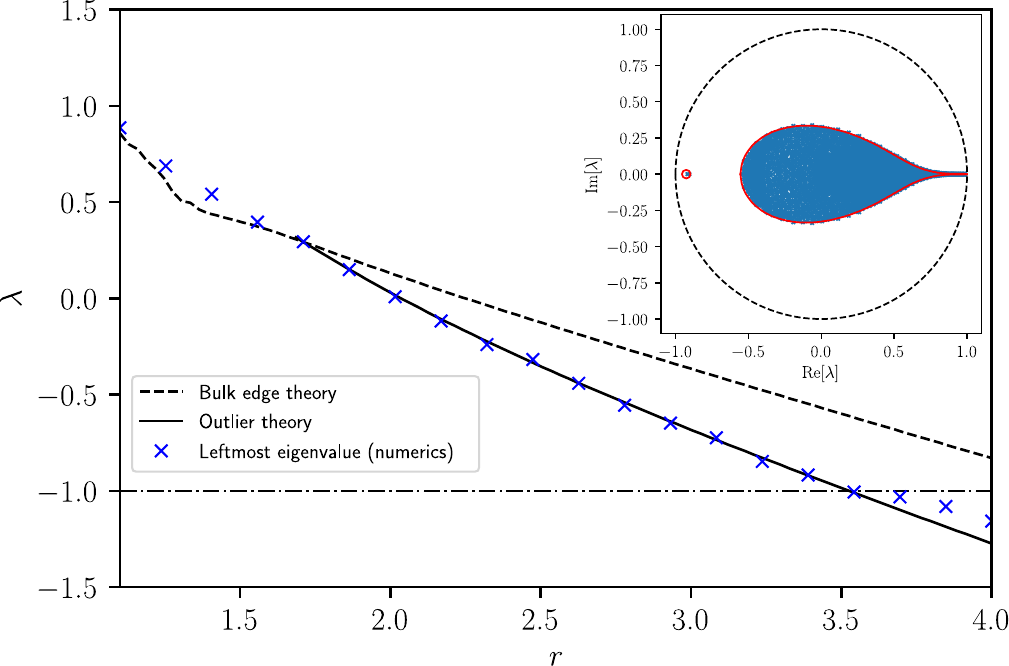}
		\caption{ Leftmost eigenvalue vs $r$ for fixed $\mu = 0.6$, $\sigma = 0.2$ and $N = 4000$. Inset: Example eigenvalue spectrum for $N = 4000$, $\mu = 0.6$, $\sigma = 0.2$, and $r= 3.4$ (red colour shows the theoretical predictions for bulk and outlier eigenvalues; the dashed line is the unit circle)}
\label{fig:leftmost_vs_r}
\end{figure}

If there is an outlier eigenvalue $\lambda_\mathrm{outlier}$, we find that it is given by the solution to 
\begin{align}
     A_1(\lambda_\mathrm{outlier})+ \frac{\sigma^2A'_{1}(\lambda_\mathrm{outlier})A_{2}(\lambda_\mathrm{outlier})}{1-\sigma^2A'_{2}(\lambda_\mathrm{outlier})} &= -\frac{1}{\mu}, \label{gsol}
\end{align}
with
\begin{align}
    A_{\beta}(\lambda) &= \int_0^1 dx \,P(x) \frac{r x^\beta}{\lambda -1 + (1-\mu)r x} ,\nonumber \\
    A_{\beta}'(\lambda) &=\int_0^1 dx \,\frac{d P(x| h)}{dh} \vert_{h = 0} \frac{r x^\beta}{\lambda-1 + (1-\mu)r x}.
\end{align}
$P(x| h)$ is given by a modification of Eq.~(\ref{abundancedistribution_mu}), in which $1-\frac{1}{r}-\mu M^*$ is replaced by $1-\frac{1}{r}-\mu M^*+ h$. The predictions of this theory are tested in Fig.~\ref{fig:leftmost_vs_r}.

\subsection*{Fluctuations about fixed points}
To characterise the dynamical character of the two types of linear instability, we use dynamic mean field theory and follow the lines of \cite{opper1992phase}. On each time step, we add a centred Gaussian random number $\xi_i(t)$ with $\langle\xi_i(t)\xi_j(t')\rangle=\delta_{ij}\delta_{tt'} r^2 T [x_i(t)]^\kappa$, where we consider $\kappa=0, 1, 2$. We then define $y_i(t)=x_i(t)-x_i^*$ and linearise the mean-field process about the fixed point of the noiseless system. We then compute $\langle \vert \tilde y(\omega) \vert^2\rangle_S$, where the tilde indicates the Fourier transform and $\langle \cdot \rangle_S$ an average over the external noise and the set of surviving species.

Expanding for small angular frequencies we find 
\begin{align}
\langle \vert \tilde y(\omega) \vert^2 \rangle_S \approx \begin{cases}
        \frac{T \pi rP(0^+)/(2\vert\omega\vert)}{(1 - \phi \sigma^2 )+ \frac{\sigma^2\vert\omega\vert \pi P(0^+)}{2r}} \hspace{0.2cm} \mathrm{for}\hspace{0.2cm} \kappa = 0,\\  ~\\
        \frac{T  P(0^+) \log \vert r/\omega\vert}{(1 - \phi \sigma^2 )+ \frac{\sigma^2\vert\omega\vert \pi P(0^+)}{2r}}\hspace{0.2cm} \mathrm{for}\hspace{0.2cm} \kappa = 1,\\~\\
	    \frac{T \phi}{(1 - \phi \sigma^2 )+ \frac{\sigma^2\vert\omega\vert \pi P(0^+)}{2r}}\hspace{0.2cm} \mathrm{for}\hspace{0.2cm} \kappa = 2.
	\end{cases} 
\end{align}
At the May--Wigner instability $\phi \sigma^2 = 1$, and the power-spectrum diverges differently depending on the kind of noise that we introduce. 

On the other hand, we can also expand for small $|\omega-\pi|$ and instead obtain $
		\left\langle \vert \tilde y(\omega) \vert^2 \right\rangle_S \approx TI^{(\kappa)}_1/[(1- I^{(2)}_1 \sigma^2) + (\omega-\pi)^2I^{(2)}_2],$
with $I^{(\kappa)}_1 =\int_0^1 dx^*\, P(x^*) \frac{r^2 (x^*)^\kappa}{(2-r x^*)^2 }$ and $
I^{(\kappa)}_2=\int_0^1\, dx^* P(x^*) \frac{r^2 (x^*)^\kappa[1-(2-r x)]}{(2-r x^*)^4 }$. At the OC instability $I^{(2)}_1 \sigma^2 = 1$, and thus $\left\langle \vert \tilde y(\omega) \vert^2 \right\rangle_S \propto |\omega-\pi|^{-2}$. We thus find qualitatively different behaviour on the approach to the May--Wigner and OC instabilities.

    \subsection*{Numerical methods}
   The largest Lyaponov exponents in Figs.~\ref{fig:phase_diagram} and \ref{fig:modified_phase_diagram} are computed following a standard approach \cite{sandri1996numerical}. For a particular a realisation of the $a_{ij}$ we run a trajectory of the map to the stationary state. Then, we create a copy, and apply a small perturbation. We then monitor the subsequent distance between the perturbed and unperturbed systems,  periodically resetting the distance to a fixed magnitude. From this we determine the LLE. We add a small immigration rate to the map in Eq.~(\ref{eq:dis_log_map}) to avoid premature absorption of components at $x_i=0$ \cite{sm}.

     To compute the Kaplan-Yorke (KY) dimension, we require the full spectrum of Lyapunov exponents $\{\Lambda_i\}$. These are obtained via successive QR decompositions of the Jacobian matrix \cite{sandri1996numerical}. We then have $
        D_{KY} = j - \sum_{i= 1}^j\Lambda_i/\Lambda_{j+1}$, 
    where $\Lambda_1> \Lambda_2> \cdots>\Lambda_N$ and $j$ is the largest integer such that $\sum_{i= 1}^j\Lambda_i>0$.
    
\begin{figure}[t]
		\centering  
		\includegraphics[width=0.9\linewidth]{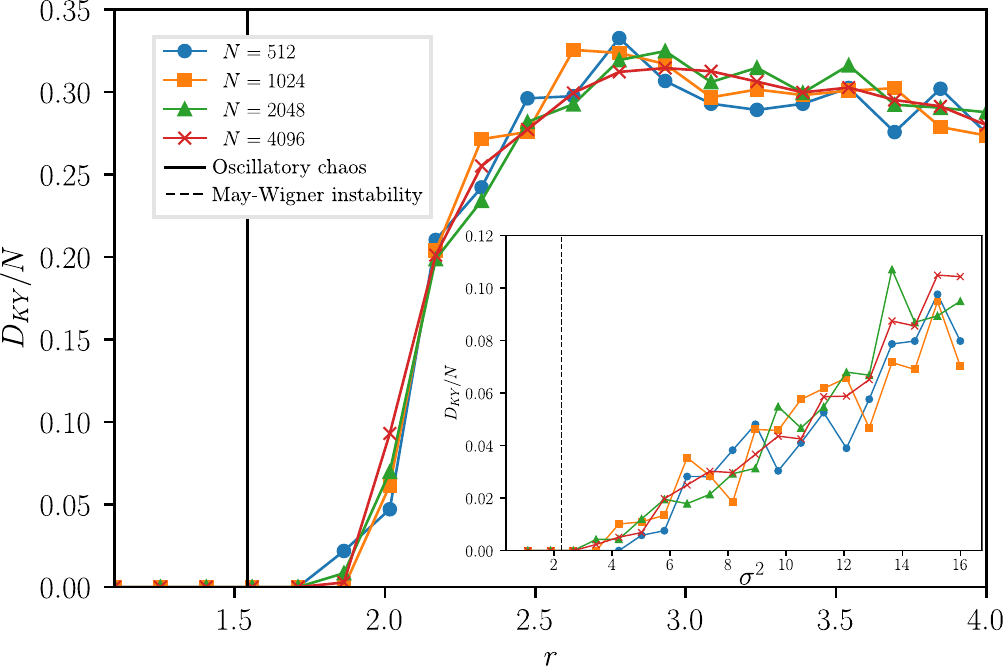}
		\caption{Scaled Kaplan-Yorke dimension vs $r$ for fixed $\mu = 0$ and $\sigma^2 = 1.5$. The attractor dimension is extensive in $N$ as the OC instability is crossed. Inset: Scaled KY dimension vs $\sigma^2$ for fixed $\mu = 0$ and $r = 1.2$. The attractor dimension is also extensive beyond the May--Wigner instability. }
		\label{fig:DKY}
	\end{figure}

    To compute Jacobian numerically it is advantageous to use a softened function $H(u) = \frac{1}{2}\left\{ 1+  \epsilon \log[\frac{\cosh( x/\epsilon)}{\cosh[(1 - x)/\epsilon]}]\right\}$. In the limit $\epsilon \to 0$ this reduces to the ramp function given below Eq.~(\ref{eq:dis_log_map}). In simulations we use $\epsilon=0.05$. The data in Fig.~\ref{fig:DKY} demonstrates that the KY dimension becomes extensive as either the May--Wigner or the OC transition is crossed in the model with $\mu=0$.

\begin{figure}[h]
		\centering  
		\includegraphics[width=0.9\linewidth]{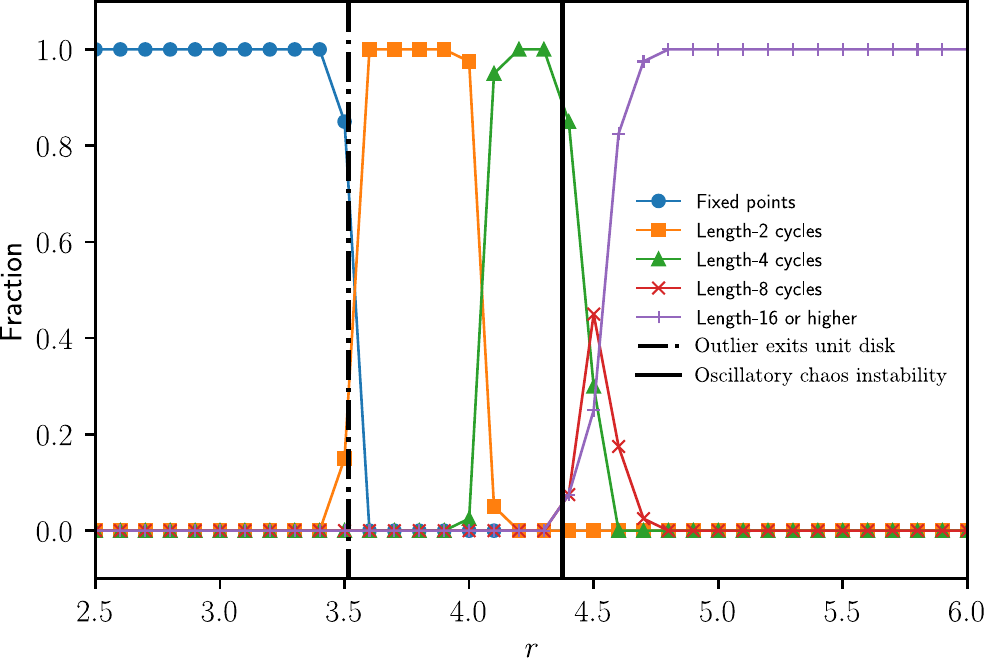}
		\caption{Fraction of  fixed points, 2-cycles, 4-cycles, 8-cycles and higher-order attractors. Each data point is from 40 independent runs ($\mu=0.6, \sigma=0.2$, $N=4000$). The vertical lines are the locations of the period-doubling transition (dot-dashed) and OC transition (solid), predicted from the theory. }
		\label{fig:fractions}
	\end{figure}

    We have also conducted simulations to identify the proportions of samples that converge to fixed points, 2-cycles, 4-cycles, 8-cycles or higher-order attractors. This classification is based on time averaged distances $d(\tau)=\avg{N^{-1}\sum_i [x_i(t)-x_i(t-\tau)]^2}_t$ for $\tau\in\{1,2,4,8\}$, see \cite{sm} for details. Figure~\ref{fig:fractions} is a horizontal cut in the phase diagram in Fig.~\ref{fig:modified_phase_diagram} at fixed $\mu= 0.6$. We see that when the outlier eigenvalue exits the unit disk at $-1$ (with the bulk remaining inside), 2-cycles subsequently dominate. As $r$ is increased further, another period doubling occurs, and 4-cycles dominate. The period-doubling cascade is curtailed when the bulk of the eigenvalue spectrum also exits the unit disk, and oscillatory chaos ensues.

\end{document}


\title{Supplemental Material: \\ Long-range eigenvalue density correlations of non-Gaussian and structured random matrices}
	
	\title{The disordered logistic map\\ \medskip \medskip{\Large--- Supplemental Material ---}}
	\author{Joseph W. Baron}
	\email{jwb96@bath.ac.uk}
	\affiliation{Department of Mathematical Sciences, University of Bath, Bath, BA2 7AY, UK}
	
	\author{Tobias Galla}
	\email{tobias.galla@ifisc.uib-csic.es}
	\affiliation{Instituto de F{\' i}sica Interdisciplinar y Sistemas Complejos IFISC (CSIC-UIB), 07122 Palma de Mallorca, Spain}

	\maketitle

	\onecolumngrid
	
	\tableofcontents

	\medskip
	
	This document provides a more detailed exposition of the calculations and simulations leading to the  results of the main text. First, we briefly summarise some key features of the conventional disordered map (Sec.~\ref{sec:conventional}). We use this to highlight the drastic effects of disorder, which are detailed in the main paper and in the later sections of this Supplement. In Sec.~\ref{section:fixedpoint}, we describe in detail how dynamic mean-field theory (DMFT) can be used to characterise the fixed-point regime of the disordered logistic map model. In Sec.~\ref{section:rmt}, we use random matrix theory to  determine the stability of the fixed-point solution, identifying two distinct types of instability. We then return to the dynamic mean-field theory in Sec.~\ref{sec:dmft_instab_sec}, and characterise the nature of the two transitions by examining the power spectrum of fluctuations in the presence of external noise. In Sec.~\ref{section:homogeneousinteraction}, we examine the modified model with an additional homogeneous interaction term, performing a new random matrix theory analysis to find both the modified bulk spectrum and the new outlier eigenvalue.  Sec.~\ref{sec:numerics} contains further details of the numerical procedures used to obtain the results in the main paper, and some additional simulations. Finally, in Sec.~\ref{sec:henon}, we show that the disordered H\'enon map displays similar behaviour to the disordered logistic map. 
	
	\section{Phenomenology of the logistic map without disorder}\label{sec:conventional}
	\subsection{Basic definition}
	The logistic map without disorder is defined by \cite{may1976simple, feigenbaum1978, strogatz2024nonlinear}, 
	\begin{align}
		x(t+1) = r \,x(t) [1- x(t)],
	\end{align}
	where $0<r\leq 4$. For this range of $r$, $x(t+1)$ lies in the interval $[0,1]$ as long as $x(t)$ is in this internal. Thus, for $0\leq x(t=0) \leq 1$ all subsequent $x(t)$, $t=1,2\dots$ are in $[0,1]$. In contrast, the disordered logistic map in Eq.~(1) of the main paper is given by 
	\begin{align}
		x_i(t+1) = H \left[ r x_i(t) \{1-x_i(t)+ \sum_{j} a_{ij} x_j(t) \}\right], \label{eq:dis_log_map_sm}
	\end{align}
	where the $a_{ij}$ are independent quenched random variables, constituting the disorder. Specifically, we use $\avg{a_{ij}}=0$, and $\avg{(a_{ij})^2}=\sigma^2/N$, where we write $\langle\cdots\rangle$ for the average over the disorder. The function $H(\cdot)$ is given by
	\begin{align}\label{eq:hdef}
		H(u) = \begin{cases}
			0 \hspace{1cm}\mathrm{if~} u\leq 0 ,  \\
			u \hspace{1cm}\mathrm{if~} 0\leq u \leq1, \\
			1 \hspace{1cm}\mathrm{if~} u\geq 1.
		\end{cases}
	\end{align}
	In the absence of disorder ($\sigma=0$, i.e., $a_{ij}=0$ for all $i,j$), the set of maps in Eq.~(\ref{eq:dis_log_map_sm}) reduces to $N$ uncoupled copies of the usual logistic map.
	
	\subsection{Stability analysis and period-doubling route to chaos}
	The behaviour of the logistic map without disorder is well-established and covered in standard textbooks such as \cite{strogatz2024nonlinear}. We briefly reiterate the main points, as this serves as a baseline for our analysis of the disordered logistic map.
	
	\medskip
	
	The logistic map without disorder has fixed points $x^\star = 0$ and $x^\star = 1 - \frac{1}{r} $. The trivial fixed point $x^\star = 0$ is linearly stable for $r<1$, whereas the non-zero fixed point $1-1/r$ is then unstable. When $1<r<3$, the non-trivial fixed point becomes stable and $x^\star=0$ is unstable. The eigenvalue at the non-zero fixed point is $\lambda  = 2-r$. Thus, at $r = 3$, this eigenvalue crosses the point $-1$, indicating an oscillatory instability. Cyclic behaviour emerges. 
	
	Beyond the point $r = 3$, one makes a two-cycle ansatz, seeking values of $x^\star$ that remain unchanged under two iterations of the map \cite{strogatz2024nonlinear}. One finds solutions $x^\star =0$, $x^\star = 1-1/r$, and 
	$x^\star_{\pm} = [(r+1) \pm \sqrt{(r-3)(r+1)}]/(2r)$. The solutions $x^\star =0$ and $x^\star = 1-1/r$ describe unstable fixed points of the map for $r>3$. The two solutions $x^\star_{\pm}$ constitute the stable two-cycle in a finite range of $r$ above $r=3$. The appropriate eigenvalue is given by linearising the twice-iterated map about the fixed points $x_\pm^\star$. The eigenvalue of the second iterate is found as $\lambda' = 4 + 2r - r^2$. As $r$ is increased, the two-cycle solution becomes unstable when this eigenvalue hits either $-1$ or $+1$. The former is seen to happen first, at $r = 1 + \sqrt{6} \approx 3.449$. At this point, another period-doubling instability occurs, and four-cycles emerge. This pattern continues as one increases $r$. This is the well-known period-doubling cascade of the logistic map \cite{feigenbaum1978, strogatz2024nonlinear}. Eventually, this gives way to chaos at a value of $r \approx 3.56995$. This behaviour is in stark contrast to the sudden onsets of chaos described in the main text, which avoid the period-doubling cascade entirely.

	\section{Dynamic mean-field theory part 1: characterising the fixed-point solution}\label{section:fixedpoint}
	We now study the disordered logistic map in Eq.~(1) of the main paper, in the limit $N \to \infty$. Using dynamic mean-field theory, we are able to characterise the fixed-point solution of this model. An analogous analysis can be performed for Eq.~(4) of the main text (i.e. when we involve the homogeneous interaction term). We describe this in Section \ref{section:homogeneousinteraction} of this document.
	
	\subsection{Effective process}
	In the limit $N\to\infty$ we can replace the interaction term in Eq.~(1) of the main text with a noise term and a memory term. This is referred to as dynamic mean field theory (DMFT), and can formally be derived via the cavity method or via generating functionals, and is a standard procedure for disordered dynamical systems, used to study spin glasses, neural networks, game learning and models in ecology. The general procedure is described for example in \cite{mezard1987, froy,galla2024generating}.  
	
	\medskip
	
	Recalling that the $a_{ij}$ are {\em iid} with $\langle a_{ij} \rangle = 0$, $\langle a_{ij}^2 \rangle = \sigma^2/N$ we find
	\begin{align}
		x(t+1) = H \left[r x(t) \left[1 - x(t)  +\eta(t)\right]\right],  \label{effproc}
	\end{align}
	where the Gaussian noise variables $\eta(t)$ have statistics
	\begin{align}\label{eq:eta_stat}
		\langle \eta(t) \rangle_\eta = 0, \hspace{2cm} \langle \eta(t) \eta(t')\rangle_\eta = \sigma^2 \langle x(t) x(t')\rangle_\eta.
	\end{align}
	The notation $\langle\cdots\rangle_\eta$ in the last equation stands for an average over realisations of $\eta$. As is standard in DMFT, the statistics of the trajectories $x(t)$ in the effective dynamics are identical to the statistics of the $\{x_i(t)\}$ across components $i$ in the original problem, in the limit $N \to \infty$ \cite{mezard1987, froy, galla2024generating}.
	
	As in the original map in Eq.~(1) of the main paper, the state $x=0$ is absorbing. That is to say, if $x(t)=0$ at some finite time $t$, then $x(t')$ is zero for all $t'\geq t$.
	
	\subsection{Fixed-point solution}\label{sec:fp_sol}
	\subsubsection{Fixed-point ansatz in the effective dynamics, and possible solutions}
	We now assume that the disordered system reaches a fixed point (this is what is observed in simulations for sufficiently small $r$ and $\sigma$). That is, we assume $x(t) = x^\star$ and $\eta(t) = \eta^\star$. Using this in Eq.~(\ref{effproc}) gives
	\begin{align}\label{eq:fixed_point}
		x^\star = H\left\{ r x^\star \left[ 1 - x^\star + \eta^\star\right]\right\},
	\end{align}
	and from Eq.~(\ref{eq:eta_stat}) we have the self-consistency relation $\langle (\eta^\star)^2 \rangle_\star = \sigma^2 \langle (x^\star)^2\rangle_\star$, where $\langle\cdots\rangle_\star$ is an average over $\eta^\star$. 
	
	\medskip
	
	We now need to determine $x^\star$ as a function of $\eta^\star$ from Eq.~(\ref{eq:fixed_point}). There are multiple possible solutions:
	\begin{itemize}
		\item For any $\eta^\star$, there is a trivial solution $x^\star = 0$.
		\item If $r \eta^\star >1$, then $x^\star = 1$ is a possible solution. The argument inside $H$ in Eq.~(\ref{eq:fixed_point}) exceeds one, and the function $H$ thus reaches the upper saturation limit [see Eq.~(\ref{eq:hdef})]. 
		\item When  $0<1 - \frac{1}{r} + \eta^\star<1$ we have the solution
		\begin{align}\label{eq:fp_sol}
			x^\star = 1 - \frac{1}{r} + \eta^\star.
		\end{align}
		
	\end{itemize}
	
	In line with \cite{opper1992phase, bunin2016interaction, bunin2017, Galla_2018, galla2024generating}, the solution in Eq.~(\ref{eq:fp_sol}) is the solution that is physically realised for a given value of $\eta^*$ whenever the expression on the right of Eq.~(\ref{eq:fp_sol}) is between zero and one. The solutions $x^\star=0$ or saturation $x^\star=1$ are only realised when they are the sole possibility. In other words, we find $x^\star=0$ only when $1-1/r+\eta^\star<0$, and $x^\star=1$ only when $1-1/r+\eta^\star>1$ [equivalent, $r\eta^\star>1$]. The physically relevant solution can be summarised as follows
	\begin{align}\label{eq:fp_sol_summary}
		x^\star = H\left( 1 - \frac{1}{r} + \eta^\star\right).
	\end{align}
	
	Degrees of freedom $x^\star=0$ or $x^\star=1$ do not contribute to the stability or instability of any fixed points similar to what is observed in \cite{opper1992phase, bunin2016interaction, Galla_2018, sidhomgalla}. We mean this in the sense that these degrees of freedom do not respond to small external perturbations. These variables are thus `frozen'. Conversely, we will call a variable $x$ (or $x_i$) `non-frozen' when it takes a value strictly between zero and one. 
	
	\subsubsection{Species abundance distribution and fraction of extinct and saturated species}

\begin{figure}
		\centering   
		\includegraphics[width=0.48\linewidth]{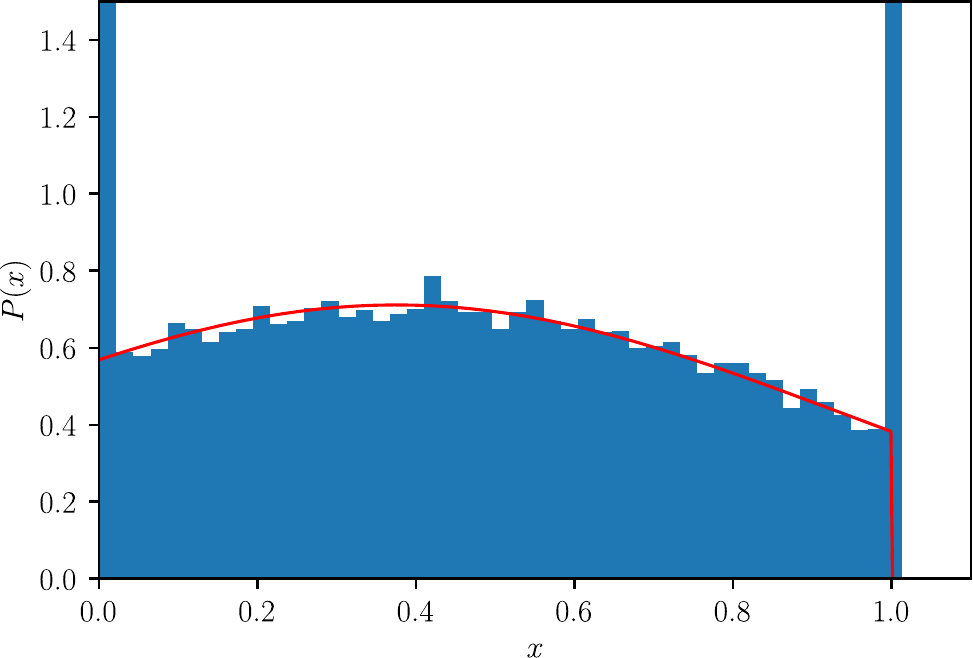}
		
		\caption{Distribution of the $x_i^\star$ for $r=1.6$ and $\sigma=1$. The histogram (vertical bars) is obtained from a 20 realisations of the system with $N=4000$. The line is from Eq.~(\ref{abundancedistribution}), with $q$ obtained from Eq.~(\ref{qeq}).}
		\label{fig:sad}
	\end{figure}

	From Eq.~(\ref{eq:fp_sol_summary}) we find that $x^\star$ is a random variable with a truncated Gaussian distribution of the form
	\begin{align}
		P(x) = \frac{1}{\sqrt{2\pi\sigma^2 q}} e^{-\frac{\left(x-1+1/r\right)^2}{2\sigma^2 q}} \Theta\left( x\right)\Theta\left(1- x\right)+ \psi_0\delta(x) + \psi_1 \delta(x-1), \label{abundancedistribution}
	\end{align}
	where we do not write out the asterisk and where $\Theta(\cdot)$ is the Heaviside function [$\Theta(u)=1$ for $u\geq 0$, and $\Theta(u)=0$ for $u<0$]. The first term on the right describes components  $0<x^\star<1$. The second and third terms reflect the fact that fractions $\psi_0$ and $\psi_1$ of components take the values $x^\star=0$ and $x^\star=1$ respectively. We find
	\begin{align}\label{eq:phi0phi1}
		\psi_0& = \int_{-\infty}^0 dx \frac{1}{\sqrt{2\pi\sigma^2 q}} e^{-\frac{\left(x-1+1/r\right)^2}{2\sigma^2 q}} , \nonumber \\
		\psi_1 &= \int^{\infty}_1 dx \frac{1}{\sqrt{2\pi\sigma^2 q}} e^{-\frac{\left(x-1+1/r\right)^2}{2\sigma^2 q}} .
	\end{align}
	For later purposes we also introduce 
	\begin{equation}
		\phi=1-\psi_0-\psi_1.
	\end{equation}
	This is the fraction of degrees of freedom that are not frozen. The parameter $q$ in Eq.~(\ref{abundancedistribution}) is to be obtained self-consistently as $q = \langle (x^\star)^2 \rangle_\star$, where $\langle\dots\rangle_\star$ is an average over the random variable $\eta^\star$ (or equivalently over $x^\star$). The self-consistency relation for $q$ can be written as
	\begin{align}
		q &= \int_0^1 dx\, \frac{x^2}{\sqrt{2\pi\sigma^2 q}} e^{-\frac{\left(x-1+1/r\right)^2}{2\sigma^2 q}} + \psi_1 \label{qeq}
	\end{align}
	The first term reflects the contributions of non-frozen components, and the second term that of components with $x^\star=1$. Components with $x^\star=0$ do not contribute. The right-hand side of Eq.~(\ref{qeq}) can be written in terms of exponential and error functions. Thus, for given model parameters $r$ and $\sigma$, Eq.~(\ref{qeq}) can be solved numerically for $q$. Using Eqs.~(\ref{eq:phi0phi1}) and (\ref{abundancedistribution}), one then obtains the distribution for $x^\star$. An example is shown in Fig.~\ref{fig:sad}. As seen in the figure, the theory describes simulation results well.
	
	\medskip
	
	In the next section, we analyse the stability of the fixed-point solution using random matrix theory.
	
	\section{Random matrix theory}\label{section:rmt}
	\subsection{Jacobian matrix and reduced interaction matrix}
	To understand the stability of the fixed-point solution in Sec.~\ref{section:fixedpoint}, we need only consider components with $0<x_i^\star<1$. This means that stability is determined by the `reduced' Jacobian matrix (the Jacobian, restricted to the variables which are not frozen). We label this matrix $\underline{\underline{J}}$. For large $N$, this matrix is of size $(\phi N)\times (\phi N)$. After some minor algebra its elements are found to be 
	\begin{equation}\label{eq:jstar}
		J_{ij}= 
		\delta_{ij} + r x_i^\star (a_{ij} -\delta_{ij}).
	\end{equation}

	\medskip
	
	The fixed-point solution becomes unstable when any of the eigenvalues of $\underline{\underline{J}}$ has a magnitude greater than 1. We now demonstrate that the eigenvalues of $\underline{\underline{J}}$ are confined to a finite region of the complex plane. By computing the boundary of this region, we are able to deduce the values of the model parameters $r$ and $\sigma$ for which the fixed point becomes unstable. 
	
	\subsection{Boundary of the eigenvalue spectrum of the reduced Jacobian matrix} \label{section:eigenvaluespectrum}
	\subsubsection{The Jacobian as a finely structured random matrix}
	To obtain the region of the complex plane to which the eigenvalues of $\underline{\underline{J}}$ are confined, we use the eigenvalue-potential method described in Refs.~\cite{sommers, haake, baron2020dispersal, baron2022eigenvalue baron2026lecture}. One notes that, given a set of $x_i^\star$, the off-diagonal  elements of $J_{ij}$ can be treated as independent (but not identically distributed) Gaussian random variables with variance
	\begin{align}
		\langle (J_{ij})^2 \rangle = \frac{r^2(x^\star_i)^2\sigma^2}{N}, \label{jvar}
	\end{align}
	where we use the fact that the correlation between $x_i^\star$ and $a_{ij}$ is subleading in $1/N$ and can therefore be neglected for our purposes. Such random matrices fall into the class of `finely-structured' random matrices studied in Ref.~\cite{poley2024eigenvalue}. However, for the present system we also have the non-trivial modification of varying diagonal elements, $J_{ii} = 1- rx_i^\star$.

	\subsubsection{Obtaining the spectral density from the eigenvalue potential}
	To compute the eigenvalue spectrum of $\underline{\underline{J}}$, we evaluate the eigenvalue potential, defined by
	\begin{align}
		\Phi(\lambda,\lambda^\star) &= -\frac{1}{N}\ln\mathrm{det}\left[\epsilon^2 \underline{\underline{\id}} +  (\lambda^\star\underline{\underline{\id}} -\underline{\underline{J}}^T) (\lambda\underline{\underline{\id}} -	\underline{\underline{J}}) \right],\label{phidef}
	\end{align}
	where $\langle\cdots\rangle$ denotes an average over the quenched randomness and $\lambda^\star$ is the complex conjugate of $\lambda$. The quantity $\epsilon$ is a small regulariser, which is eventually taken to zero \textit{after} the limit $N \to\infty$, so as to avoid singularities. For a detailed exposition of the eigenvalue potential method and random matrix analyses in general, see Ref. \cite{baron2026lecture}. In what follows, we give a more brief account of the full calculation, which is a variation of other similar calculations in the literature. 
	
	The so-called disorder-averaged resolvent (or matrix Green's function) can be extracted from the eigenvalue potential via  the following expression,
	\begin{align}
		G(z,z^\star) &\equiv \frac{1}{N} \mathrm{Tr} \left\{ (z^\star \underline{\underline{\id}} -\underline{\underline{J}}^T ) \left[(z \underline{\underline{\id}} -\underline{\underline{J}} )(z^\star \underline{\underline{\id}} -\underline{\underline{J}}^T ) + \epsilon^2 \underline{\underline{\id}}  \right]^{-1}\right\} =-\frac{\partial}{\partial z } \Phi(z,z^\star), 
	\end{align}
	The eigenvalue density in the complex plane is then extracted via
	\begin{align}
		\rho(z) &\equiv \frac{1}{N}\sum_{\nu} \delta(\lambda - \lambda_\nu)= \frac{1}{\pi}  \lim_{\epsilon\to 0}\frac{\partial}{ \partial z^\star} G(z,z^\star) =-\frac{1}{\pi}  \lim_{\epsilon\to 0}\frac{\partial^2}{\partial z \partial z^\star} \Phi(z,z^\star) .\label{densityfromres}
	\end{align}
	where here the $\{\lambda_\nu\}$ are the eigenvalues of a single instance of the matrix $\underline{\underline{J}}$. Therefore, if we can compute the eigenvalue potential $\Phi(\lambda, \lambda^\star)$, we will be able determine the values of $\lambda$ for which the eigenvalue density $\rho(\lambda)$ is non-zero. 
	
	\medskip
	
	The evaluation of $\Phi(\lambda, \lambda^\star)$ is accomplished using the replica method \cite{mezard1987}. It has been shown for example in Refs.~\cite{sommers, haake, edwardsjones} that the replicas `decouple' in problems of this type, meaning that the logarithm and the ensemble average in Eq.~(\ref{phidef}) commute (a discussion can also be found in the Supplemental Material of~\cite{baron2022eigenvalues}). This means that we can write
	\begin{align}
		\Phi(\lambda, \lambda^\star) =- \frac{1}{N} \ln \left\langle\mathrm{det}\left[\epsilon^2 \underline{\underline{\id}} +  (\lambda^\star\underline{\underline{\id}} -\underline{\underline{J}}^T) (\lambda\underline{\underline{\id}} -	\underline{\underline{J}}) \right] \right\rangle .
	\end{align}
	After a Hubbard-Stratonovich transformation \cite{hubbard}, the determinant in this expression can be written as a Gaussian integral 
	\begin{align}
		\mathrm{det}\left[\epsilon^2 \underline{\underline{\id}} +  (\lambda^\star\underline{\underline{\id}} -\underline{\underline{J}}^T) (\lambda\underline{\underline{\id}} -	\underline{\underline{J}}) \right]^{-1} = &\int \prod_i \frac{d^2z_i d^2y_i}{2 \pi^2} \exp\left[ - \sum_i (y_i^\star y_i + \epsilon^2 z_i^\star z_i)\right] \nonumber \\
		&\times\exp\left[ i \sum_{ij} z_i^\star (J_{ji} - \lambda^\star \delta_{ij})y_j\right] \exp\left[ i \sum_{ij} y_i^\star (J_{ij} - \lambda \delta_{ij})z_j\right] .
	\end{align}
	Performing the ensemble average using Eq.~(\ref{jvar}) and noting the expression in Eq.~(\ref{eq:jstar}), we obtain 
	\begin{align}
		\exp\left[ -N \Phi(\lambda) \right] &= \int \prod_{i} \left( \frac{d^2z_i d^2y_i}{2 \pi^2}\right) \exp\left[ - \sum_{i} (y_i^\star y_i + \epsilon^2 z_i^\star z_i)\right]\nonumber \\
		\times & \exp\left[ -i \sum_{i} \left[z^{\star}_i y_i  (\lambda^\star + r x_i -1)+ z_i y^{\star}_i  (\lambda + r x_i -1) \right]  \right]\nonumber \\
		\times & \exp\Bigg[ - \frac{r^2 \sigma^2 }{2N} \sum_{ij} x_i^2(z_i^{\star} y_j + z_i y_j^{\star})^2\Bigg] , \label{averaged1}
	\end{align}
	where we have written fixed-point components at $x_i$ and not as $x_i^\star$, again to avoid confusion with complex-conjugation. 
	
	\medskip
	
	We now imagine that we coarse-grain by grouping components $i$ according to their values of $x_i$. That is, we split the interval $[0,1]$ into $1/\Delta$ bins of width $\Delta$. The centre of each bin is given by a value $x_\alpha$, where $\alpha$ runs from $0$ to $1/\Delta$. We approximate the values $x_i$ with the bin centre $x_\alpha$, such that there are $N_\alpha \sim {\cal O}(N)$ values of $x_i$ approximated by the value $x_\alpha$. We will first take the limit $N\to \infty$ and then the limit $\Delta \to 0$. The problem then reduces to that of finding the eigenvalue spectrum of a random matrix with block-structured statistics. This enables us to follow along the lines of previous works that have studied the eigenvalue spectra of block-structured random matrices \cite{baron2020dispersal, poley2024eigenvalue, patil2024spectral}. 
	
	We begin by rewriting Eq.~(\ref{averaged1}) as
	\begin{align}
		\exp\left[ -N \Phi(\lambda) \right] &= \int \prod_{\alpha i} \left( \frac{d^2z^\alpha_i d^2y^\alpha_i}{2 \pi^2}\right) \exp\left[ - \sum_{i} ( y^{\alpha\star}_i y^\alpha_i + \epsilon^2 z^{\alpha\star}_i z^\alpha_i ) \right]\nonumber \\
		\times & \exp\left[ -i \sum_{\alpha i} \left[z^{\alpha\star}_i y^\alpha_i  (\lambda^\star + r x_\alpha -1)+ z^\alpha_i y^{\alpha\star}_i  (\lambda + r x_\alpha -1) \right]  \right]\nonumber \\
		\times & \exp\Bigg[ - \frac{r^2 \sigma^2 }{2N} \sum_{\alpha \beta ij} x_\alpha^2(z_i^{\alpha\star} y^\beta_j + z^\alpha_i y_j^{\beta\star})^2\Bigg],  \label{averagedalpha}
	\end{align}
	where the indices $i$ and $j$ now only run over the components in the groups $\alpha$ and $\beta$ respectively. We then introduce the following order parameters
	\begin{align}
		u &= \frac{1}{N}\sum_{\alpha i} x_\alpha^2 z^{\alpha\star}_i z^\alpha_i, \,\,\,\,\, v_\alpha = \frac{1}{N_\alpha}\sum_{i }  y^{\alpha\star}_i y^\alpha_i, \nonumber \\
		w_\alpha &=  \frac{1}{N_\alpha}\sum_i z^{\alpha\star}_i y^\alpha_i, \,\,\,\,\, w_\alpha^\star = \frac{1}{N_\alpha} \sum_{i } y^{\alpha\star}_i z^\alpha_i.\label{orderparameters}
	\end{align}
	We impose these definitions in the integral in Eq.~(\ref{averagedalpha}) using Dirac delta functions in their complex exponential representation. We can thus rewrite Eq.~(\ref{averagedalpha}) in the form
	\begin{align}
		\exp\left[ -N \Phi(\lambda) \right] &= \int \mathcal{D}\left[ \cdots \right]\exp\left[N(\Psi + \Theta + \Omega)\right] ,\label{saddlesetup}
	\end{align}
	where $\mathcal{D}\left[ \cdots \right]$ denotes integration over all of the order parameters and their conjugate (`hatted') variables, and where
	\begin{align}
		\Psi &= i  \hat u u  + i \sum_{\alpha} \gamma_\alpha(\hat v_\alpha  v_\alpha +\hat w_\alpha  w^\star_\alpha  +   \hat w^\star_\alpha  w_\alpha   ), \nonumber \\
		\Theta &= -\sum_\alpha v_\alpha - r^2 \sigma^2 u \sum_\alpha \gamma_\alpha  v_\alpha - i \sum_\alpha \gamma_\alpha [w_\alpha (\lambda^\star + r x_\alpha -1) + w_\alpha^\star (\lambda + r x_\alpha -1)] , \nonumber \\
		\Omega &=  \sum_\alpha \gamma_\alpha \ln  \Bigg[\int   \left( \frac{d^2z^\alpha d^2y^\alpha}{2 \pi^2}\right)  \exp\bigg\{ -i  [(\hat u x^2_\alpha+ i \epsilon^2) z^{\alpha\star} z^\alpha  + \hat v_\alpha   y^{\alpha\star} y^\alpha   + \hat w_\alpha   y^{\alpha\star} z^\alpha + \hat w_\alpha^{\star}  z^{\alpha\star} y^\alpha  ] \bigg\} \Bigg] . \label{argument}
	\end{align}
	We have defined $\gamma_\alpha = N_\alpha/N$. We note that in the limit $N\to \infty$ and $\Delta \to 0$, this quantity will be given by the fixed-point distribution $\gamma_\alpha \approx \Delta \times P(x_\alpha)$. We further note that the integrals over $y_i$ and $z_i$ in Eq.~(\ref{argument}) are uncoupled for different values of $i$ as a result of introducing the order parameters in Eq.~(\ref{orderparameters}). We highlight also that we have neglected terms involving $N^{-1} \sum_{i} (y^\alpha_i)^2$, $N^{-1} \sum_{i} (z^\alpha_i)^2$, $N^{-1} \sum_{i} z^\alpha_i y^\alpha_i$ and similar terms involving the complex conjugates of $z^\alpha_i$ and $y^\alpha_i$, which do not contribute in the thermodynamic limit (see Refs.~\cite{edwardsjones, sommers, haake, baron2022eigenvalues} for further discussion).
	
	Carrying out the integrals over the variables $y_i$ and $z_i$ in the expression for $\Omega$ in Eq.~(\ref{argument}) one obtains
	\begin{align}
		\Omega &= -\sum_\alpha \gamma_\alpha \ln\left[\hat w_\alpha  \hat w_\alpha^\star  - (x_\alpha^2 \hat u+ i \epsilon^2) \hat v_\alpha  \right].
	\end{align}
	We now suppose that $N \gg 1$, and we carry out the integral in Eq.~(\ref{saddlesetup}) in the saddle-point approximation. To do this, we extremise the expression $\Psi + \Theta + \Omega$. Extremising in turn with respect to the conjugate variables $\hat u, \hat v, \hat w_\alpha$ and $\hat w_\alpha^\star$, and subsequently taking the limit $\epsilon \to 0$, we find
	\begin{align}
		i u &=  - \sum_\alpha\gamma_\alpha x^2_\alpha  \frac{\hat v_\alpha}{f_\alpha} , \,\,\,\,\, iv_\alpha = -  x_\alpha^2 \frac{\hat u  }{f_\alpha}  , \,\,\,\,\,  i w_\alpha = \frac{\hat w_\alpha}{f_\alpha } , \,\,\,\,\,  i w_\alpha^\star = \frac{\hat w_\alpha^\star}{ f_\alpha} , \nonumber \\
		f_\alpha &= \hat w_\alpha  \hat w_\alpha^\star - x_\alpha^2 \hat u \hat v_\alpha  .\label{diffwrthatted}
	\end{align}
	Next, we extremise $\Psi + \Theta + \Omega$ in Eq.~(\ref{saddlesetup}) with respect to  $u, v, w_\alpha$ and $w_\alpha^\star$, obtaining
	\begin{align}
		i\hat u &= r^2\sigma^2 v , \,\,\,\,\, i\gamma_\alpha\hat v_\alpha = 1+ r^2\sigma^2 \gamma_\alpha  u , \nonumber \\
		i\hat w_\alpha &=  i (\lambda + r x_\alpha - 1), \,\,\,\,\, i\hat w_\alpha^\star = i(\lambda^\star + r x_\alpha - 1) , \label{diffwrtunhatted}
	\end{align}
	where we introduce $v = \sum_\alpha \gamma_\alpha  v_\alpha$. In a similar fashion to Ref.~\cite{baron2020dispersal}, we observe that the quantities $w_\alpha$ and $w_\alpha^\star$, when evaluated at the saddle point, are related to the resolvent via
	\begin{align}
		G(\lambda) &= \frac{\partial \Phi(\lambda, \lambda^\star)}{\partial \lambda} = i\sum_\alpha  \gamma_\alpha w^\star_\alpha , \nonumber \\
		G^\star(\lambda) &= \frac{\partial \Phi(\lambda, \lambda^\star)}{\partial \lambda^\star} =  i \sum_\alpha \gamma_\alpha w_\alpha . \label{resolventatsaddlepoint}
	\end{align}
	Thus, if we can solve Eqs.~(\ref{diffwrthatted}) and (\ref{diffwrtunhatted}) for $i\sum_\alpha  \gamma_\alpha w^\star_\alpha $ as a function of only $\lambda$ and $\lambda^\star$, we can obtain the eigenvalue density using Eq.~(\ref{densityfromres}).
	
	\medskip
	All that remains is to solve Eqs.~(\ref{diffwrthatted}) and (\ref{diffwrtunhatted}) for the quantities $w^\star_\alpha$, and we will have obtained the resolvent [and therefore eigenvalue density -- see Eq.~(\ref{densityfromres})] that we desire. We proceed by first eliminating $\hat u$. One then sees that 
	\begin{align} 
		v = r^2\sigma^2 v \sum_\alpha \gamma_\alpha \frac{x_\alpha^2 }{f_\alpha} . \label{unhatted}
	\end{align}
	
	\noindent This equation has two solutions. 
	\medskip
	
	\noindent\underline{First solution:}\\
	One solution is $v =0$, implying $\hat u = 0$ [see Eq.~(\ref{diffwrtunhatted})], and hence $f_\alpha =  \hat w_\alpha \hat w^\star_\alpha = -1/( w_\alpha w_\alpha^\star)$. We therefore obtain from Eqs.~(\ref{diffwrthatted}) and (\ref{diffwrtunhatted})
	\begin{align}
		1 &=    i w_\alpha^\star (\lambda + r x_\alpha - 1)  , \nonumber \\
		1 &=   i  w_\alpha(\lambda^\star + r x_\alpha - 1)   . \label{resolventoutside}
	\end{align}
	From this, one can solve for the resolvent $G(\lambda) = \frac{1}{N}\sum_\alpha  \gamma_\alpha w_\alpha^\star$, which we see is an analytic function of $\lambda$ in this case. Using Eq.~(\ref{densityfromres}), this means that the eigenvalue density vanishes in regions of the complex plane for which $v=0$ is the only valid solution of Eq.~(\ref{unhatted}).
	
	\medskip
	
	\noindent\underline{Second solution:} \\
	The other solution to Eq.~(\ref{unhatted}) is 
	\begin{align}
		r^2\sigma^2\sum_\alpha \gamma_\alpha \frac{ x_\alpha^2}{f_\alpha} = 1 . \label{conditionforinside}
	\end{align}
	Now, we see from the expression for $iu$ and $i \hat v_\alpha$ in Eqs.~(\ref{diffwrthatted}) and (\ref{diffwrtunhatted}) that when Eq.~(\ref{conditionforinside}) is satisfied, $u\to \infty$. This means that $i\hat v_\alpha \to r^2\sigma^2 x_\alpha^2 u$, and hence
	\begin{align}
		\hat v_\alpha \hat u \to  x_\alpha^2 g(\lambda, \lambda^\star),  
	\end{align}
	where $g(\lambda, \lambda^\star)$ is a function to be found that is independent of $\alpha$. Hence, we obtain the following simultaneous equations, which enable us to find the resolvent $G(\lambda, \lambda^\star) = \sum_{\alpha} \gamma_\alpha (i w_\alpha^\star)$:
	\begin{align}
		\frac{1}{r^2\sigma^2} &= \sum_\alpha \gamma_\alpha \frac{x_\alpha^2}{f_\alpha} , \nonumber \\
		- w_\alpha w_\alpha^\star &= 1/f_\alpha + x_\alpha^2 \frac{g(\lambda,\lambda^\star)}{f_\alpha^2}, \nonumber \\
		iw_\alpha  &=   \frac{\lambda + rx_\alpha - 1}{f_\alpha},\nonumber \\ 
		iw_\alpha^\star &= \frac{\lambda^\star + r x_\alpha-1}{f_\alpha} . \label{simfordens}
	\end{align}
	In principle, one can solve these along with Eq.~(\ref{conditionforinside}) to find $g(\lambda, \lambda^\star)$, $iw_\alpha$ and $iw_\alpha^\star$ as functions of $\lambda$ and $\lambda^\star$. In this case, the resolvent is no longer an analytic function of $\lambda$. Therefore, in the region of the complex plane where Eq.~(\ref{conditionforinside}) is satisfied, the eigenvalue density is non-zero.
	
	\subsubsection{Results for the spectral boundary}
	
\begin{figure}
		\centering   
		\includegraphics[width=0.8\textwidth]{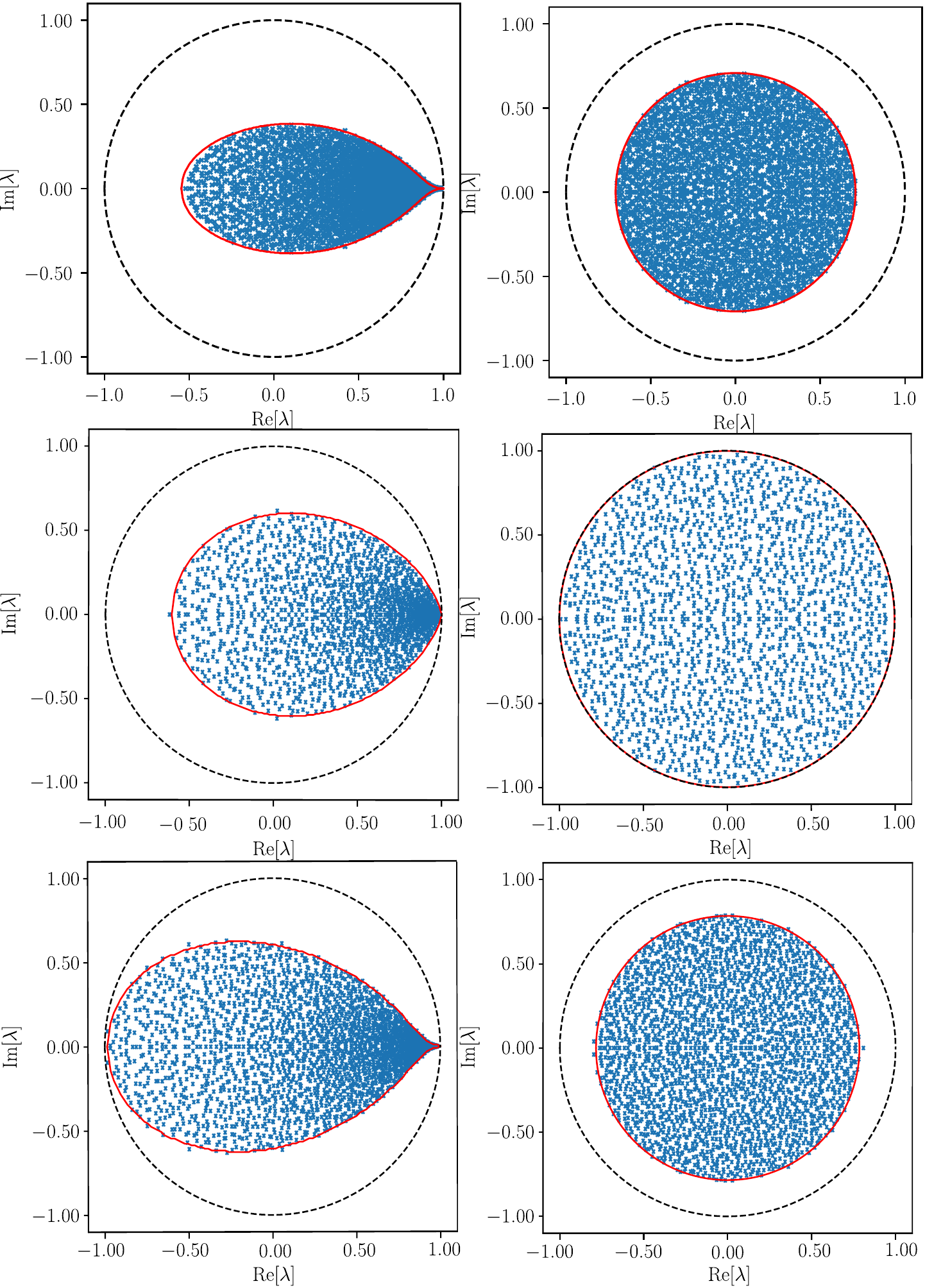}   
		\caption{{\bf Eigenvalue spectra of the reduced  Jacobian (left) and reduced interaction matrix (right)}. Markers are from simulations ($N=4000$), the red lines from the theory [red lines on the left are from Eq.~(\ref{boundaryj}), those on the right are a circle around the origin with radius $\sigma\sqrt{\phi}$, see Sec.~\ref{sec:red_int}]. The black dashed line in all panels is the unit circle. Upper row ($r=1.4$, $\sigma=0.8$): Spectra for parameters well inside the stable phase . The spectrum of the reduced Jacobian comes arbitrarily close to the point $\lambda=1$, indicating that fixed points are marginally stable. The eigenvalues of the reduced interaction matrix are well inside the unit disk (top right panel). Middle row ($r = 1.2$, $\sigma = 1.5$): Situation close to the onset of the May--Wigner instability (instability at $\lambda=1$). Lower row ($r = 1.6$, $\sigma = 1$) : Eigenvalue spectra close to the onset of the oscillatory instability at $\lambda=-1$. The spectral edge of the reduced Jacobian matrix approaches $-1$, while the spectrum of the reduced interaction matrix is contained inside the unit disk.  }
		\label{fig:spectra_grid}
	\end{figure}

	For our purposes, it is most important to understand where the region of non-zero eigenvalue density resides in the complex plane, rather than to calculate the precise density itself. The boundary of the eigenvalue spectrum is given by the set of values of $\lambda$ where both solutions for the resolvent (analytic and non-analytic) are satisfied simultaneously. Combining the analytic solution $f_\alpha = -1/(w_\alpha w^\star_\alpha) = \vert\lambda + r x_\alpha - 1\vert^2$ with Eq.~(\ref{conditionforinside}), one obtains
	\begin{align}
		\sum_\alpha \gamma_\alpha \frac{r^2 \sigma^2 x_\alpha^2}{\vert\lambda + r x_\alpha - 1\vert^2} = 1.
	\end{align}
	Now, taking the limit $\Delta \to 0$ and using $\gamma_\alpha = \Delta \times P(x_\alpha)$, this becomes
	\begin{align}
		\int_0^1 dx\, P(x) \frac{r^2 \sigma^2 x^2}{(\lambda_x + r x -1)^2 + \lambda_y^2} = 1 , \label{boundaryj}
	\end{align}
	where $\lambda = \lambda_x + i \lambda_y$, and where $P(x)$ is given by Eq.~(\ref{abundancedistribution}). This is Eq.~(3) in the main text. The expression describes the boundary of the eigenvalue spectrum by providing $\lambda_y$ as a function of $\lambda_x$ (or vice versa). The validity of this expression is verified in the left-hand panels of Figs. \ref{fig:spectra_grid}, as well as in Fig.~2 of the main text.
	
	\subsection{Instabilities}
	The fixed-point solution is unstable iff there are eigenvalues of the reduced Jacobian outside the unit disk in the complex plane. Through numerical evaluation of Eq.~(\ref{boundaryj}), we find that the eigenvalue spectrum only ever leaves the unit disk either at $\lambda=1$ or at $\lambda=-1$; the spectrum does not exit at any other points on the unit circle. As a next step, we therefore determine the largest and smallest real eigenvalues of the reduced Jacobian matrix.  
	
	\subsubsection{Largest real eigenvalue and the May--Wigner instability}\label{sec:largest_eigenvalue}
	We now proceed to find the location of the right-hand edge of the eigenvalue spectrum of the reduced Jacobian. We first write the integral in Eq.~(\ref{boundaryj}) as follows
	\begin{align}\label{eq:lambda1_sing}
		&\int_0^1 dx~ P(x) \frac{r^2 \sigma^2 x^2}{(\lambda_x + r x -1)^2 + \lambda_y^2} \nonumber \\
		&= \sigma^2 \phi+ \sigma^2\int_0^1 dx P(x)\left[ \frac{(1-\lambda_x)}{r}\frac{2 r (rx -1 +\lambda_x)}{(\lambda_x + r x -1)^2 + \lambda_y^2} + \frac{(1-\lambda_x)^2- \lambda_y^2}{(\lambda_x + r x -1)^2 + \lambda_y^2}\right], 
	\end{align}
	where we recall that the integral is over $0<x<1$ (the boundaries are excluded) and $\phi=\int_{0<x<1} dx\, P(x)$ is the fraction of degrees of freedom that are not frozen (see Sec.~\ref{sec:fp_sol}). 
	\\
	
	\noindent {\em Right edge of the spectrum for $\phi \sigma^2<1$:}
	\\
	
	The right-hand edge of the spectrum lies on the real axis, and we therefore consider the limit of small $|\lambda_y|$ in Eq.~(\ref{eq:lambda1_sing}). In this limit and if $0<(1-\lambda_x)/r<1$, the remaining integrals are dominated by the behaviour around the pole $x=(1-\lambda_x)/r$. We consider the function $P(x)$ to be effectively constant around this point, and we can thus take this out of the integral. We can thus evaluate the remaining integrals to obtain
	\begin{align}
		\int_0^1 dx P(x) \frac{r^2 x^2}{(\lambda_x + r x -1)^2 + \lambda_y^2} &\approx  \phi + P\left(\frac{1-\lambda_x}{r}\right)\frac{(1-\lambda_x)}{r} \ln\left[ \frac{(r- 1 + \lambda_x)^2+\lambda_y^2}{( 1 - \lambda_x)^2+\lambda_y^2}\right] \nonumber \\
		&+ P\left(\frac{1-\lambda_x}{r}\right)\frac{(1-\lambda_x)^2- \lambda_y^2}{r \vert\lambda_y\vert}\left[\arctan\left(\frac{r-1+\lambda_x}{\vert\lambda_y\vert} \right)+ \arctan\left(\frac{1-\lambda_x}{\vert\lambda_y\vert} \right)\right] .
	\end{align}
	Now, using the condition in Eq.~(\ref{boundaryj}) and after some algebra, we recognise that as $\lambda_y\to 0$, we must have $\lambda_x \to 1$, with the scaling $\lambda_y \sim(1-\lambda_x)^2$. As a consequence, we then also have $\lambda_y^2\ll (1-\lambda_x)^2$ and $(1-\lambda_x) \ll r$. Therefore, we arrive at 
	\begin{align}
		\vert\lambda_y\vert \approx \frac{\pi\sigma^2(1-\lambda_x)^2P\left(\frac{1-\lambda_x}{r}\right)}{r(1-\sigma^2 \phi) + 2 \sigma^2(1-\lambda_x)\ln\left[ \frac{1-\lambda_x}{r}\right] P\left(\frac{1-\lambda_x}{r}\right)}. \label{marginalspectrum} 
	\end{align}
	
	This approximation is valid for small $\lambda_y$, and connects $\lambda_x$ and $\lambda_y$ on the border of the eigenvalue spectrum of the reduced Jacbobian, near the right-hand edge of the spectrum. In the calculation, we have made the assumption $0<(1-\lambda_x)/r<1$. The inequality $(1-\lambda_x)/r<1$ is fulfilled for $\lambda_x$ close to one (as $r$ is fixed). In addition, we must require $\lambda_x<1$ to satisfy $(1-\lambda_x)/r>0$. Eq.~(\ref{marginalspectrum}) is therefore valid only for $\lambda_x$ just below one. 
	
	\medskip
	
\begin{figure}[h]
		\centering   
		\includegraphics[width=0.5\linewidth]{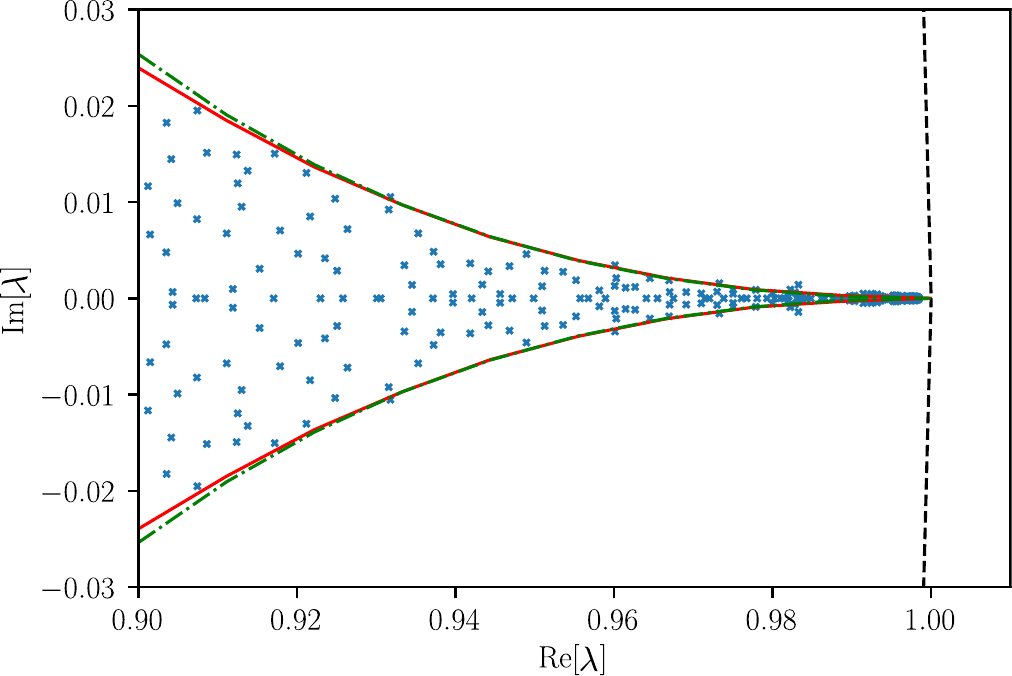}
		\caption{Comparison of the approximation in Eq.~(\ref{marginalspectrum}) (dot-dashed green line) to the full theory in Eq.~(\ref{boundaryj}) (solid red line) at the right-most edge of the spectrum. The boundary of the spectrum always touches the point $\lambda = 1$. Blue crosses are the results of numerical realisations of the reduced Jacobian matrix for system parameters $r = 1.6$, $\sigma = 0.8$ and $N = 4000$. The black dashed line indicates the unit circle in the complex plane. }
		\label{fig:zoomedspectrum}
	\end{figure}

	We also note that $1-\phi\sigma^2$ must be positive in order for Eq.~(\ref{marginalspectrum}) to be meaningful.  Otherwise the right-hand side of the equation becomes negative for $\lambda_x\uparrow 1$, inconsistent with the fact that the left-hand side is non-negative. Provided that this condition holds, the relation in Eq.~(\ref{marginalspectrum}) describes the edge of the eigenvalue spectrum of the reduced Jacobian to the left of the point $(1,0)$ in the complex plane. For $\lambda_y \to 0$ we find $\lambda_x \uparrow 1$. Therefore, for $\phi\sigma^2<1$ the rightmost eigenvalue of the reduced Jacobian is always $\lambda_x = 1$. We test the approximation in Eq.~(\ref{marginalspectrum}) in Fig. \ref{fig:zoomedspectrum}. \\

	\noindent {\em Instability condition $\phi \sigma^2=1$:}\\
	
	When the model parameters ($r$ and $\sigma^2$) are varied so that we reach the point where $\phi \sigma^2 = 1$, the assumptions that lead to Eq.~(\ref{marginalspectrum}) become invalid. The eigenvalue spectrum ceases to have its boundary at $\lambda_x = 1$. The leading eigenvalue now strays past the point $\lambda_x = 1$ for $\lambda_y = 0$ [making the assumption that $0<(1-\lambda_x)/r<1$ invalid], and the system becomes unstable. We therefore identify $\phi\sigma^2=1$ as the condition for the onset of the May--Wigner transition. As discussed further in Sec.~\ref{sec:red_int}, this instability can also be identified from the spectrum of the so-called reduced interaction matrix.

	\subsubsection{Left-most eigenvalue and the oscillatory instability}
	Assuming that the left-most eigenvalue is real-valued and such that $(1-\lambda_x)/r>1$, we can set $\lambda_y=0$ in Eq.~(\ref{boundaryj}), and find that the integral converges. One must then solve the following numerically for the leftmost eigenvalue $\lambda_L$
	\begin{align}
		\int_0^1 dx\, P(x) \frac{r^2 \sigma^2 x^2}{(\lambda_L + r x -1)^2} = 1. \label{leftmost}
	\end{align}
	An oscillatory instability occurs when $\lambda_L=-1$, i.e., when the following criterion is satisfied by $\sigma$ and $r$
	\begin{align}
		\int_0^1 dx\, P(x) \frac{r^2 \sigma^2 x^2}{( r x -2)^2} = 1. \label{2cyclecriterion}
	\end{align}
	We note that the assumption $(1-\lambda_x)/r>1$ translates to $r<2$ for $\lambda_x=1$. For $r>2$ the fixed point is unstable for all $\sigma^2>0$. 
	
	\subsection{Further observations and remarks}
	\subsubsection{Further discussion and relation to reduced interaction matrix}\label{sec:red_int}
	To make what happens at the point $\phi \sigma^2 = 1$ more clear, we turn to the reasoning of Ref.~\cite{baron2023breakdown} (see also \cite{stone}), and consider the reduced interaction matrix (the interaction matrix $a_{ij}$ restricted to non-frozen components). The eigenvalues of this matrix also yield information about stability. Briefly, this can be seen as follows. We wish to find the point in parameter space at which an eigenvalue of the reduced Jacobian $\underline{\underline{J}}$ first becomes greater than 1. This point coincides with the determinant $\det\left(\underline{\underline{J}} - \underline{\underline{\id}}\right)$ changing sign. Given that we only consider $i$ for which $x_i^\star>0$, we see from Eq.~(\ref{eq:jstar}) [$    J_{ij} - \delta_{ij}= 
	r x_i^\star (a_{ij} -\delta_{ij})$] that $\det\left( \underline{\underline{J}} - \underline{\underline{\id}}\right)$ can only change sign when $\det\left(\underline{\underline{a}} - \underline{\underline{\id}}\right)$ changes sign, where $\underline{\underline{a}}$ here means the interaction matrix with rows and columns corresponding to frozen components removed. Therefore, an eigenvalue of $\underline{\underline{J}}$ becoming greater than one means that an eigenvalue of the reduced interaction matrix $\underline{\underline{a}}$  becoming greater than one. Similar to \cite{baron2023breakdown}, the eigenvalues of the reduced interaction matrix are confined to a circle with radius $ \sqrt{\phi} \sigma$. This is also verified in Fig.~\ref{fig:spectra_grid}. The system therefore becomes unstable when 
	\begin{align}
		\phi \sigma^2 = 1, \label{opperinstability}
	\end{align}
	in keeping with what we found above. 
	
	We note that the reduced interaction matrix cannot inform us about the oscillatory chaotic instability, which corresponds to the bulk of the eigenvalue spectrum leaving the unit circle at $\lambda = -1$. This is because the transition is instead detected via a sign change of $\det\left(\underline{\underline{J}} + \underline{\underline{\id}}\right)$. Unlike, $\underline{\underline{J}} - \underline{\underline{\id}}$, the matrix $\underline{\underline{J}} + \underline{\underline{\id}}$ cannot be factorised as a simple product involving the positive definite diagonal matrix with entries $x_i^\star$.

	\subsubsection{Leftmost eigenvalue in the absence of a cutoff on the $x_i$}\label{section:nocutoff}
	In our model the function $H(u)$ takes the value $H=1$ for $u\geq 1$. This imposes an upper cutoff on the $x_i$, i.e., all $x_i(t)$ take values of at most one. The effect of this cutoff on stability becomes clear at this point. Suppose that there were no upper bound [i.e., $H(u)=u$ for all $u\geq 0$]. Points $\lambda=\lambda_x+i\lambda_y$ on the edge of the spectrum then satisfy
	\begin{align}
		\int_0^\infty dx\, P(x) \frac{r^2 \sigma^2 x^2}{(\lambda_x + r x -1)^2 + \lambda_y^2} = 1,
	\end{align}
	where the integral now extends to infinity rather than to one as in Eq.~(\ref{boundaryj}). For $\lambda_y = 0$, the pole of the integrand is at $x=(1-\lambda_x)/r$. This is now within the integration range $x>0$ for all values of $\lambda_x<1$. Assuming $\lambda_x<1$, we therefore approximate the integral for small $|\lambda_y|$ by assuming that it is is dominated by the maximum of the integrand. We find again Eq.~(\ref{marginalspectrum}), which we repeat here for convenience
	\begin{align}\label{marginalspectrum2}
		\vert\lambda_y\vert \approx \frac{\pi\sigma^2(1-\lambda_x)^2P\left(\frac{1-\lambda_x}{r}\right)}{r(1-\sigma^2 \phi) + 2 \sigma^2(1-\lambda_x)\ln\left[ \frac{1-\lambda_x}{r}\right] P\left(\frac{1-\lambda_x}{r}\right)}. 
	\end{align}
	Since Eq.~(\ref{marginalspectrum2}) is now also valid when $(1-\lambda_x)/r>1$, and we note now that $\lim_{u\to\infty} u^2 P(u)=0$ [since the stationary distribution of the components $P(x)$ is now a Gaussian with no upper cutoff], this means that solutions of Eq.~(\ref{marginalspectrum2}) with $|\lambda_y|\ll 1$ can be found for $\lambda_x \ll -1$. There is therefore no leftmost eigenvalue -- the spectrum extends indefinitely into the left half-plane. For this reason, in the thermodynamic limit, the fixed point is always unstable for $\sigma>0$ in the model without upper cutoff on the $x_i$. Such a cutoff is therefore a necessary requirement to have stability in the presence of disordered couplings. 
	
	\subsubsection{$\sigma \to 0^+$ in the case with upper cut-off}\label{section:rc2}
	We have noted already that, in the presence of the cut-off on $x$ (i.e. when components saturate at $x_i =1$), the support of the eigenvalue spectrum can only be such that $0<(1-\lambda_x)/r<1$. We have also shown above that even minimal disorder leads to instability for all $\sigma>0$ when there is no cutoff. Let us now understand instability onset in the case where $\sigma \to 0^+$ in the presence of the cutoff. We do this by considering the leftmost eigenvalue $\lambda_L$.
	
	One might be tempted to assume that the integral in Eq.~(\ref{leftmost}) is dominated by the contribution at $x = 1-1/r$ in this case, since $P(x)$ has its peak there [see Eq.~(\ref{abundancedistribution})]. If we proceed along these lines, one finds that Eq.~(\ref{leftmost}) yields
	\begin{align}
		(r-1)^2 \sigma^2 \approx (\lambda_L + r - 2)^2,
	\end{align}
	which gives $\lambda_L = 2-r$, as is the case for the usual logistic map. However, this value of $\lambda_L$ would mean that the integrand of Eq.~(\ref{leftmost}) would have a pole at $x= (1-\lambda_L)/r = (r-1)/r$ (which is in the integration range $0<x<1$), and the integral would no longer be dominated by the peak of $P(x)$ -- i.e. the assumption breaks down. 
	
	If we instead assume that the integrand Eq.~(\ref{leftmost}) does indeed have a pole in the integration range, one simply obtains Eq.~(\ref{marginalspectrum2}) again with the replacements $\lambda_y = 0$ and $\lambda _x = \lambda_L$. The solution to this is given by $P\left( \frac{1-\lambda_L}{r}\right) = 0$, which yields $ \frac{1-\lambda_L}{r} = 1 \Rightarrow \lambda_L = 1-r$. This again yields a contradiction, because the supposed pole at $x = (1-\lambda_L)/r$ is not within the integration range.
	
	We thus see that for small $\sigma$, $\lambda_L$ must be such that Eq.~(\ref{leftmost}) comes close enough to having a pole in the integration range that the integrand is dominated by contributions where $(\lambda_L + r x - 1)$ is small, but it does not quite enter the range of integration. One therefore has 
	\begin{align}
		\sigma^2 \sim (\lambda_L + r -1)^2 ,
	\end{align}
	for small $\sigma$. Equivalently, one has $\lambda_L \approx 1-r$ for $\sigma\to 0^+$, and the critical value at which the transition occurs is $r_c = 2$. This is consistent with the numerical solution of Eq.~(\ref{leftmost}), which was used to produce the solid line in Fig.~1 of the main text.

	\section{Dynamic mean-field theory part 2: power spectra of fluctuations}\label{sec:dmft_instab_sec}
	
	We now compute the power spectrum of fluctuations about the fixed-point solution when we add time-dependent noise to the dynamics. These power spectra characterise the dynamic nature of the two instabilities. 
	
	\subsection{Derivation of instabilities from dynamic mean-field theory}
	\subsubsection{Conditions for the onset of linear instability}\label{sec:dmft_instab}
	We follow the ideas of \cite{opper1992phase} and add a small dynamic noise component to the map,
	\begin{align}
		x_i(t+1) = H \left[ r x_i(t) \bigg\{1-x_i(t)+ \sum_{j} a_{ij} x_j(t) \bigg\}+\xi_i(t)\right], \label{eq:dis_log_map_sm_noise}
	\end{align}
	where the $\xi_i(t)$ represent the dynamic noise, which is assumed to be white, and set to be independent across components (i.e. $\xi_i$ and $\xi_j$ are not correlated). The noise does not have to be additive, as we will discuss below. The effective process in Eq.~(\ref{effproc}) therefore becomes
	\begin{align}
		x(t+1) = H\left\{rx(t)\left[1 - x(t) + \eta(t) \right] + \xi(t)\right\}, 
	\end{align}
	where $\xi(t)$ is Gaussian noise, and where we have again $\langle \eta(t) \eta(t') \rangle_\eta = \sigma^2 \langle x(t) x(t') \rangle_\eta$ [see Eq.~(\ref{eq:eta_stat})].
	
	\medskip
	
	\noindent We study three different types of white noise $\xi(t)$: 
	\begin{enumerate}
		\item[(1)] We follow Opper and Diederich \cite{opper1992phase} and consider noise with correlator $\langle \xi(t) \xi(t') \rangle_\xi = T r^2x^2(t)\delta_{t, t'}$, where $T>0$ is a fixed parameter. 
		\item[(2)] We consider demographic noise along the lines of Altieri et al \cite{altieri2021properties}, where one has $\langle \xi(t) \xi(t') \rangle_\xi = T r^2 x(t)\delta_{t, t'}$. 
		\item[(3)] We consider purely additive noise: $\langle \xi(t) \xi(t') \rangle_\xi = T r^2\delta_{t, t'}$.
	\end{enumerate}
	All three types of noise can be written in the form $\langle \xi(t) \xi(t') \rangle_\xi = T r^2 [x(t)]^\kappa\delta_{t, t'}$, with $\kappa=2, 1$ or $0$ in the three different cases.  
	
	\medskip
	
	Assuming the existence of a fixed point in absence of noise ($T=0$), we consider the small-noise limit ($T\ll 1$), and linearise the process expanding about the fixed point solution. For small $T$, the deviations from the fixed point are small. Writing $y(t) = x(t) - x^\star$ and $\delta\eta(t) = \eta(t) - \eta^\star$, we have 
	\begin{align}\label{eq:linearised}
		y(t+1) = y(t) + rx^\star \left[ -y(t) + \delta\eta(t) \right] + \xi^\star(t) 
	\end{align}
	for non-frozen components, where $\langle \xi^\star(t) \xi^\star(t') \rangle_\xi = T r^2 (x^\star)^\kappa\delta_{t, t'}$ is the approximate noise close to the fixed point (i.e., we make a linear-noise approximation \cite{gardiner2009stochastic}, as indicated by the asterisk in $\xi^\star$). We now introduce the discrete-time Fourier transform and its inverse
	\begin{align}
		\tilde g(\omega) &= \sum_{t= 0}^{n-1} e^{if t} g(t),  \nonumber \\
		g(t) &= \frac{1}{n}\sum_{m = 0}^{n-1} e^{ \frac{2 \pi i m t}{n}} \tilde g\left( \frac{2 \pi i m }{n}\right),
	\end{align}
	where $\omega = 2 \pi m/n$, $m = 0, 1, 2, \dots n-1$ and $n$ is the total number of time points in the time series we consider. For the purposes of the further calculation, we take the limit $n \to \infty$. Applying the Fourier transform in Eq.~(\ref{eq:linearised}), we thus have for the power spectrum of fluctuations about the fixed point
	\begin{align}
		\langle \vert \tilde y(\omega) \vert^2 \rangle_S &= B_2 \sigma^2 \langle \vert \tilde y(\omega) \vert^2 \rangle_S + B_\kappa,   \label{powerspectrumgeneral1}
	\end{align}
	where we define
	\begin{align}\label{eq:A1A2}
		B_\ell &= \left \langle r^2 (x^\star)^\ell \left\vert (e^{i\omega}-1) + r x^\star \right\vert^{-2}\right\rangle_S
	\end{align}
	for $\ell=0,1,2,3$ with $\langle \cdots\rangle_S$ denoting an average over non-frozen components and the external noise $\xi(t)$. We can re-write Eq.~(\ref{powerspectrumgeneral1}) as 
	\begin{align}\label{powerspectrumgeneral2}
		\langle \vert \tilde y(\omega) \vert^2 \rangle_S &= \frac{ T B_\kappa}{1 - B_2 \sigma^2}.
	\end{align}
	If the denominator in Eq.~(\ref{powerspectrumgeneral2}) $1-B_2 \sigma^2$ becomes zero for a particular frequency $\omega$, the corresponding entry in the power spectrum $\langle \vert \tilde y(\omega) \vert^2\rangle_S$ diverges, signalling the onset of instability. We find that this can occur either at $\omega=0$ or $\omega=\pi$, and we will discuss these cases in turn.
	
	\medskip
	
	\noindent\underline{Instability at $\omega=0$}\\
	For $\omega = 0$, we have $B_2 = \phi$, and the power spectrum diverges at $\omega = 0$ if
	\begin{align}
		\phi \sigma^2 = 1.
	\end{align}
	One notes that this is independent of the kind of noise that we add (i.e. independent of $\kappa$). This is the instability first identified in disordered continuous-time replicator models by Opper and Diederich \cite{opper1992phase}, and corresponds to the condition that we obtained in Eq.~(\ref{opperinstability}), where an eigenvalue crosses at $\lambda = 1$.
	
	\medskip
	
	\noindent\underline{Instability at $\omega=\pi$}\\
	In our model, we also find instabilities for $\omega = \pi$. Setting the denominator in Eq.~(\ref{powerspectrumgeneral2}) to zero (for $\omega=\pi$) this occurs when
	\begin{align}\label{eq:pi_instab}
		\left\langle\frac{(x^{\star})^2}{ \left\vert  r x^\star - 2 \right\vert^2}\right\rangle_S = \frac{1}{\sigma^2}
	\end{align}
	This is identical to the condition signaling that the left-most eigenvalue of the reduced Jacobian becomes equal to $-1$ [see Eq.~(\ref{2cyclecriterion})]. For $r<2$ the denominator in Eq.~(\ref{eq:pi_instab}) has no zero in the integration range $0<x<1$, and the integral is finite. For a given $r<2$ the instability thus occurs at a non-zero value of $\sigma^2$. For $\sigma^2$ smaller than this value the fixed point is stable, above it the fixed point is unstable.
	
	\medskip
	
	We note that the denominator of $B_\ell$ has a quadratic pole at $x^\star=2/r$ when $\omega = \pi$. For $r<2$, this pole is not in the integration range $0<x^\star<1$. However, as we approach $r = 2$ from below, the integrals $B_\ell$ become dominated by the behaviour of the integrand near pole (which is now located just outside the integration range). So, even for very small $\sigma$, we have that $B_\ell \to \infty$ for $\omega=\pi$ as $r\uparrow 2$. This means that we cannot possibly have a stable fixed point solution, since $\langle \vert \tilde y(\omega=\pi)\vert^2 \rangle_S$ in Eq.~(\ref{powerspectrumgeneral2}) would become negative for $\omega=\pi$. Hence, the system is unstable against perturbations with $\omega=\pi$ for any non-zero $\sigma^2$ when $r>2$. This is in agreement with the earlier finding in Section \ref{section:rc2} that $r_c = 2$ is the critical value at which the oscillatory transition takes place for $\sigma \to 0^+$. 
	
	\subsubsection{Comment on the model without upper cutoff}
	The condition for the onset of instability at $\omega=\pi$ in the model without upper cutoff is also given by Eq.~(\ref{eq:pi_instab}), except that the average over $x$ now extends over the entire range $x>0$. This means that the zero of the denominator ($x=2/r$) is within the integration range for all $r>0$, and thus the left-hand side of Eq.~(\ref{eq:pi_instab}) diverges. In the model without a cutoff, the fixed point is thus unstable against perturbations with $\omega=\pi$ for all $r>0, \sigma>0$. This is again in agreement with our earlier findings in Section \ref{section:nocutoff} using the random-matrix approach.

	\subsubsection{Connection between random-matrix theory approach and dynamic mean-field theory}
	The condition for the onset of any instability is $B_2\sigma^2=1$, or in other words
	\begin{equation}\label{eq:help1}
		\int_0^1 dx\, P(x) \frac{\sigma^2 r^2 x^2}{\vert (e^{i\omega}-1) + r x \vert^2}=1.
	\end{equation}
	For clarity, we note again that the boundaries $x=0$ and $x=1$ themselves are excluded from the integration range. Eq.~(\ref{eq:help1}) can be rewritten as
	\begin{equation}
		\int_0^1 dx\, P(x) \frac{\sigma^2 r^2 x^2}{(\cos\omega-1+rx)^2+\sin^2\omega}=1.
	\end{equation}
	
	Comparison with Eq.~(\ref{boundaryj}) then reveals the correspondence $\lambda_x=\cos\omega$ and $\lambda_y=\sin\omega$. This can be understood as follows. Suppose we start with a combination of model parameters for which the model has a stable fixed point. As the model parameters are varied the onset of instability occurs when the spectrum of the reduced Jacobian is no longer contained in the unit disk. This departure can, in principle, occur at any point $\lambda_d$ on the unit circle ($d$ stands for `departure'). Such an eigenvalue can then be written in the form $\lambda_d=\cos\omega_d+i\sin\omega_d$, with a unique $\omega_d\in[0,2\pi)$. In the analysis of the power spectrum of fluctuations $\langle|\tilde y(\omega)|^2\rangle_S$ along the lines of Sec.~\ref{sec:dmft_instab}, a divergence would then occur at $\omega=\omega_d$. In our model, the relevant instabilities occur at $\lambda_d=1$ (corresponding to $\omega_d=0$) and $\lambda_d=-1$ ($\omega_d=\pi$). 
	
	We have not attempted to prove analytically that one of these instabilities always occurs  before any linear instability at  points $\lambda_d\not\in\{-1,1\}$ (equivalently $\omega_d\not\in\{0,\pi\}$). However, we have not found any evidence of any other first departure points. It is possible though to construct variants of the H\'enon map (Sec.~\ref{sec:henon}) with other departure points. This can be achieved for negative values of $b$ in Eq.~(\ref{eq:henon_disordered}), for example.

	\subsection{Fluctuation spectra near the May--Wigner instability}\label{sec:one_over_f}
	In this subsection and the next, we analyse the shape of the power spectrum close to the two instabilities more carefully. We start with the instability at $\omega = 0$ (the May--Wigner instability), and discuss the oscillatory instability at $\omega=\pi$ in Sec.~\ref{sec:one_over_f_squared}. 
	
	Our goal is first to approximate the expression in Eq.~\ref{powerspectrumgeneral2} for values of $\omega$ close to zero. To do this, we follow the procedure first devised in \cite{opper1992phase}, and also described in \cite{froy}. We have for small $\omega$ 
	\begin{align}
		B_2 &\approx \left\langle\frac{(r x^\star)^2}{  \omega^2 + (r x^\star)^2 }\right\rangle_S = \int_0^1 dx P(x) \left[1 - \frac{\omega^2/r^2}{  \omega^2/r^2 +  x^2 }\right] \approx \phi - \frac{\vert\omega\vert \pi}{2r} P(0^+) ,
	\end{align}
	where we have used $\lim_{\epsilon\to 0}\epsilon/(\epsilon^2 + x^2) = \pi \delta(x)$. 
	
	We also need to compute $B_\kappa$ [see Eq.~(\ref{eq:A1A2})] for $\kappa=0$ and $\kappa=1$. For the case $\kappa = 1$, we have
	\begin{align}
		B_{1} &\approx \left\langle\frac{r^2 x^\star}{  \omega^2 + (r x^\star)^2 }\right\rangle_S \approx P(0^+) \log\left\vert\frac{r}{\omega}\right\vert.
	\end{align}
	That is, there is a logarithmic divergence at $\omega = 0$. In the case $\kappa = 0$, one has
	\begin{align}
		B_{0} \approx \left\langle\frac{r^2 }{  \omega^2 + (r x^\star)^2 }\right\rangle_S \approx \frac{\pi r}{2\vert\omega\vert}P(0^+).
	\end{align}
	We therefore have from Eq.~(\ref{powerspectrumgeneral2}), 
	\begin{align}
		\langle \vert \tilde y(\omega) \vert^2 \rangle_S \approx \begin{cases}
			\frac{T \pi rP(0^+)/(2\vert\omega\vert)}{(1 - \phi \sigma^2 )+ \sigma^2 \vert\omega\vert \pi P(0^+)/(2r)} \hspace{0.5cm} \mathrm{for}\hspace{0.2cm} \kappa = 0,\\  ~\\
			\frac{T  P(0^+) \log \vert r/\omega\vert}{(1 - \phi \sigma^2 )+ \sigma^2\vert\omega\vert \pi P(0^+)/(2r)}\hspace{0.5cm} \mathrm{for}\hspace{0.2cm} \kappa = 1,\\~\\
			\frac{T \phi}{(1 - \phi \sigma^2 )+ \sigma^2\vert\omega\vert \pi P(0^+)/(2r)}\hspace{0.5cm} \mathrm{for}\hspace{0.2cm} \kappa = 2.
		\end{cases} \label{powerspectrumapproxopper}
	\end{align}
	These expressions show that the shape of the power spectrum around $\omega = 0$ depends on the form of the input noise (i.e., on $\kappa$). This is because the integrands in the definition of $B_\ell$ [Eq.~(\ref{eq:A1A2})] have singularities near $x=0$ for $\omega\to 0$. A similar singularity occurs in the calculation of the eigenvalue spectrum near $\lambda=1$, see e.g Eq.~(\ref{eq:lambda1_sing}).
	
	When $\phi \sigma = 1$, and focusing on the case $\kappa=2$, we obtain the $1/f$ noise reported by Opper and Diederich \cite{opper1992phase} (see also Ref. \cite{froy}). However, this form of the fluctuation spectrum is not robust to changes in the noise. We do not find $1/f$ noise for $\kappa=0$ and $\kappa=1$ at the onet of instability. For $\kappa=0$ for example, we observe $1/\omega^2$ behaviour when $\phi\sigma^2=1$.

	\subsection{Oscillatory instability and $1/f^2$ noise}\label{sec:one_over_f_squared}
	We now adapt this approach to the case $\omega \approx \pi$. Expanding for small $\vert \pi - \omega\vert$, we have 
	\begin{align}
		e^{i \omega} - 1 = e^{i \pi} e^{i (\omega -\pi)} -1 \approx - 2 - i (\omega - \pi) + \frac{1}{2}(\omega - \pi)^2, 
	\end{align}
	and therefore
	\begin{align}\label{eq:help2}
		&B_2\approx \left\langle\frac{(r x)^2}{  (\omega-\pi)^2 + (2-r x - [\omega-\pi]^2/2)^2 }\right\rangle_S   .
	\end{align}
	In the model with cutoff which we discuss here, the instability at $\omega=\pi$ can only occur at a non-zero value of $\sigma^2$ for $r<2$ (for $r>2$ the model is unstable for all $\sigma>0$). Since the point $x^\star = 2/r$ is outside the range $0<x<1$ for $r<2$, there is no singularity in the denominator of Eq.~(\ref{eq:help2}) as $\omega \to \pi$. This means that we have
	\begin{align}
		&B_2\approx \int_0^1 dx P(x) \frac{(r x)^2}{(2-r x)^2 } -  (\omega-\pi)^2\int_0^1 dx P(x) \frac{(r x)^2[1-(2-r x)]}{(2-r x)^4 } \equiv I^{(2)}_1 - (\omega-\pi)^2 I^{(2)}_2.
	\end{align}
	We can approach the integral $B_\kappa$ for $\kappa\neq 2$ in a similar fashion, and we find
	\begin{align}
		&B_\kappa\approx \int_0^1 dx P(x) \frac{r^2 (x)^\kappa}{(2-r x)^2 } -  (\omega-\pi)^2\int_0^1 dx P(x) \frac{r^2 (x)^\kappa[1-(2-r x)]}{(2-r x)^4 } \equiv I^{(\kappa)}_1 - (\omega-\pi)^2 I^{(\kappa)}_2.
	\end{align}
	Expanding the expression in Eq.~(\ref{powerspectrumgeneral2}) for small $(\omega - \pi)^2$, we therefore have
	\begin{align}
		\langle \vert \tilde y(\omega) \vert^2 \rangle_S &= \frac{TI^{(\kappa)}_1}{(1- I^{(2)}_1 \sigma^2) + (\omega-\pi)^2I^{(2)}_2} . \label{powerspectrum2cycle}
	\end{align}
	Since $I^{(\kappa)}_\beta$ are just constants in $\omega$, the phenomenology around $\omega \approx \pi$ is very different from that near $\omega=0$ studied above. This can be attributed to the difference in behaviour of the eigenvalue spectrum at $\lambda = 1$ and $\lambda = -1$. As discussed earlier, the eigenvalue spectrum always (for all model parameters in the stable phase) comes arbitrarily close to $\lambda=1$. In contrast, again strictly within the stable phase, the left-most eigenvalue is separated from $-1$ by a finite distance. 
	
	The transition to instability occurs at $\sigma^2 I^{(2)}_1 = 1$. At this onset of instability, one always finds $1/f^2$ noise, regardless of the type of input noise (i.e., of the value of $\kappa$).
	
	\subsection{Comparison against simulations}
\begin{figure}
		\centering   
		\includegraphics[width=0.48\linewidth]{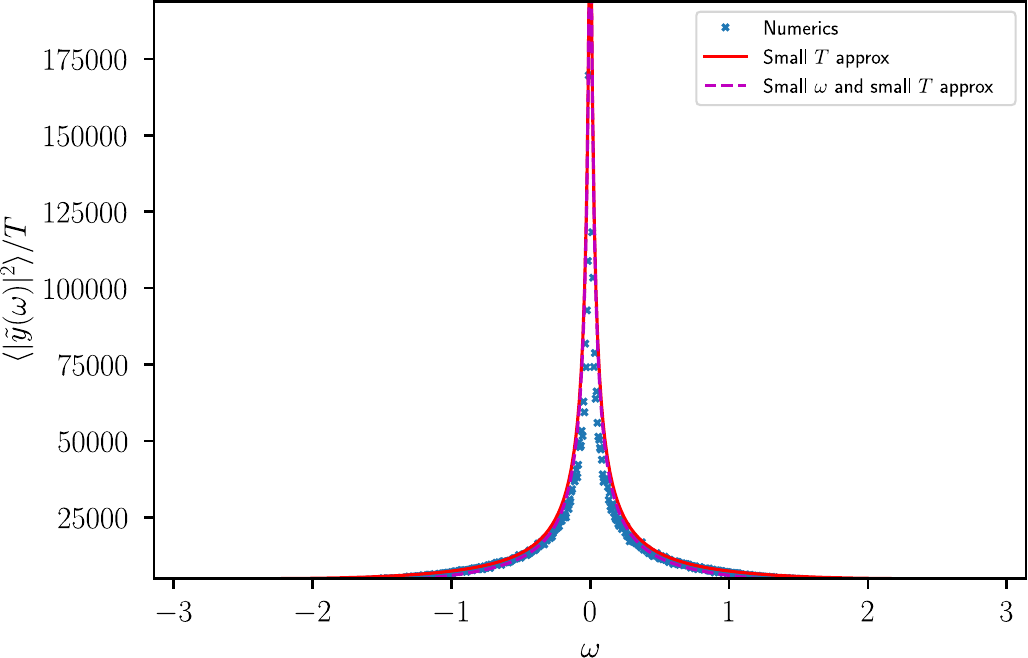}
		\includegraphics[width=0.45\linewidth]{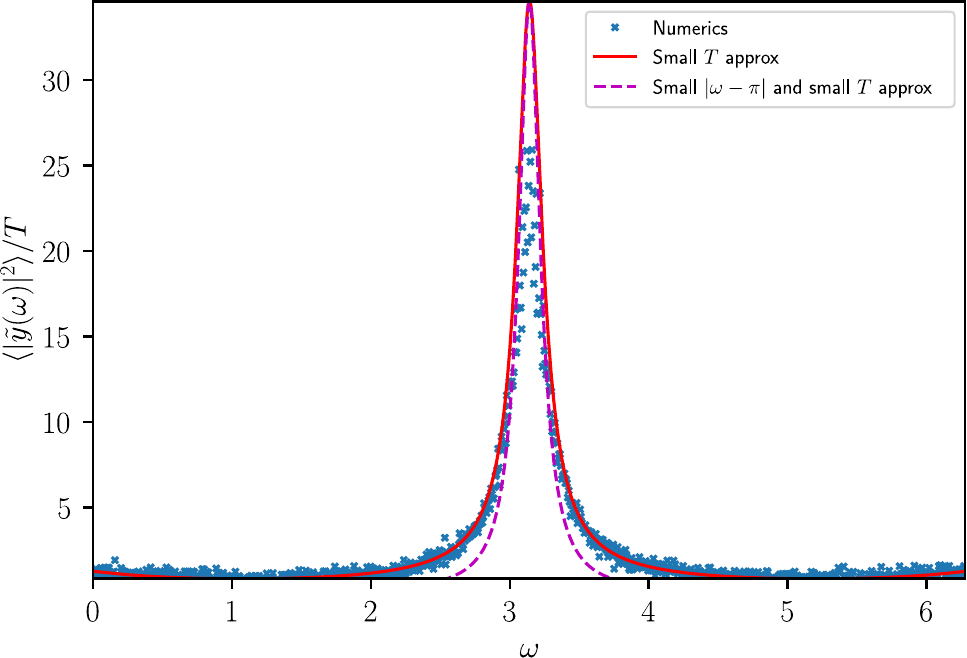}
		
		\caption{Power spectra close to the instability lines. The results of numerical simulation are compared with the predictions in Eqs.~(\ref{powerspectrumgeneral1}), (\ref{powerspectrumapproxopper}) and (\ref{powerspectrum2cycle}) in the case $\kappa = 2$. The solid red line is the prediction in Eq.~(\ref{powerspectrumgeneral1}), which is valid for small $T$ (since the linear-noise approximation is used). The dashed purple lines are the predictions of Eqs.~(\ref{powerspectrumapproxopper}) and (\ref{powerspectrum2cycle}), which presume a sharp peak at $\omega = 0$ and $\omega = \pi$ respectively (i.e. that one is close to the transition). Panel (a): Power spectrum in the case $r = 1.2$, $\sigma = \sqrt{2.1}$, $T = 0.02$, close to the onset of disorder-induced chaos. There is a peak at $\omega = 0$ and we observe a $\sim1/\omega$ decay close to the peak. Panel (b): Power spectrum in the case $r = 1.8$, $\sigma = 0.59$, $T = 10^{-6}$, close to the onset of quasi-periodic chaos. There is now a peak at $\omega = \pi$ and we observe a $\sim1/\vert\omega -\pi\vert^2$ decay close to this peak. }
		\label{fig:powerspectra}
	\end{figure}

	The predictions for the power spectrum are compared against simulations in Fig.~\ref{fig:powerspectra} in this Supplement.

	\section{Model with a homogeneous interaction term}\label{section:homogeneousinteraction}
	We now examine the case where we include an homogeneous interaction term, whose strength is dictated by the parameter $\mu$, as in Eq.~(4) of the main text, which we repeat here for convenience
	\begin{equation}
		x_i(t+1)=H\bigg[rx_i(t)[1-x_i(t)+\mu\{x_i(t)-M(t)\}+\sum_j a_{ij}x_j(t)\bigg],
	\end{equation}
	and we write $M(t) = N^{-1}\sum_i x_i(t)$. If $\sigma^2=0$ (i.e., all $a_{ij}=0$) and for homogeneous initial conditions $x_i(0)=x_0$, we have $x_i(t)=M(t)$ for all $i$ and $t$, and thus the model again reduces to the conventional logistic map.
	
	\medskip

	\subsection{Fixed point solution}
	Repeating the analysis of Section \ref{section:fixedpoint}, we now find that the stationary mean-field analysis yields
	\begin{equation}
		x^\star=H\bigg[r x^\star[1-(1-\mu)x^\star-\mu M^\star+\eta^\star]\bigg],
	\end{equation}
	where again $\langle (\eta^\star)^2 \rangle_\eta = \sigma^2 \langle (x^\star)^2 \rangle$. We note again that $x^\star=0$ is always a solution. Similarly, we have $x^\star=1$ if the argument inside $H(\cdot)$ is greater than one. Non-frozen solutions $0<x^\star<1$ are of the form
	\begin{equation}
		x^\star=\frac{1}{1-\mu}\left[1-\frac{1}{r}-\mu M^\star+\eta^\star\right].
	\end{equation}
	Following a similar reasoning to the case $\mu  = 0$ presented earlier, we conclude in general 
	\begin{equation}
		x^\star=H\left[\frac{1}{1-\mu}\big(1-\frac{1}{r}-\mu M^\star+\eta^\star\big)\right].
	\end{equation}  
	This means that the distribution of components at a fixed point takes the form
	\begin{align}
		P(x) &= \frac{1}{\sqrt{2\pi\sigma^2 q/(1-\mu)^2}} \exp\left[-\frac{\left(x-\frac{1}{1-\mu}\left(1-1/r-\mu M^*\right)\right)^2}{2\sigma^2 q/(1-\mu)^2}\right]\Theta\left( x\right)\Theta\left(1- x\right) \nonumber \\
		&+ \psi_0\delta(x) + \psi_1 \delta(x-1), \label{abundancedistribution_mu}
	\end{align}
	with
	\begin{align}\label{eq:phi0phi1_mu}
		\psi_0& = \int_{-\infty}^0 \frac{1}{\sqrt{2\pi\sigma^2 q/(1-\mu)^2}} \exp\left[-\frac{\left(x-\frac{1}{1-\mu}\left(1-1/r-\mu M^*\right)\right)^2}{2\sigma^2 q/(1-\mu)^2}\right] , \nonumber \\
		\psi_1 &= \int^{\infty}_1 dx  \frac{1}{\sqrt{2\pi\sigma^2 q/(1-\mu)^2}} \exp\left[-\frac{\left(x-\frac{1}{1-\mu}\left(1-1/r-\mu M^*\right)\right)^2}{2\sigma^2 q/(1-\mu)^2}\right] .
	\end{align}
	The self-consistency relations are 
	\begin{eqnarray}
		M^*&=&\int_0^1 dx\, P(x) x + \psi_1 \nonumber \\
		q&=& \int_0^1 dx\, P(x) x^2+\psi_1,\label{mandqmodmu}
	\end{eqnarray}
	where the boundaries are {\em not} included in the integrals on the right.
	
	\subsection{Modified reduced Jacobian and structure of the eigenvalue spectrum}
	Now that we have the distribution of the $x_i$ at the fixed poinr, we may also repeat the calculation of Section \ref{section:eigenvaluespectrum}. First, we require the modified expression for the reduced Jacobian. This is given by [c.f. Eq.~(\ref{eq:jstar})]
	\begin{equation}
		J_{ij}= 
		\delta_{ij} + r x_i^\star \left[a_{ij}' - \frac{\mu}{N} -\delta_{ij}\left(1-\mu \right)\right],
	\end{equation}
	where we recall that we restrict the indices $i$ and $j$ to those components that have $0<x_i^\star<1$. We are careful now to distinguish the original interactions $a_{ij}$ from the reduced interactions $a_{ij}'$. The interaction matrix elements, conditioned on survival, have correlations that are relevant for what follows, but which we could safely ignore before when $\mu = 0$, as we explain below.
	
	\medskip

	The introduction of homogeneous interactions (a non-zero $\mu$) has two main effects on the Jacobian, compared to the case $\mu=0$. First, the diagonal elements pick up a factor $1-\mu$. Second, the homogeneous interactions introduce a low-rank perturbation to the original interaction matrix $\underline{\underline{a}}$, and also to the reduced Jacobian. As is well known, low-rank perturbations can result in the production of outlier eigenvalues, while leaving the bulk region unchanged \cite{orourke, baron2022eigenvalue}.
	
	Our approach will therefore be as follows. First, we compute the bulk region to which the majority of the eigenvalues of the reduced Jacobian are confined. We do this following the procedure in Section \ref{section:eigenvaluespectrum}, considering the matrix with entries $J_{ij} - rx_i^\star \mu N^{-1}$.  Second, we consider the single outlier eigenvalue that can emerge due to the low-rank perturbation. This is accomplished using the approach of Ref. \cite{baron2023breakdown}. We note that for this outlier eigenvalue, we must consider carefully the intricate correlations of the Jacobian matrix elements that emerge when we condition on the components being non-frozen. These correlations are irrelevant for the bulk calculation, which is why we have been able to ignore them until now. 
	
	\subsection{Modified bulk region}
	As mentioned above, we may consider the matrix with entries $J_{ij} - rx_i^\star \mu N^{-1}$ to find the bulk region of the eigenvalue spectrum to which most of the eigenvalues are confined. In doing so, we need only make trivial modifications to the calculation in Section \ref{section:eigenvaluespectrum}. Ultimately, we obtain the following expression for the boundary of the bulk region [c.f. Eq.~(\ref{boundaryj})]
	\begin{align}
		\int_0^1 dx\, P(x) \frac{r^2 \sigma^2 x^2}{(\lambda_x + r x(1-\mu) -1)^2 + \lambda_y^2} = 1 , 
	\end{align}
	where now, $P(x)$ is given by Eq.~(\ref{abundancedistribution_mu}). This is tested in the inset of Fig.~5 of the main text (see End Matter).

	\subsection{Outlier eigenvalue}
	We now turn our attention to the outlier. This calculation is non-trivial, since one must take into account the detailed statistics of the surviving interactions $a_{ij}'$ (we recall that we are here careful to distinguish these from the original interactions $a_{ij}$). Due to there being a non-trivial interdependency of the set of frozen species and the values of $a_{ij}$, higher-order correlations exist amongst the reduced interaction matrix coefficients $a_{ij}'$ than in the original interaction matrix \cite{bunin2016interaction, bunin2017, baron2023breakdown}. We must therefore be careful to take these higher-order statistics into account in order to correctly locate the outlier eigenvalue. In the following, we modify the calculation of Ref. \cite{baron2023breakdown} for our present case. We provide only a summary of the full calculation here.
	
	\medskip
	
	We first observe that the outlier eigenvalue $\lambda_{\mathrm{outlier}}$ of the reduced Jacobian matrix by definition must obey
	\begin{align}
		\det\left( \lambda_{\mathrm{outlier}} \id_{N_\ess} - \underline{\underline{J}} \right) =0,
	\end{align}
	where $ \id_{N_\ess}$ is the identity matrix of size $N_\ess\times N_\ess$, where $N_\ess = \phi N$ is the number of non-frozen components. We recall that the reduced Jacobian matrix has entries 
	\begin{align}
		J_{ij} =\delta_{ij}+ r x_i [a'_{ij} - \mu N^{-1} -\delta_{ij} (1-\mu)].
	\end{align}
	
	Suppose we introduce a uniform vector $\underline{u}$ with all entries equal to $1$. Using Sylvester's determinant identity, one finds
	\begin{align}
		\det\left( \id_{N_\ess} -r \nu N_\ess^{-1} \underline{\underline{X}}\,\underline{u} \,\underline{u}^T \underline{\underline{G}}\right) =  1 - \frac{\nu}{N_\ess} \sum_{ij} G_{ij} r x_j = 0, \label{outliercalcgeneral}
	\end{align}
	where we have introduced the resolvent matrix $\underline{\underline{G}}  = [ \lambda_{\mathrm{outlier}} \id_{N_\ess} - (\underline{\underline{J}}  -  r\nu N_\ess^{-1} \underline{\underline{X}}\,\underline{u} \,\underline{u}^T )]^{-1}$ and the matrix $(\underline{\underline{X}})_{ij} = \delta_{ij} x_i$. Thus, to find the outlier eigenvalue, one has to find the resolvent matrix and solve Eq.~(\ref{outliercalcgeneral})  for $\lambda_{\mathrm{outlier}}$. We stress that all elements of the resolvent matrix are required in Eq.~(\ref{outliercalcgeneral}), not only the diagonal entries.
	
	We note that we have the freedom to choose the value of $\nu$, as long as it is non-zero and $ \lambda_{\mathrm{outlier}} \id_{N_\ess} - (\underline{\underline{J}}'  - r \nu N_\ess^{-1} \underline{\underline{X}}\,\underline{u} \,\underline{u}^T )$ remains invertible. We exploit this freedom to simplify the calculation of the resolvent, and we choose $\nu = - \phi \mu$ in Eq.~(\ref{outliercalcgeneral}). We thus see that the disorder-averaged resolvent matrix that we must evaluate to find the outlier can be expressed as the following series
	\begin{align}
		\mathcal{G}[\omega] &\equiv \left\langle N_\ess^{-1}\sum_{i,j \in \mathcal{S}}  G_{ij} r x_j \right\rangle  = \left\langle N_\ess^{-1}  \sum_{i,j \in \mathcal{S}} \left[ \delta_{ij}f_i^{-1}  - x_{i} a_{ij}'\right]^{-1} rx_j \right\rangle \nonumber \\
		&= \left\langle N_\ess^{-1}\sum_{ij} \left[ f_i \delta_{ij} \theta_i \theta_j + \theta_i \theta_j f_i x_i a_{ij}'f_j x_j + \theta_i \theta_j\sum_k \theta_k f_i x_i a'_{ik} f_k x_k a'_{kj} f_j x_j  + \cdots \right] \right\rangle, \label{resolventexpansion0}
	\end{align}
	where $f_i^{-1} =r^{-1}[ \omega -1 + (1-\mu) r x_i]$ and sums over $i,j \in \mathcal{S}$ denote a sum over the reduced interaction matrix elements, whereas sums over $ij$ indicate a sum over all elements of the original interaction matrix. The quantities $\theta_i = \Theta(x_i^\star)\Theta(1-x_i^\star)$ are zero for frozen species and equal to 1 otherwise. 
	
	To find the terms of this series, we now construct the following generating functional (an MSRJD-type path integral \cite{altlandsimons}), from which we can extract all of the terms of the series in Eq.~(\ref{resolventexpansion0})
	\begin{align}
		Z[\bpsi, \blambda] &= \int D[\mathbf{x}, \mathbf{\hat x}] \exp \left( i \sum_{i,t} \left[ \hat x_i(t) \left( \frac{ x_i(t+1)}{x_i(t)} -H\left\{r\left[ 1 - (1-\mu)x_i(t) - \mu M(t) + g_i(t) + h_i(t)\right] \right\} \right) \right] \right) 
		\nonumber \\
		&\times \exp\left( i \int dt \,\sum_i\hat g_i(t)\left[ g_i(t) - \sum_j a_{ij} x_j(t)\right]\right) \nonumber \\
		&\times \exp\left( -i \int dt \, \sum_{ij} \lambda_{ij}(t) a_{ij}  \theta_i(t) \theta_j(t)\right)\exp\left(i \sum_i \int dt x_i(t) \psi_i(t) \right) . \label{genfunct}
	\end{align}
	This generating functional has the usual form \cite{galla2024generating}, with the addition of another source term containing the auxiliary variables $\lambda_{ij}(t)$, which we introduce in this step. The dynamics of $x_i(t)$ are thus constrained to follow the disordered Logistic map dynamics, and by functionally differentiating with respect to $\lambda_{ij}(t)$, we can obtain the terms in the series in Eq.~(\ref{resolventexpansion0}). The path integral measure $D[\mathbf{x}, \mathbf{\hat x}]$ indicates a sum over all possible trajectories of the dynamic variables and their conjugate `momenta', subject to the normalisation $Z[\psi = 0, \lambda = 0] = 1$. The purpose of considering the generating functional is that all possible statistical information about the trajectories is available from it via differentiation. We use this fact to our advantage below.
	
	We now find for the disorder-averaged resolvent [from which we can find the outlier eigenvalue via Eq.~(\ref{outliercalcgeneral})]
	\begin{align}
		\mathcal{G}[\omega] =  \left\langle N_\ess^{-1}\sum_{ij} \lim_{t \to \infty}\left[f_i \delta_{ij} \theta_i \theta_j + i\theta_i \theta_j \frac{\delta Z}{\delta \lambda_{ij}(t)} f_i x_i f_j x_j - \theta_i \theta_j  \sum_k \frac{\delta ^2 Z}{\delta \lambda_{ik}(t) \delta \lambda_{kj}(t)} f_i x_i f_k x_k f_j x_j + \cdots \right] \right\rangle \bigg\rvert_{\psi = 0, \lambda = 0} .\label{resolventexpansion}
	\end{align}
	
	To find the terms of the series in Eq.~(\ref{resolventexpansion}), we begin by calculating the following average that appears in the expression for $\left\langle Z[\psi, \blambda]\right\rangle$
	\begin{align}
		A &= \left\langle \exp \left( -i \sum_i \int dt \left[  \sum_j a_{ij} \hat g_i(t) x_j(t) \right] \right)\exp\left( -i \int dt \, \sum_{ij} \lambda_{ij}(t) a_{ij} \theta_i(t) \theta_j(t)\right)\right\rangle \nonumber \\
		&=  \exp\left(- \frac{\sigma^2}{2 N} \sum_{ij} \left[\int dt \, \,\hat g_i(t) x_j(t) + \lambda_{ij}(t) \theta_i(t) \theta_j(t) \right]^2 \right).
	\end{align}
	One thus finds that the derivatives in Eq.~(\ref{resolventexpansion}) can be written as, for example,
	\begin{align}
		\left\langle\frac{\delta ^2 Z}{\delta \lambda_{ik} \delta \lambda_{kj}}\right\rangle  \bigg\vert_{\psi = 0, \lambda = 0} =  \left\langle \frac{1}{A} \frac{\delta ^2 A}{\delta \lambda_{ik} \delta \lambda_{kj}} \right\rangle_D ,
	\end{align}
	where we note that there are two kinds of averages here: an average over realisations of the interaction coefficients represented by $\langle \cdots \rangle$ (i.e. without subscript) and an average over the dynamics enforced by the disorder averaged generating functional denoted by $\langle \cdots \rangle_D$, such that
	\begin{align}
		\langle \cdots \rangle_D &\equiv \int D[\mathbf{x}, \mathbf{\hat x}] \left[ \cdots\right]\exp \left( i \sum_{i,t} \left[ \hat x_i(t) \left( \frac{ x_i(t+1)}{x_i(t)} -H\left\{r\left[ 1 - (1-\mu)x_i(t) - \mu M(t) + g_i(t)\right] \right\} \right) \right] \right) 
		\nonumber \\
		&\times \exp\left( i \int dt \,\sum_i\hat g_i(t) g_i(t) \right) \exp\left(- \frac{\sigma^2}{2 N} \sum_{ij} \left[\int dt \, \,\hat g_i(t) x_j(t) \right]^2 \right)
	\end{align}
	
	Let us now begin to construct the series for the resolvent. Consider the derivatives of $A$:
	\begin{align}
		B_{ij}(t) &\equiv \frac{1}{A}\frac{\delta A}{\delta \lambda_{ij}(t)} \nonumber \\
		&= - \Bigg[ \theta_i(t) \theta_j(t) \frac{\sigma^2}{N} \sum_{t'} \hat g_i(t') x_j(t') +  \theta_i(t) \theta_j(t) \frac{\sigma^2}{N} \sum_{t'}\theta_i(t') \theta_j(t') \lambda_{ij}(t') \Bigg] , \nonumber \\
		\frac{\delta^2 A}{\delta \lambda_{ik}(t)\delta \lambda_{kj}(t)} &=  B_{ik} B_{kj} A, \nonumber \\
		\frac{\delta^3 A}{\delta \lambda_{ik}(t)\delta \lambda_{kl}(t)\lambda_{lj}(t)} &=  B_{ik} B_{kl}B_{lj}  A.\label{derivatives}
	\end{align}
	
	Taking into account similar considerations for the higher-order terms in $B_{ij}$, we obtain the following series for the full resolvent
	\begin{align}
		\mathcal{G}[\omega] =  N_\ess^{-1}\sum_{ij} \lim_{t \to \infty}\left[\langle f_i \delta_{ij} \theta_i \theta_j \rangle_D + \langle i\theta_i \theta_j  f_i x_i B_{ij}f_j x_j\rangle_D - \sum_k \langle \theta_i \theta_j  \theta_k f_i x_i B_{ik} f_k x_k B_{kj}f_j x_j \rangle_D+ \cdots \right]  . \label{simplifiedseries}
	\end{align}
	To understand how this series may be evaluated, let us take the specific example of the third term. We have
	\begin{align}
		&-  \frac{ 1}{N\phi } \sum_{ijk } \langle \theta_i \theta_j  \theta_k f_i x_i B_{ik} f_k x_k B_{kj}f_j x_j \rangle_D\nonumber \\
		&= -  \frac{ \sigma^4}{N^3\phi } \sum_{ijk } \bigg[ \left\langle \theta_i(t) \theta_k(t) \theta_j(t) \sum_{t', t''} f_i  x_i(t) \hat g_i(t') x_k(t) x_k(t') \hat g_k(t'') f_k x_j (t) x_j(t'') f_j\right\rangle_D  \bigg]. \label{exampleterm}
	\end{align}
	First, we observe that since the different species decouple in the thermodynamic limit, the sums factorise. Then taking the limit $t \to \infty$ and assuming that time-translational invariance applies, we find 
	\begin{align}
		&-  \frac{ 1}{N\phi } \sum_{ijk } \langle \theta_i \theta_j  \theta_k f_i x_i B_{ik} f_k x_k B_{kj}f_j x_j \rangle\vert_{\lambda = 0}= \frac{ \sigma^4 }{\phi} A_{\chi_1} A_{\chi_2} A_q ,
	\end{align}
	where we define 
	\begin{align}
		A_{\chi_1} &= \frac{d}{dh}\int_0^1 dx \,\frac{d P(x; h)}{dh} \vert_{h = 0} \frac{r x}{\omega-1 + (1-\mu)r x} , \nonumber \\
		A_{\chi_2} &= \int_0^1 dx \,\frac{d P(x; h)}{dh} \vert_{h = 0} \frac{r x^2}{\omega-1 + (1-\mu)r x} , \nonumber \\
		A_{q} &= \int_0^1 dx \,P(x; h= 0) \frac{r x^2}{\omega-1 + (1-\mu)r x} , \nonumber \\
		A_{M} &= \int_0^1 dx \,P(x; h = 0) \frac{r x}{\omega -1 + (1-\mu)r x} , \label{Aeqs}
	\end{align}
	where $P(x;h)$ is the stationary distribution of the $\{x_i\}$ in the presence of an external field
	\begin{align}
		P(x) = \frac{1}{\sqrt{2\pi\sigma^2 q/(1-\mu)^2}} \exp\left(-\frac{\left(x-\frac{1}{1-\mu}\left(1-1/r-\mu M^* + h\right)\right)^2}{2\sigma^2 q/(1-\mu)^2}\right) .
	\end{align}
	The constants $q$ and $M^\star$ are as given in Eq.~(\ref{mandqmodmu}). Evaluating each of the terms in Eq.~(\ref{simplifiedseries}) in a similar way, the full series for the resolvent is then
	\begin{align}
		\mathcal{G}[\omega] &= \frac{1}{\phi}A_M + \frac{\sigma^2}{\phi}A_{\chi_1} A_{q} + \frac{\sigma^4}{\phi}A_{\chi_1}  A_{\chi_2}A_{q} + \frac{\sigma^6}{\phi}A_{\chi_1}  A_{\chi_2}^2A_{q} + \cdots \nonumber \\
		&=\frac{1}{\phi} A_M+ \frac{\sigma^2A_{\chi_1}A_{q}}{\phi(1-\sigma^2A_{\chi_2})} .\label{resummedseriesoutlier}
	\end{align}
	Finally, now that we have the function $\mathcal{G}[\omega]$, the outlier eigenvalue we seek is then given by the solution $\lambda_{\mathrm{outlier}}$ to [c.f. Eq.~(\ref{outliercalcgeneral})] 
	\begin{align}
		\mathcal{G}[\lambda_{\mathrm{outlier}}] &= -\frac{1}{\mu \phi}. \label{gsol}
	\end{align}
	This expression for the outlier eigenvalue is tested against numerics in Fig.~5 in the main text (see End Matter).
	
	\medskip
	
	In summary, to find the outlier eigenvalue for given $\mu$, $\sigma$ and $r$, one must first find $M^\star$, and $q$ [see Eq.~(\ref{mandqmodmu})]. Subsequently, one evaluates the quantities $A_{\chi_1}$, $A_{\chi_2}$, $A_{q}$ and $A_M$ in Eq.~(\ref{Aeqs}) for a range of values of $\omega$ to obtain $\mathcal{G}[\omega]$. One then searches for the argument of $\mathcal{G}[\cdot]$ that satisfies Eq.~(\ref{gsol}). 
	
	\subsection{Outlier eigenvalue approximation for small $\sigma$ and recovery of usual logistic map instability}
	Following similar arguments to Section \ref{section:rc2}, we find that for small $\sigma$ the leftmost edge of the bulk of the Jacobian matrix spectrum is at 
	\begin{align}
		\lambda_L = 1-(1-\mu)r.
	\end{align}
	Let us now understand whether the supposed outlier eigenvalue calculated in the previous section is to the left or to the right of this value. 
	
	For $\sigma \to 0^+$, we can concentrate on the first term in Eq.~(\ref{resummedseriesoutlier}), and we obtain
	\begin{align}
		\mu A_M +1 = 0.
	\end{align}
	Let us examine the conditions under which the integrand for $A_M$ has a pole. One requires that $x_c = (1-\lambda_\mathrm{outlier})/[r(1-\mu)]$ be in the integration range $0<x_c<1$ in order for there to be a pole. This means that there is a pole when 
	\begin{align}
		\lambda_L < \lambda_\mathrm{outlier} < 1.
	\end{align}
	Conversely, if indeed $\lambda_\mathrm{outlier}$ takes a value outside of the bulk region, then there is no pole. 
	
	Assuming that the outlier is indeed outside the bulk region, we can approximate the integral over $x$ in the expression for $A_M$ by replacing $x$ with the value at which $P(x)$ is maximum. Since all species are essentially equivalent in the limit $\sigma\to 0^+$, we have then that $x_\mathrm{max} =1-1/r$. This then yields
	\begin{align}
		\frac{A_M}{\phi} \approx \frac{r-1}{\lambda_\mathrm{outlier}-1 + (1-\mu)(r-1)} = -\frac{1}{\mu}, 
	\end{align}
	where $\phi \to 1$ as $\sigma\to 0^+$. Rearranging, one obtains
	\begin{align}
		\lambda_\mathrm{outlier} \approx 2 - r.
	\end{align}
	We therefore see that for $ \mu\gtrsim 1/ r$, we ought to see a protruding outlier for small $\sigma$. Furthermore, this protruding outlier gives rise to a period-doubling instability when $r \approx 3$. Thus, when $\mu$ is sufficiently large and $\sigma$ is sufficiently small, we recover at least some part of the usual period doubling cascade.
	
	\section{Details of numerical methods and further numerical results}\label{sec:numerics}
	\subsection{Simulation of the map}
	Simulating the map in Eq.~(1) of the main paper is straightforward. The following aspect should be noted however. In the discrete-time disordered logistic map, and with the strict ramp function [$H(u)=0$ for $u<0$, $H(u)=u$ for $0\leq u\leq 1$ and $H(u)=1$ for $u>1$] a component $x_i(t)$ can become zero at a finite times, and it then stays zero indefinitely. This is different from the continuous-time generalised Lotka--Volterra equations \cite{bunin2017,Galla_2018}, for example, where the extinction of a degree of freedom can only occur asymptotically as $t\to\infty$. In ecology, the degrees of freedom $x_i$ are species abundances. Convergence to $x_i=0$ at long times thus corresponds to (asymptotic) species extinction.
	
	The possibility to reach zero at finite times in the map leads to the following problem. It is possible for the disordered logistic map to have a stable fixed point at which a particular component $x_i^\star$ is strictly positive, but where this  component is absorbed at $x_i=0$ at a finite time during the dynamics. Whether or not this happens for a particular component will depend on the initial condition. To prevent this from happening we introduce a small `immigration rate' $\rho$, and simulate 
	\begin{equation}
		x_i(t+1)=H\left[r x_i(t) \bigg\{1-x_i(t)+\sum_j a_{ij} x_j(t)\bigg\}\right]+\rho.
	\end{equation}
	In simulations we use values of the order of $\rho=10^{-6}$. Any stable fixed point of the map with small immigration rate will be close to that of the disordered logistic map with $\rho=0$, but we avoid the problem of premature absorption of components.

	\subsection{Measurement of Lyapunov exponents}
	To perform the measurements leading to the estimates of the Lyapunov exponents in Figs.~1 and 3 of the main paper, we use the following procedure (for a fixed choice of model parameters):
	\begin{enumerate}
		\item Draw the interaction coefficients $a_{ij}$. Also draw {\em iid} initial conditions $x_i(0)$ ($i=1\dots,N$) from the interval $(0,1]$. Then iterate the map for the components $x_i(t)$ for $t_{\rm eq}$ steps, $t=1,\dots,t_{\rm eq}$.
		\item Start a second copy of the system, with identical interaction coefficients, and initial conditions $y_i(t_{\rm eq})=x_i(t_{\rm eq})+v_i$, where $v_i=D\times u_i/\sqrt{N^{-1}\sum_i u_i^2}$ with {\em iid} Gaussian random numbers $u_i$ of mean zero and variance one. This means that the $y_i$ are obtained by applying a (small) perturbation to the $x_i$. The model parameter $D \ll 1$ sets the amplitude of the perturbation.
		\item Iterate one copy of the map for the $x_i$ and another copy for the $y_i$, with identical interaction coefficients (those from item 1). Once every $t_{\rm meas}$ time steps, adjust the $y_i$ such that the distance between $\mathbf{x}$ and $\mathbf{y}$ is reset to $D$. More precisely, compute $d(t)=\sqrt{N^{-1}\sum_i [y_i(t)-x_i(t)]^2}$, and then  set 
		\begin{equation}
			y_i(t)=x_i(t)+D\times\frac{y_i(t)-x_i(t)}{d(t)}.
		\end{equation}
		This reset is carried out for all $i=1,\dots,N$ simultaneously.
		\item Every time the distance is reset (step 3), record the ratio $d(t)/D$. This is the relative amount by which the perturbation has grown or shrunk since the last reset (which will have occurred $t_{\rm meas}$ time steps ago). 
		\item Repeat until the end time $t_{\rm end}$ of the simulation. This will produce $(t_{\rm end}-t_{\rm eq})/t_{\rm meas}$ values of $d(t)/D$.
		\item At the end of the run, the estimate for the largest Lyapunov exponent is obtained as
		\begin{equation}
			\lambda=\frac{1}{t_{\rm meas}}\left<\ln\frac{d(t)}{D}\right>_t,
		\end{equation}
		where $\langle\cdots\rangle_t$ is the average over the $[t_{\rm end}-t_{\rm eq}]/t_{\rm meas}$ values of $\ln \frac{d(t)}{D}$ recorded along the trajectory.
	\end{enumerate}
	For each combination of the model parameters ($N, r, \sigma^2$ and $\mu$), this procedure is repeated for $S$ independent samples of the interaction matrix $a_{ij}$, and the $S$ estimates of the Lyapunov exponents are then averaged. Typical parameters we used are $t_{\rm eq}=3000$, $t_{\rm end}=5000$, $t_{\rm meas}=10$, $D=10^{-5}$, $S=10$. 
	
	\medskip
	
\begin{figure}
		\centering   
		\includegraphics[width=0.8\linewidth]{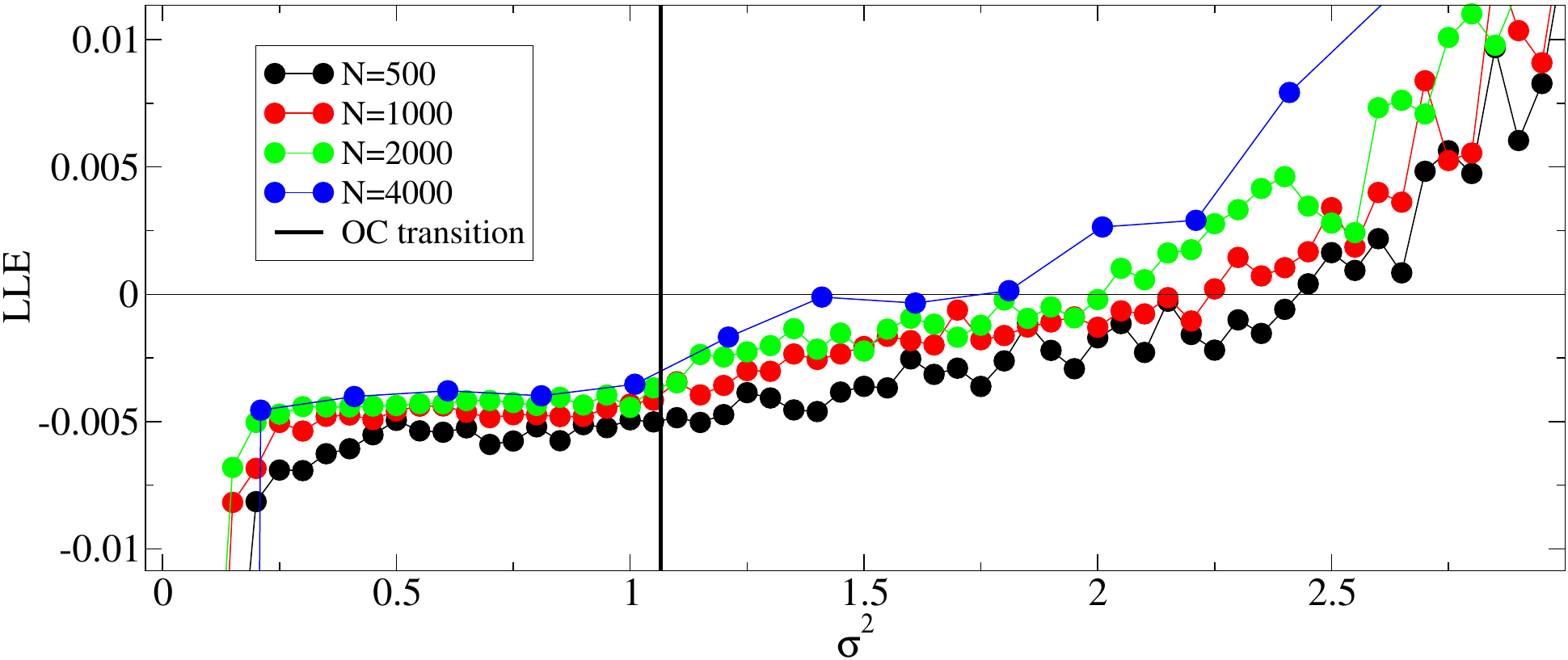}
		\caption{{\bf Largest Lyapunov exponent across the oscillatory chaotic (OC) transition.}  Markers show estimates of the largest Lyapunov exponent from simulations at constant $r=1.6$, and for different system sizes $N$ as indicated. The vertical solidline is the location of the OC instability, as predicted from the theory. Invasion rate is $\rho=10^{-6}$. Each data point is from 20-50 independent runs (depending on system size).}
		\label{fig:LLE_finite_size}
	\end{figure}

	The dependence of the LLE on the size of the system in simulations is illustrated in Fig.~\ref{fig:LLE_finite_size}. This is for the model with $\mu=0$ and fixed $r=1.6$, increasing $\sigma^2$ across the OC transition (vertical cut across the solid line in Fig.~1 of the main paper). For each fixed value of $N$, the LLE crosses from negative to positive as $\sigma^2$ is increases. The point at which this happens, $\sigma_c^2(N)$, is found to decrease with $N$, and approaches the onset of instability predicted by the theory (vertical line). This confirms that the observation of negative LLEs just above the predicted onset of instability in Fig.~1 can be attributed to finite-size effects. The size of the light blue area above the transition lines reduces with increasing system size.
	
	\medskip
	
	We also note that, in simulations, the LLE comes out as negative in the stable phase (to the left of the vertical line in Fig.~\ref{fig:LLE_finite_size}). The spectrum of the reduced Jacobian is here predicted to come arbitrarily close to the point $\lambda=1$ in the complex plane (see Fig.~2 in the main paper, and Sec.~\ref{sec:largest_eigenvalue} of this SM). This prediction is a consequence of the fact that, in the thermodynamic limit, the components $x_i^\star$ fill the entire unit interval $[0,1]$. In the stable phase, one would therefore expect a LLE of zero for $N\to\infty$. The small negative values in Fig.~\ref{fig:LLE_finite_size} can be attributed to finite-size effects and to the fact that we use a finite invasion rate of $10^{-6}$ in simulations. No fixed-point components $x_i^\star$ can be smaller than this value in simulations, and hence the interval $[0,1]$ is not fully populated. 
	
	\subsection{Classification of dynamics}
	\subsubsection{Algorithm}

	\begin{figure}[t]
		\centering       \includegraphics[width=0.6\linewidth]{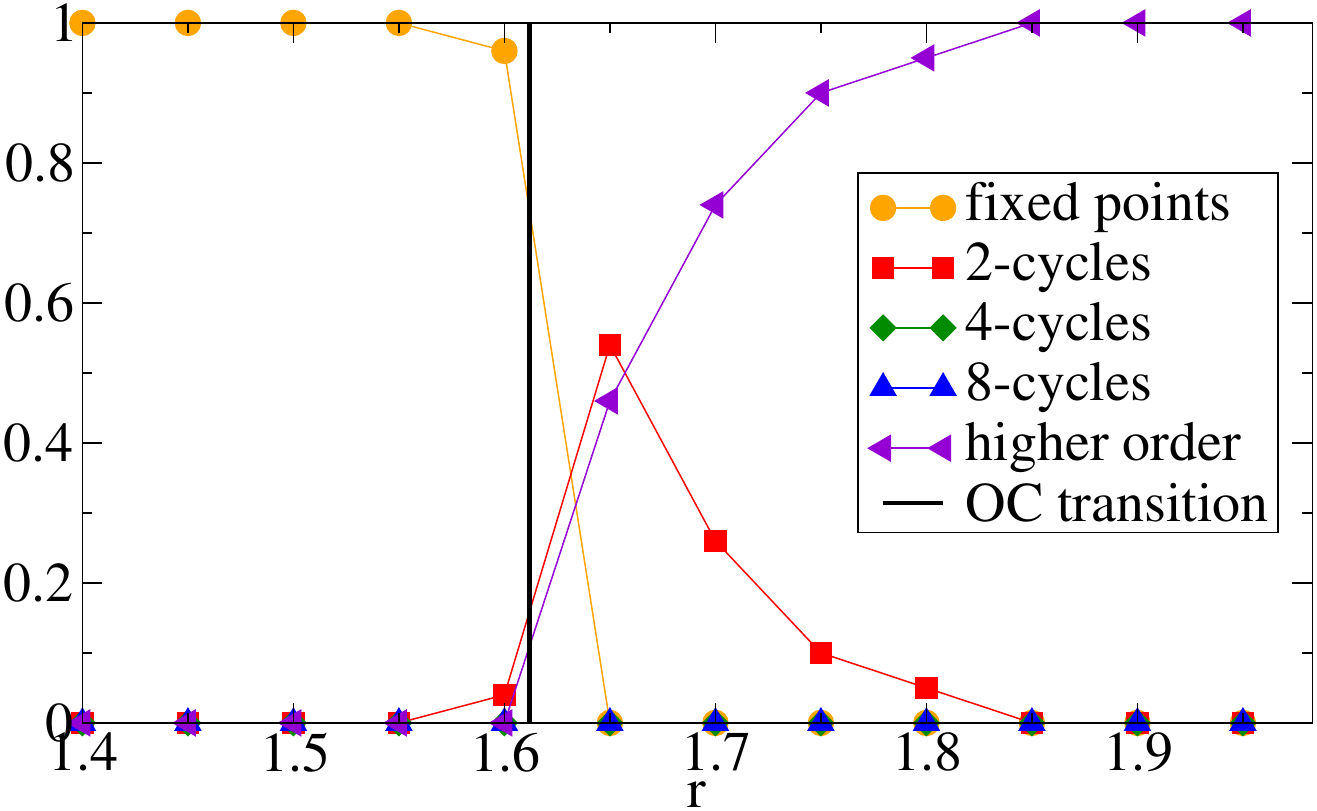}   
		\caption{{\bf Different types of dynamics when crossing the oscillatory chaotic (OC) transition}. We plot the fraction of samples classified as fixed points, 2-cycles, 4-cycles, 8-cycles and higher-order attractors, for fixed $\mu=0$ and $\sigma^2=1$. System size is $N=4000$, each data point is from 100 independent simulation runs. Vertical solid line is the location of the OC transition predicted from the theory ($r\approx 1.612$)
			\label{fig:classify1}}
	\end{figure}
	\begin{figure}[t]
		\centering        
		\includegraphics[width=0.6\linewidth]{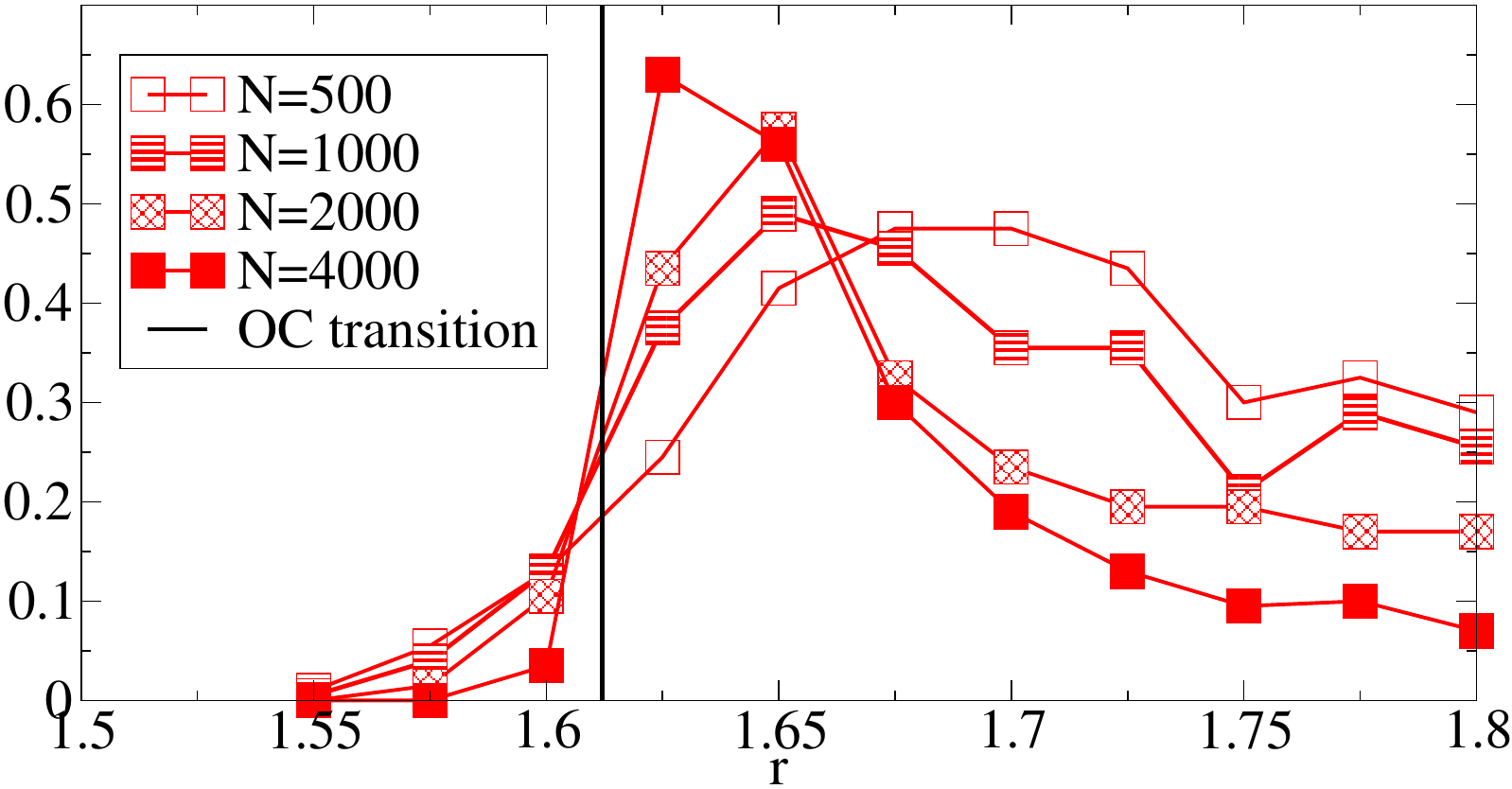}  
		\caption{{\bf Cycles are a finite-size effect when crossing the oscillatory chaotic (OC) transition.} Comparison of the fraction of samples identified as 2-cycles for different system sizes $N$ (each data point is an average over 200 runs). Same parameters as in Fig.~\ref{fig:classify1}. Vertical solid line is the location of the OC transition predicted from the theory ($r\approx 1.612$).
			\label{fig:classify2}}
	\end{figure}
	We have also carried out a classification of the different types of dynamics that can occur. To do this, we draw a sample of the interaction matrix, then run the system until some end time (typically $t_{\rm end}=2000$). From the second half of the trajectory we measure the quantities
	\begin{equation}
		d_\tau=\left\langle \frac{1}{N}\sum_i [x_i(t)-x_i(t-\tau)]^2\right\rangle_t,
	\end{equation}
	for $\tau=1,2,4,8$, where $\langle\cdots\rangle_t$ is a time average. Thus, $d_\tau$ indicates by how much components change from time step $t$ to $t+\tau$. We then use these quantities to broadly classify each run as either a fixed point, a two-cycle, a four-cycle etc. For example, for fixed points, all $d_\tau$ will be zero. For two-cycles $d_1$ will be non-zero, but $d_2, d_4$ and $d_8$ will all be zero. In practice, we set a threshold $\vartheta$, and then say:
	\begin{enumerate}
		\item If $d_1, d_2, d_4$ and $d_8$ are all lower than or equal to $\vartheta$, the run is classed as having converged to a fixed point.
		\item If $d_1>\vartheta$, but $d_2, d_4, d_8\leq \vartheta$, the run is classified as a two-cycle.
		\item If $d_1$ and $d_2$ are larger than $\vartheta$, but $d_4$ and $d_8$ are lower than or equal to $\vartheta$, the run is classified as a four-cycle.
		\item If $d_1, d_2$ and $d_4$ are all larger than $\vartheta$, but $d_8\leq \vartheta$ the run is classified as an eight-cycle.
		
		\item If none of the above conditions are fulfilled, the sample is recorded as having a higher-order attractor. This would include cycles of length 16 or more, and irregular motion. 
	\end{enumerate}
	For each fixed parameter set ($r,\sigma^2$, $\mu$ and $N$), this then results in a (estimated) proportion of samples that are classed as fixed points, two-cycles, four-cycles and so on. The exact numerical values for these proportions will depend on the choice of the threshold $\vartheta$, and we recognize that we are not identifying cycles of lengths that are not powers of two. This is because we are primarily concerned with identifying a possible period-doubling scenario. We have not attempted to distinguish between 16-cycles, 32-cycles etc., and non-periodic motion (this appears beyond the limitations of the method).
	
	We now briefly describe our results from this classification as the three different types of instability are crossed.
	\subsubsection{Transition controlled by an isolated outlier eigenvalue}
	In Fig.~7 of the main paper (End Matter) we show the observed proportions of the different types of attractor for fixed $\mu=0.6, \sigma=0.2$ as we increase $r$ to move horizontally across the dot-dashed and solid lines in Fig.~3 (main paper).  To the left of the dot-dashed line all eigenvalues of the reduced Jacobian are predicted to be inside the unit circle, and we expect stable fixed points. This is confirmed by the simulation data in Fig.~7.
	
	Between the dot-dashed and solid  vertical lines, the isolated outlier eigenvalue of the reduced Jacobian has crossed the point $-1$ in the complex plane, but the bulk part of the spectrum is still contained inside the unit disk. We expect non-chaotic periodic motion of the system, at least for values of $r$ just beyond the dot-dashed line. The absence of chaos is confirmed by the measurements of the attractor dimension in the inset of Fig.~3 in the main paper. The data in Fig.~7 confirms that, upon crossing the transition controlled by the outlier eigenvalue (dot-dashed line), we first observe a predominance of two-cycles, and then, as $r$ is increased further, 4-cycles become the only outcome. 
	
	While we see some emergence of 8-cycles, a further increase of $r$ leads to the bulk spectrum crossing the point $-1$ (OC instability, solid line in Fig.~7). High-dimensional chaotic behaviour then sets in (inset of Fig.~3). The successive crossings of the outlier eigenvalue and the bulk spectrum mean that we observe a partial period-doubling cascade.

	\subsubsection{Oscillatory chaotic transition}
	In Fig. \ref{fig:classify1}, we show the observed proportions of the different types of attractor as we cross the OC instability. In contrast to the results for the period-doubling transition (Fig.~7 main text), we do not see any evidence of a succession of a dominance of 2-cycles and a dominance of 4-cycles. In fact, the data in the figure shows no indication of any 4-cycles. Instead, at the phase boundary, we see the sudden onset of a mixture of 2-cycles and runs that are classified as higher-order attractors (chaotic trajectories would be included here), with the 2-cycles disappearing as we move away from the phase boundary. As is demonstrated in Fig. \ref{fig:classify2}, we can attribute the 2-cycles that are observed here to finite-size effects. The range of parameters for which these cycles are observed  shrinks as we increase $N$. This is in-line with the finite-size effects on the largest Lyapunov exponent near the OC transition, shown in Fig.~\ref{fig:LLE_finite_size}. 
	
	In the thermodynamic limit, we expect the OC transition to be characterised by an abrupt transition to chaos, due to the continuous nature of the bulk eigenvalue spectrum and its well-defined boundary.  Our numerical characterisations of the LLE (see above) and KY dimensions (see main panel of Fig.~6 of the main text) also support this picture.

	\subsubsection{May--Wigner transition}
\begin{figure}[t]
		\centering       \includegraphics[width=0.6\linewidth]{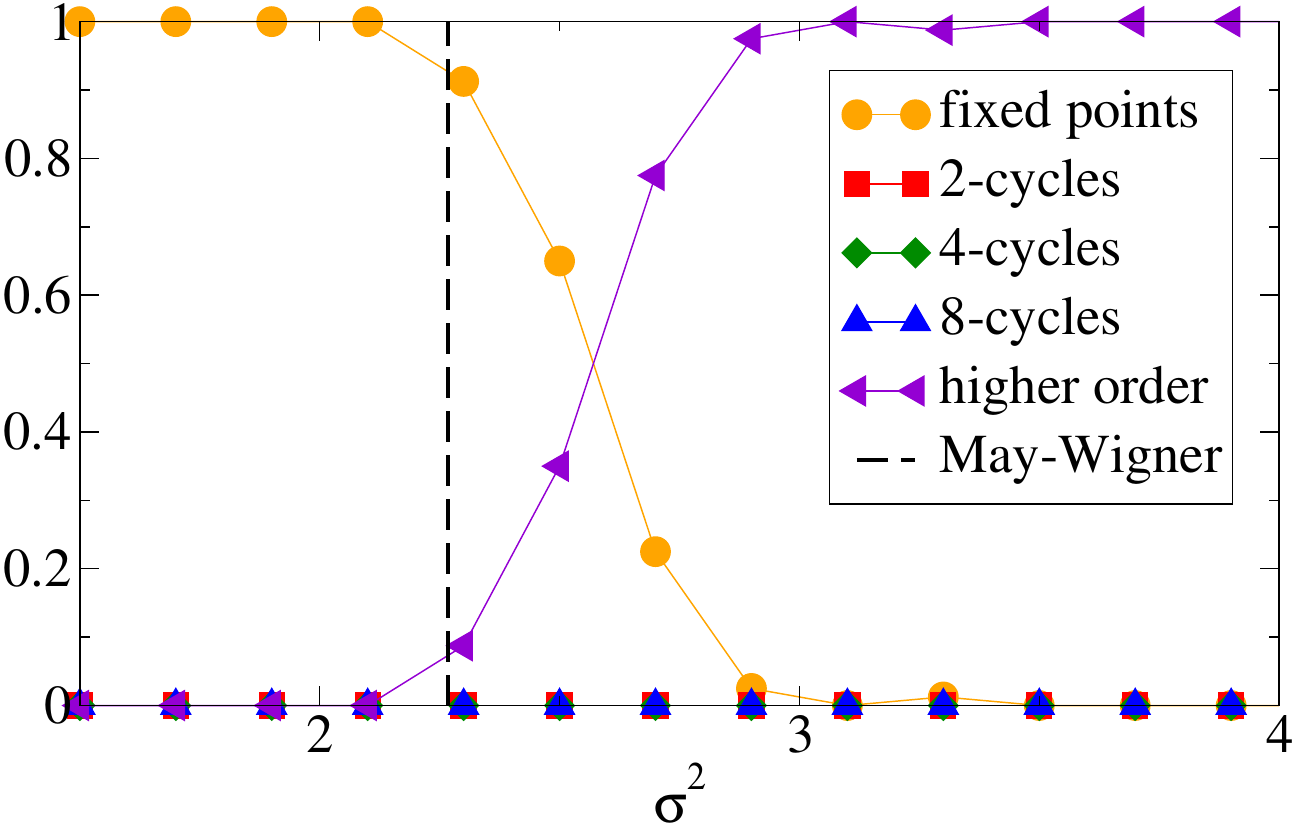}   
		\caption{{\bf Dynamical behaviour when crossing the May--Wigner transition} Fraction of samples classified as fixed points, 2-cycles, 4-cycles, 8-cycles and higher-order attractors, for fixed $\mu=0$ and $r=1.2$. System size is $N=4000$, each data point is from 40 independent simulation runs. We find no evidence of 2-cycles, 4-cycles or 8-cycles. Vertical dashed line shows the location of the May--Wigner transition as predicted from the theory ($\sigma^2\approx 2.268$).
			\label{fig:classify3}}
	\end{figure}

	We also perform a similar numerical analysis for the May--Wigner transition, the results of which are shown in Fig. \ref{fig:classify3}. In this case, since the eigenvalue spectrum does not depart the unit disk at $-1$ but rather at $1$, there is no hint of cyclic behaviour. Instead, there remain some trajectories that are classified as fixed points as the phase boundary is crossed. This is in contrast to the OC transition of Fig. \ref{fig:classify1}, where the fixed point behaviour ends abruptly. Again, we expect the transition to chaotic behaviour to become more sharp as $N$ is increased.

	\subsection{Measurement of the attractor dimension}
	The Kaplan-Yorke dimension provides yet more information on the chaotic trajectories. It reflects the number of directions in which significant growth or folding of the trajectories occurs around the chaotic attractor. To compute the KY dimension, we require the full spectrum of Lyapunov exponents $\{\Lambda_i\}$, which can be obtained via successive QR decompositions of the Jacobian matrix (evaluated at each point along the trajectory), following the method of Sandri \cite{sandri1996numerical}. One then computes the KY dimension via
	\begin{align}
		D_{KY} = j - \frac{\sum_{i= 1}^j\Lambda_i}{\Lambda_{j+1}},
	\end{align}
	where $\Lambda_1> \Lambda_2> \cdots>\Lambda_N$ and $j$ is the largest integer such that $\sum_{i= 1}^j\Lambda_i>0$.
	
	Sandri's method involves the QR decomposition of the Jacobian matrix. In this more general case, where we do assume a fixed point, the Jacobian matrix is given by 
	\BE
	J_{ii}&=& r H'\left[r x_i \left(1 - (1-\mu)x_i+  \sum_{j=1}^N (a_{ij} - \mu N^{-1}) x_j\right)\right] \bigg[1-2(1-\mu)x_i+\sum_{j=1}^N (a_{ij} - \mu N^{-1}) x_j\bigg] \nonumber \\
	J_{ij}&=& r x_i (a_{ij}-\mu N^{-1})  H'\left[r x_i \left(1 - (1-\mu)x_i + \sum_{j=1}^N (a_{ij} - \mu N^{-1}) x_j\right)\right]~\mbox{for~} i\neq j.
	\EE

	\section{The disordered H\'enon map}\label{sec:henon}
	\subsection{Model and fixed point analysis}
	To demonstrate that the phenomenology that we observed in the logistic map is also seen in other maps, we consider another paradigmatic example, the H\'enon map \cite{henon1976two, hitzl1985exploration, strogatz2024nonlinear}. We modify the conventional H\'enon map by adding a saturating disordered term 
	\begin{align}\label{eq:henon_disordered}
		x_i(t+1) &= \alpha \left[1 - a x_i^2(t) + x'_i(t) \right]+ (1-\alpha)H\bigg[\sum_{j\neq i} a_{ij} x_j(t)\bigg]\equiv F_i(\mathbf{x},\mathbf{x'}), \nonumber \\
		x'_i(t+1) &=b x_i(t)\equiv G_i(\mathbf{x},\mathbf{x'}),
	\end{align}
	where we define
	\begin{align}
		H(u) = \begin{cases}
			2 \hspace{0.5cm}& \mathrm{for}\,\, u\geq 2, \\
			u \hspace{0.5cm}& \mathrm{for}\,\, -2\leq u\leq 2, \\
			-2 \hspace{0.5cm}& \mathrm{for}\,\, u\leq- 2.
		\end{cases}
	\end{align}
	The $a_{ij}$ are again from a Gaussian distribution with mean zero and variance $\sigma^2/N$, see Eq.~(2) in the main paper. We are not considering homogeneous couplings here.
	\medskip
	
	We note that we have chosen a slightly different convention for the implementation of the ramp function $H(\cdot)$ in comparison to the disordered logistic map of the main text. This was done primarily for mathematical convenience, since there is no ecological interpretation for the H\'enon map. We recover the usual H\'enon map by setting $\alpha = 1$. If the model parameters $a$ and $b$ are such that $a<0.5$ and $b<0.5$, all $x_i(t)$ and $x'_i(t)$ remain in the interval $[-2,2]$ if the initial conditions for the $x_i, x'_i$ are between $-2$ and $2$. 
	
	\medskip
	
	We can again perform a dynamic mean-field analysis and find the effective process
	\begin{align}
		x(t+1) &= \alpha \left[1 - a x^2(t) + x'(t) \right]+ (1-\alpha)H\left[\eta(t)\right], \nonumber \\
		x'(t+1) &=b x(t),
	\end{align}
	where the $\eta(t)$ are zero-average Gaussian random variables with correlator (to be determined self-consistently),
	\begin{align}
		\langle \eta(t) \eta(t')\rangle_\eta = \sigma^2 \langle x(t) x(t') \rangle_\eta .
	\end{align}
	Making a fixed-point ansatz, we find
	\begin{align}\label{eq:x_of_eta}
		x^\star = \frac{1}{2a}\left(b - \frac{1}{\alpha} + 
		\sqrt{\left(b - \frac{1}{\alpha}\right)^2 + 4 a \left[1 + \frac{1 - \alpha}{\alpha} H(\eta^\star)\right]}\right),
	\end{align}
	where $\eta^\star$ is now a static Gaussian random variable with variance $\langle (\eta^\star)^2 \rangle_\eta = \sigma^2 \langle (x^\star)^2 \rangle_\eta$. One notes that the argument of the square root can become negative for $\alpha<1$. This happens when $a<-(b-1/\alpha)^2/[4(2-\alpha)]$ or when $a>(b-1/\alpha)^2/[4(2 - 3\alpha)]$. For such choices of the model parameters there are no fixed-point solutions. However, for fixed $\alpha$ and $b$, there is a finite range of $a$ for which a fixed-point solution is possible. 
	
	\subsection{Reduced Jacobian eigenvalue spectrum and stability}
	We write $x_i^\star, x_i'^{\star}$ for the components of a fixed point of the system. As in the case of the logistic map, the argument $\sum_j a_{ij} x_j^\star$ inside the function $H(\cdot)$ can reach saturation, and the components for which this happens are effectively `frozen out'. The reduced Jacobian matrix (the Jacobian from which all rows and columns $i$ with $|\sum_j a_{ij} x_j^\star|>2$ are removed) is block structured with elements
	\begin{align}
		J^{(11)}_{ij} &= -2a\alpha x^\star_i \delta_{ij} + (1-\alpha) a_{ij}, \nonumber \\
		J_{ij}^{(12)} &= \alpha \delta_{ij} , \nonumber \\
		J_{ij}^{(21)} &= b \delta_{ij},\nonumber \\
		J_{ij}^{(22)} &= 0,
	\end{align}
	where we write $J_{ij}^{(11)}=\partial F_i/\partial x_j$, $J_{ij}^{(21)}=\partial G_i/\partial x_j$, $J_{ij}^{(12)}=\partial F_i/\partial x'_j$ and $J_{ij}^{(22)}=\partial G_i/\partial x'_j$ with $F_i$ and $G_i$ as in Eq.~(\ref{eq:henon_disordered}). We can perform a calculation along the same lines as that in Section \ref{section:eigenvaluespectrum}. The eigenvalue potential is given by
	\begin{align}
		\exp\left[ -N \Phi(\lambda) \right] &= \int \prod_{i}\prod_{\gamma=1}^{2} \left( \frac{d^2z^{(\gamma)}_i d^2y^{(\gamma)}_i}{2 \pi^2}\right) \exp\left[ - \sum_{i\gamma} y^{(\gamma)\star}_i y^{(\gamma)}_i  \right]\nonumber \\
		\times & \exp\left[ -i \sum_{i} \left[z^{(1)\star}_i y^{(1)}_i  (\lambda^\star + 2 a \alpha x_i ) + \alpha z^{(2)\star}_i y^{(1)}_i +bz^{(1)\star}_i y^{(2)}_i +z^{(2)\star}_i y^{(2)}_i  \lambda^\star + \mathrm{c.c.} \right]  \right]\nonumber \\
		\times & \exp\Bigg[ - \frac{\sigma^2 (1-\alpha)^2 }{2N} \sum_{ij} (z_j^{(1)\star} y^{(1)}_i + z^{(1)}_j y_i^{(1)\star})^2\Bigg] . \label{averaged}
	\end{align} 
	In principle, we could perform the same coarse-graining that was discussed in Section \ref{section:eigenvaluespectrum}. However, for ease of notation, we introduce order parameters
	\begin{align}
		u^{(\alpha,\beta)}_{i} &=  z^{(\alpha)\star}_i z^{(\beta)}_i, \,\,\,\,\, v^{(\alpha,\beta)}_i =   y^{(\alpha)\star}_i y^{(\beta)}_i,\nonumber \\
		w^{(\alpha,\beta)}_i &=   z^{\alpha\star}_i y^\beta_i, \,\,\,\,\, w_i^{(\alpha,\beta)\star} =  y^{\beta\star}_i z^\alpha_i.
	\end{align}
	Performing the saddle-point approximation, one arrives at
	\begin{align}
		&i v_i^{(1,1)} = -\frac{\partial }{\partial \hat v_i^{(1,1)} }\ln (-f_i) \nonumber \\
		&=-\frac{1}{f_i}\left[\hat v_i^{(2,2)}(\hat u_i^{(1,1)} \hat u_i^{(2,2)} -\hat u_i^{(1,2)}  \hat u_i^{(2,1)} ) +\hat w_i^{(2,1)\star} ( \hat u_i^{(1,2)}   \hat w_i^{(2,2)} - \hat u_i^{(2,2)} \hat w_i^{(1,2)}  ) + 
		\hat w_i^{(2,2)\star} (\hat u_i^{(2,1)}  \hat w_i^{(1,2)}   - \hat u_i^{(1,1)}  \hat w_i^{(2,2)}  ) \right],
	\end{align}
	with
	\begin{align}
		f_i &= \det\begin{bmatrix}
			\hat u_i^{(1,1)} & \hat w_i^{(1,1)} &  \hat u_i^{(1,2)} & \hat w_i^{(1,2)}\\
			\hat w_i^{(1,1)\star} & \hat v_i^{(1,1)} & \hat w_i^{(1,2)\star} & \hat v_i^{(1,2)} \\
			\hat u_i^{(2,1)} & \hat w_i^{(2,1)} &  \hat u_i^{(2,2)} & \hat w_i^{(2,2)}\\
			\hat w_i^{(2,1)\star} & \hat v_i^{(2,1)} & \hat w_i^{(2,2)\star} & \hat v_i^{(2,2)}
		\end{bmatrix}.
	\end{align}
	More generally, we have
	\begin{align}
		iu^{(\alpha, \beta)}_i =-\frac{\partial }{\partial \hat u_i^{(\alpha, \beta)} }\ln (-f_i)  , &\,\,\,\,\,iv^{(\alpha, \beta)}_i =-\frac{\partial }{\partial \hat v_i^{(\alpha, \beta)} }\ln (-f_i)  , \nonumber \\
		i w_i^{(\alpha, \beta)} = -\frac{\partial }{\partial \hat w_i^{(\alpha, \beta)\star} }\ln (-f_i)   ,& \,\,\,\,\,   i w_i^{(\alpha, \beta)\star} = -\frac{\partial }{\partial \hat w_i^{(\alpha, \beta)} }\ln (-f_i),
	\end{align}
	as well as
	\begin{align}
		i\hat u_i^{(1,1)} &= \frac{\sigma^2 (1-\alpha)^2}{N} \sum_j v_j^{(1,1)} , \,\,\,\,\, i\hat v_i^{(1,1)} = 1+ \frac{\sigma^2(1-\alpha)^2}{N} \sum_j  u_j^{(1,1)} , \,\,\,\,\, i\hat v_i^{(2,2)} = 1\nonumber \\
		i\hat w_i^{(1,1)} &=  i (\lambda + 2a\alpha x_i), \,\,\,\,\, i\hat w_i^{(1,1)\star} = i(\lambda^\star + 2a\alpha x_i) , \nonumber \\
		i\hat w_i^{(2,2)} &=  i \lambda  , \,\,\,\,\, i\hat w_i^{(2,2)\star} = i\lambda^\star , \nonumber \\
		i\hat w_i^{(2,1)} &=  i \alpha = i\hat w_i^{(2,1)\star}, \,\,\,\,\, i\hat w_i^{(1,2)} =  i b = i\hat w_i^{(1,2)\star}.
	\end{align}
	All other conjugate variables are equal to zero. We therefore have
	\begin{align}
		i v_i^{(1,1)} &= \frac{\hat u_i^{(1,1)}\hat w_i^{(2,2)\star}    \hat w_i^{(2,2)} }{f_i}, \nonumber \\
		i\hat u_i^{(1,1)} &= \frac{\sigma^2 (1-\alpha)^2}{N} \sum_j v_j^{(1,1)}, \nonumber \\
		f_i &= \left\vert \hat w_i^{(1,1)}\hat w_i^{(2,2)}- \hat w_i^{(1,2)} \hat w_i^{(2,1)}  \right\vert^2  - 
		\hat u_i^{(1,1)}(\hat v_i^{(2,2)} \hat w_i^{(1,2)\star} \hat v_i^{(2,1)} + \hat v_i^{(1,1)} \hat w_i^{(2,2)} \hat w_i^{(2,2)\star}) .
	\end{align}
	As with the calculation in Section \ref{section:eigenvaluespectrum}, we see that there are two solutions. These are $\hat u_i^{(1,1)} = 0$ and
	\begin{align}
		1 = \frac{1}{N}\sum_i \frac{\sigma^2(1-\alpha)^2 \vert\lambda\vert^2}{f_i} .
	\end{align}
	
\begin{figure}
		\centering    \includegraphics[width=0.45\linewidth]{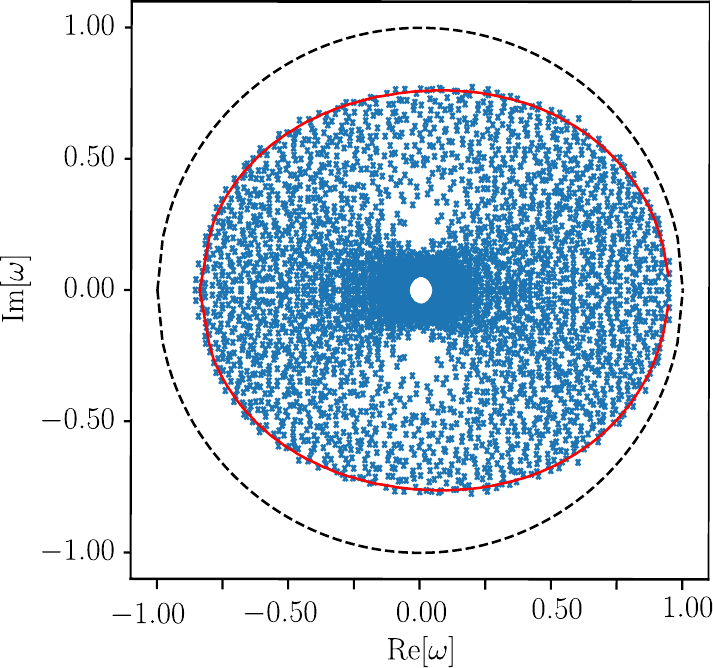}
		
		\caption{Eigenvalue spectrum of the reduced Jacobian for the disordered H\'enon map. Markers are from simulations with $N=4000$. The red line is from the theory [Eq.~(\ref{eq:henon_boundary})]. The dashed black line is the unit circle. Parameters: $a = -0.1$, $b = 0.1$, $\alpha = 0.5$ $\sigma = 0.5$.}
		\label{fig:jacobian_henon}
	\end{figure}

\begin{figure}
		\centering    \includegraphics[width=0.6\linewidth]{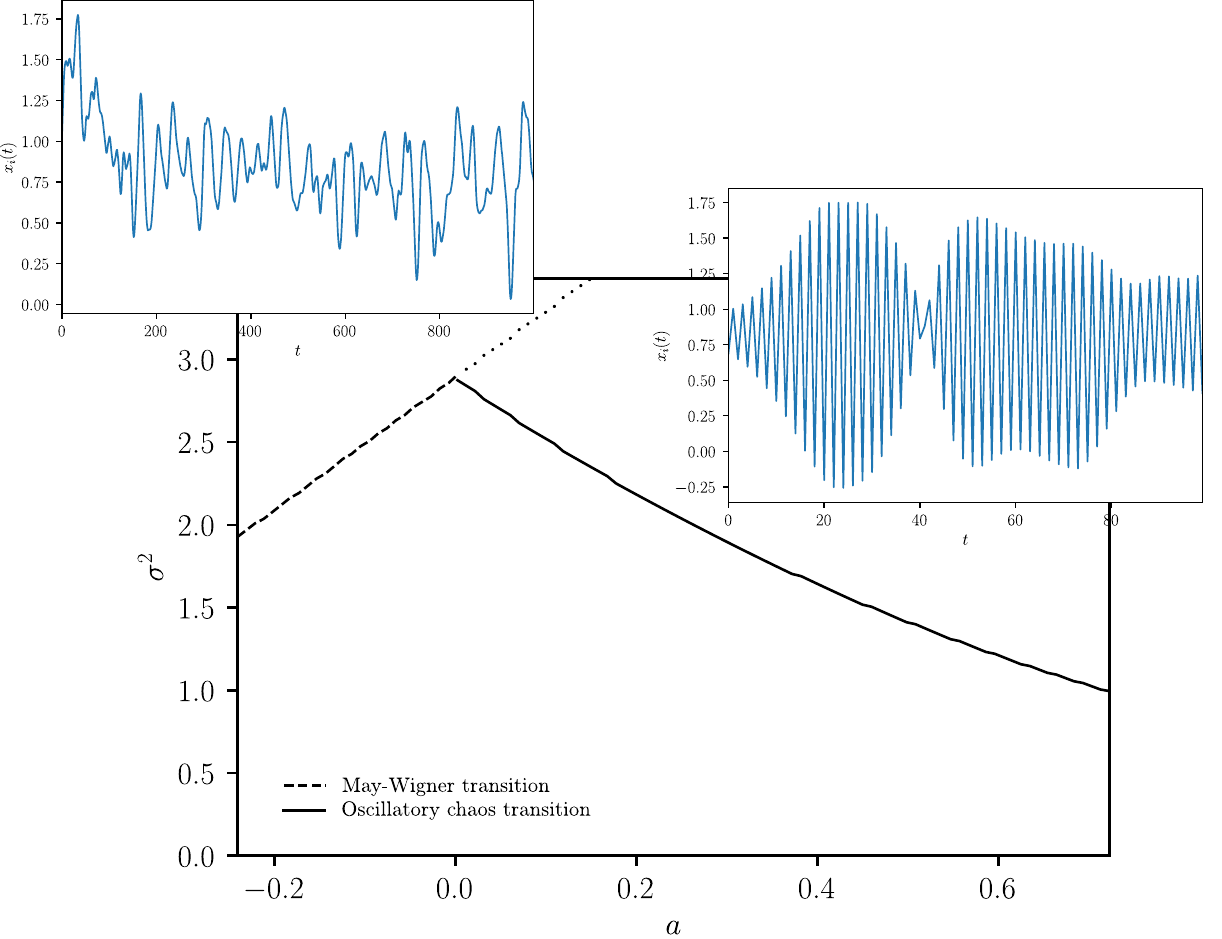}
		
		\caption{Phase diagram for the disordered H\'enon map. Parameters: $b = 0.3$, $\alpha = 0.5$. For these parameter values, the system undergoes runaway growth outside the range of $a$ depicted here.}
		\label{fig:pd_henon}
	\end{figure}

	The points $\lambda$ that satisfy both of these conditions simultaneously define the edge of the bulk eigenvalue spectrum. Given that we are not considering homogeneous coupling, there is no outlier eigenvalue. Taking the limit $N\to \infty$ so that we can replace the sum with an integral, we see that the values of $\lambda$ on the edge of the spectrum satisfy
	\begin{align}\label{eq:henon_boundary}
		1 = \int_{-2}^2 d\eta \, \frac{1}{\sqrt{2\pi q \sigma^2}}e^{-\frac{\eta^2}{2 q \sigma^2}} \frac{\sigma^2(1-\alpha)^2\vert \lambda\vert^2}{\left\vert \lambda^2 + 2 \alpha a x(\eta)\,\lambda - \alpha b \right\vert^2},
	\end{align}
	with $x(\eta)$ as given in Eq.~(\ref{eq:x_of_eta}). This is verified in Fig. \ref{fig:jacobian_henon}. Using this expression, we can find the system parameters for which instability occurs, and we obtain the phase diagram in Fig. \ref{fig:pd_henon}. We see that this diagram has a very similar structure to that of the logistic map. 
	
	\subsection{Power spectra}
	\subsubsection{General calculation and onset of divergence}
	In the same way as for the logistic map, we add white noise variables $\xi_i(t)$ to the dynamics of the H\'enon map, and we obtain
	\begin{align}\label{eq:henon_disordered_2}
		x_i(t+1) &= \alpha \left[1 - a x_i^2(t) + x'_i(t) \right]+ (1-\alpha)H\left[\xi_i(t) + \sum_{j\neq i} a_{ij} x_j(t)\right], \nonumber \\
		x'_i(t+1) &=b x_i(t).
	\end{align}
	For components that have a fixed point value $\eta^*$ with $\vert\eta^*\vert<2$, we have (after carrying out a Fourier transform)
	\begin{align}\label{eq:henonfluct}
		\left[ e^{i\omega} + 2 a \alpha x^\star - \alpha b e^{-i\omega}\right]\tilde v(\omega) &=  (1-\alpha)\widetilde{\delta\eta}(\omega) + \tilde\xi(\omega), 
	\end{align}
	where we have used $v$ for fluctuations about the $x$-component of the fixed point, and $\delta\eta$ for fluctuations about $\eta^\star$. From Eq.~(\ref{eq:henonfluct}) we have
	\begin{align}\label{eq:henonspec}
		\langle \vert \tilde v(\omega)\vert^2\rangle_S=  \frac{TA_2}{1 - (1-\alpha)^2\sigma^2 A_1},
	\end{align}
	where 
	\begin{align}
		A_1 &= \left \langle\frac{1 }{\left\vert e^{i\omega} + 2 a \alpha x^\star - \alpha b e^{-i\omega} \right\vert^2} \right\rangle_S, \nonumber \\
		A_2 &= \left \langle\frac{ (x^\star)^\kappa}{\left\vert e^{i\omega} + 2 a \alpha x^\star - \alpha b e^{-i\omega} \right\vert^2}\right\rangle_S.
	\end{align}
	In a similar fashion to the logistic map, we see that instability occurs when the denominator of the power spectrum of fluctuations becomes zero. This occurs when 
	\begin{align}
		\sigma^2(1-\alpha)^2A_1 = 1. \label{instabilityhenon}
	\end{align} 
	In principle, the condition Eq.~(\ref{instabilityhenon}) could be satisfied for many different values of $\omega$. Indeed, for $b<0$, we find that Eq.~(\ref{instabilityhenon}) may be satisfied for values of $\omega \notin \{0, \pi\}$. This is to be expected, since the H\'enon map without disorder also possesses this property \cite{hitzl1985exploration}. Restricting ourselves to $b>0$ however, we find that divergence of the power spectrum only occurs at $\omega = 0$ and $\omega = \pi$. 
	
	\subsubsection{Instability at $\omega=0$ and power spectrum of fluctuations}
	
	We investigate the instability that occurs for $\omega = 0$. We plot the points at which this occurs in parameter space in Fig. \ref{fig:pd_henon} as the dashed line. Expanding for small $\omega$, we find
	\begin{align}
		A_1 &\approx \left \langle\frac{1 }{\left[ 1  + 2 a \alpha x^\star - \alpha b  + \omega^2 (1-\alpha b)/2 \right]^2 + \omega^2(1+\alpha b)^2} \right\rangle_S \approx I_1 + I_2 \omega^2, \nonumber \\
		A_2 &\approx \left \langle\frac{ (x^\star)^\kappa }{\left[ 1  + 2 a \alpha x^\star - \alpha b  + \omega^2 (1-\alpha b)/2 \right]^2 + \omega^2(1+\alpha b)^2} \right\rangle_S \approx I_1' + I_2' \omega^2 ,
	\end{align}
	where we define
	\begin{align}
		I_1 = \left \langle\frac{1 }{\left[ 1  + 2 a \alpha x^\star - \alpha b  \right]^2 } \right\rangle_S ,& \,\,\,\,\,
		I_2 = -\left \langle\frac{2\omega^2(1+\alpha^2 b^2 + a \alpha x^\star - a \alpha^2 b x^\star) }{\left[ 1  + 2 a \alpha x^\star - \alpha b  \right]^4 } \right\rangle_S , \nonumber \\
		I_1' = \left \langle\frac{(x^\star)^\kappa }{\left[ 1  + 2 a \alpha x^\star - \alpha b  \right]^2 } \right\rangle_S, &\,\,\,\,\,
		I_2' = -\left \langle\frac{2 (x^\star)^\kappa \omega^2(1+\alpha^2 b^2 + a \alpha x^\star - a \alpha^2 b x^\star)}{\left[ 1  + 2 a \alpha x^\star - \alpha b  \right]^4 } \right\rangle_S.
	\end{align}
	The quantity $x^\star = (\alpha b -1)/(2a\alpha)$ is outside the range $\vert x^\star \vert <2$ for the values of $\alpha$ and $b$ used in Fig.~\ref{fig:pd_henon} and the range of $a$ shown in the figure. Hence, we obtain only $1/\omega^2$ noise, in the case of the H\'enon map, on the approach to the May--Wigner transition. The logistic map is special in this sense. Due to the fact that its eigenvalue spectrum always touches $\lambda = 1$, the logistic map exhibits different types of fluctuations for different types of input noise. As the example of the H\'enon map shows, this is not the case more generally for other maps.
	
	\subsubsection{Instability at $\omega=\pi$}
	We also indicate (with the solid line) the range of system parameters at which instability occurs when $\omega = \pi$ in Fig. \ref{fig:pd_henon}. We may also describe the divergence of the power spectrum on the approach to this transition. Expanding in a similar fashion to Section \ref{sec:one_over_f_squared}, one finds 
	\begin{align}
		A_1 &\approx \left \langle\frac{1 }{\left[ -1  + 2 a \alpha x^\star + \alpha b  + (\omega-\pi)^2 (1+\alpha b)/2 \right]^2 + (\omega-\pi)^2(1+\alpha b)^2} \right\rangle_S \approx H_1 + H_2 (\omega-\pi)^2, \nonumber \\
		A_2 &\approx \left \langle\frac{ (x^\star)^\kappa }{\left[ -1  + 2 a \alpha x^\star + \alpha b  + (\omega-\pi)^2 (1+\alpha b)/2 \right]^2 + (\omega-\pi)^2(1+\alpha b)^2} \right\rangle_S \approx  H_1' + H_2' (\omega-\pi)^2 ,
	\end{align}
	where we define
	\begin{align}
		H_1 = \left \langle\frac{1 }{\left[ 1  -2 a \alpha x^\star - \alpha b  \right]^2 } \right\rangle_S ,& \,\,\,\,\,
		H_2 = -\left \langle\frac{2(\omega-\pi)^2(\alpha b +\alpha^2 b^2 + a \alpha x^\star + a \alpha^2 b x^\star) }{\left[ 1  - 2 a \alpha x^\star - \alpha b  \right]^4 } \right\rangle_S , \nonumber \\
		H_1' = \left \langle\frac{(x^\star)^\kappa }{\left[ 1  - 2 a \alpha x^\star - \alpha b  \right]^2 } \right\rangle_S, &\,\,\,\,\,
		H_2' = -\left \langle\frac{2 (x^\star)^\kappa \omega^2(\alpha b+\alpha^2 b^2 + a \alpha x^\star + a \alpha^2 b x^\star)}{\left[ 1  - 2 a \alpha x^\star - \alpha b  \right]^4 } \right\rangle_S.
	\end{align}
	We therefore see that, analogously to the logistic map, the transition dominated by frequencies at $\omega = \pi$ diverges as $1/(\omega-\pi)^2$ close to the transition.